\documentclass[a4paper,11pt]{article}
\usepackage{jheppub} 
\usepackage{lineno}
\usepackage{physics}
\usepackage{multirow}
\usepackage{longtable} 
\usepackage{comment}

\title{Higgs Physics with the XFEL Compton $\boldsymbol{\gamma\gamma}$ Collider Concept at $\boldsymbol{\sqrt{s}=125}$ GeV}

\author[a,b]{Umar Sohail Qureshi,}
\author[b]{Tim Barklow,}
\author[b, c]{and Ariel Schwartzman}
\affiliation[a]{Department of Physics, Stanford University, Stanford, CA, 94305}
\affiliation[b]{SLAC National Accelerator Laboratory, Menlo Park, CA, 94025}
\affiliation[c]{Department of Particle Physics and Astrophysics, Stanford University, Stanford, CA, 94305}

\emailAdd{uqureshi@cern.ch}

\abstract{We investigate single Higgs production in $\sqrt{s}=125$ GeV $\gamma\gamma$ collisions at the X-Ray Free Electron (XFEL) Compton Collider (XCC) concept and present an analysis targeting the major hadronic, semi-leptonic, and fully leptonic final states of the Higgs boson, including $H\to s\overline{s}$. 
In addition to studying Higgs production at a novel collider concept, our approach couples a novel set transformer-based deep learning framework that acts on particle-flow object point clouds with a genetic algorithm optimizer for signal-background discrimination, yielding significantly higher sensitivity than traditional methods. Our results demonstrate that an XFEL $\gamma\gamma$ collider can probe the Higgs sector with extremely high precision and enable new physics opportunities, complementary to proposed $e^+e^-$ machines. 
}

\begin{document}
\maketitle
\flushbottom

\section{Introduction and Motivation}\label{sec:intro}
The origin of the electroweak symmetry breaking and mass generation within the standard model (SM) \cite{EnglertBrout1964, GuralnikHagenKibble1964, Weinberg1967, Salam1968} is among the most pressing and unsettling problems in theory of elementary particle physics. The Higgs mechanism \cite{Higgs1964a, Higgs1964b} offers a succinct resolution by introducing a neutral, spin-zero field whose vacuum expectation value endows gauge bosons and fermions with mass. The 2012 discovery of the Higgs boson by ATLAS \cite{ATLAS2012Higgs} and CMS \cite{CMS2012Higgs} collaborations at CERN's Large Hadron Collider (LHC) transformed this idea from hypothesis to  reality, shifting the experimental focus 
to precision measurements of its couplings, widths, and quantum numbers\cite{Cepeda2019HLHE}. 

Thus, a central objective of the post-discovery Higgs program, for both present LHC and future colliders, is to probe the scalar's interactions via sub-percent to percent–level determinations of partial widths and couplings in a framework that remains agnostic to beyond-the-SM (BSM) degrees of freedom \cite{Cepeda2019HLHE, ILCTDR2013, FCCeeCDR2019, CEPC_CDR2018, CLIC2018Summary}. This is of particular interest because BSM phenomena can modify the Higgs couplings to SM particles in several ways. For example, $H b\overline{b}$ and $H\tau\tau$ couplings are sensitive to models with additional Higgs doublets \cite{Branco2012TwoHDM}; $Hb\overline{b}$ to supersymmetric particles with left-right mixing \cite{HallRattazziSarid1994}; $HWW$ and $HZZ$ couplings to any singlet that mixes with a Higgs field \cite{RobensStefaniak2015Singlet}; $Hgg$ and $H\gamma\gamma$ to models with vector-like quarks \cite{DawsonFurlan2012VLQ}; and $Ht\overline{t}$ to composite Higgs and top quark models \cite{PanicoWulzer2016Composite}. More importantly, the magnitude of these modifications is restricted to $\mathcal{O}(m_H/M)$ by Haber's Decoupling Theorem \cite{HaberHempfling1993, GunionHaber2003}, where $m_{H}\approx 125$ GeV is the mass of the Higgs boson and $M$ is mass of any new particle that modifies the Higgs coupling. Since searches at the LHC have largely excluded BSM phenomena below the TeV scale \cite{PDG2024}, deviations due to new physics are expected to be limited to percent or even sub-percent-level.

The ATLAS and CMS experiments have an extensive program \cite{ATLAS:2024fkg, Elmetenawee:2023tnc} to measure the aforementioned Higgs properties with precision; however, the hadron collider environment imposes fundamental limitations that are expected to persist even with the data volumes anticipated at the HL-LHC \cite{HLLHCTDR2020, Cepeda2019HLHE}. Outside the relatively clean $H\to \gamma \gamma$ and $H\to 4\ell$ decay modes, Higgs signals are extracted from background-dominated samples (typical  $S/B\sim \mathcal{O}(1/10)$), relying on multivariate classifiers whose performance is ultimately bounded by control of backgrounds and detector effects. Pile-up at the level of $\expval{\mu}=200$ interactions per bunch crossing drives trigger thresholds, contaminates global event observables, and degrades the performance of physics object and event reconstruction. Furthermore, jet-energy scale and resolution, ($b/c$)-tagging calibrations (including mistag asymmetries), missing-energy tails, and luminosity determination set hard floors on systematics. In addition, theory uncertainties from PDFs, $\alpha_s$, renormalization and factorization scales, higher-order QCD/EW corrections, parton-showering, hadronization, underlying-event modeling and interference with continuum backgrounds propagate nontrivially and do not cancel in the ratios often proposed to tame them \cite{rojo2016pdf4lhcrecommendationsrunii}.

Moreover, several key channels are intrinsically challenging: $H\to WW^*$ lacks a narrow mass peak; $H\to b\overline{b}$ and $H\to c\overline{c}$ compete with overwhelming QCD backgrounds; $H\to \mu \mu$ and $H\to Z\gamma^*$, although clean, are statistics-limited with small branching fractions; and $t\overline{t}$ and $VH$ measurements mix production-mode and decay-mode systematics. Most importantly, at hadron colliders, absolute couplings are entangled with the total width: without a decay-independent normalization of the production rate, $\kappa$-fits \cite{CaolaMelnikov2013, KauerPassarino2012} require assumptions about unseen or exotic decays, and indirect width constraints (e.g., off-shell $gg\to ZZ$) rest on theoretical hypotheses equating on and off-shell couplings. Therefore, even the most precise HL-LHC Higgs coupling projections are at best a  few percent. As such, the levels of precision offered by the HL-LHC are largely insensitive to new phenomena, and any apparent anomaly would remain vulnerable to alternative background or modeling interpretations. These considerations motivate complementary measurements in cleaner initial states, especially $e^+e^-$ and $\gamma \gamma$, which provide absolute normalizations (via $\sigma(e^+e^-\rightarrow ZH)$ and $\sigma(e^-\gamma \rightarrow e^-H)$, respectively), polarization control, and lower systematics,  enabling model-independent extractions of Higgs couplings
with much improved precision. 

A much-studied
option 
for $e^+e^-$ linear collider programs is operation in a real $\gamma\gamma$ configuration \cite{Ginzburg1983, Telnov1990, Asner2001PhotonColliderCP}. Instead of the conventional head-on collisions of high-energy $e^+e^-$ beams, one can convert the facility into a photon–photon collider by Compton back-scattering intense laser pulses off the primary lepton beams. This mode offers several advantages, foremost among them the ability to access directly the loop-induced $H\gamma\gamma$ interaction and determine the Higgs charge conjugation and parity (CP) composition through control of the photon polarization. Until recently, these $\gamma\gamma$ collider designs have almost exclusively utilized optical wavelength lasers, which have suffered from a weaker physics case compared to their $e^+e^-$ counterparts.
In these optical $\gamma\gamma$ colliders, the center-of-mass energy of the electron–photon system is traditionally constrained to $x< 4.82$, where $x=4E_e\omega_0/m_e^2 $, $m_e$ is the electron mass and $E_e$ $(\omega_0$) is the electron (laser photon) energy. Larger $x$ values are conventionally considered problematic due to the linear QED thresholds of $x=4.82$ (8.00) for the processes $\gamma \gamma_0 \to e^+e^-$ ($e^-\gamma_0 \to e^- e^+e^-$), where $\gamma$ and $\gamma_0$ refer to the Compton-scattered and laser photon, respectively. However, as shown in Refs.~\cite{Barklow2023XCC, Castelazo:2026iuu}, larger $x$ values, such as those produced by the XCC, carry significant advantages. As $x$ is increased, the $E_{\gamma\gamma}$ luminosity distribution with respect to center-of-mass energy is more sharply peaked near the maximum center-of-mass energy value. Such a distribution significantly increases the production rate of a narrow resonance relative to background processes when the peak is tuned to the resonance mass. 

\begin{figure}
\centering
    \begin{minipage}{0.495\linewidth}
    \centering
        \includegraphics[width=\linewidth]{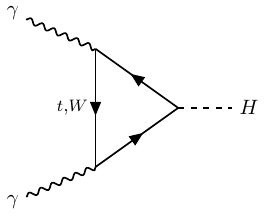}
    \end{minipage}
    \begin{minipage}{0.485\linewidth}
    \centering
        \includegraphics[width=\linewidth]{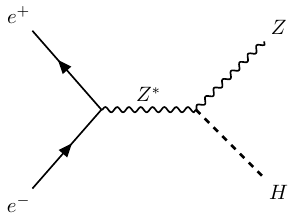}
    \end{minipage}
    \label{fig:feynmandig}
    \caption{Representative Feynman diagrams for (left) the $t/W$ loop-induced production of a Higgs boson in $\gamma\gamma$ collisions, and (right) the Higgsstrahlung process $e^+e^- \to ZH$ via a virtual $Z$ boson.}
\end{figure}

In this study, we devise an analysis strategy for a single SM Higgs boson produced in $\gamma\gamma$ collisions through a loop of virtual top quarks and $W$ bosons \cite{Berger:2025ijd} as shown in the left panel of Fig.~\ref{fig:feynmandig}, and consider all of its major hadronic, leptonic, and semi-leptonic decay modes.  Since the $E_{\gamma\gamma}$ luminosity distribution is sharply peaked near the Higgs resonant mass of $125$ GeV, it is produced copiously, with an expected 10-year yield of about 1.1 million $\gamma\gamma \to H$ events. Furthermore, its energetic decay products lead to final states that can readily be detected in the detector's central regions since the Higgs is produced in isolation, without an associated vector boson. This is in contrast to the tree-level Higgsstrahlung process $e^+e^-\to ZH$ that necessarily includes an on-shell $Z$ boson in the final state as shown in the right panel of Fig.~\ref{fig:feynmandig}. While the recoiling $Z$ in Higgsstrahlung provides a crucial hook for model-independent measurements of the total Higgsstrahlung cross-section and invisible partial width, 
it also reduces the cross section due to phase space suppression,  saps away energy-momentum contributions from the co-produced Higgs -- resulting in softer decay products in less central regions of the detector-- and generally complicates the analysis of Higgs decays.

A key component of this study is also the development of an analysis strategy that utilizes a novel set transformer-based deep learning architecture. The algorithm operates on unordered, event-level sets (point clouds) of final state particle particles containing kinematic, trajectory, and particle identification information. This method has been shown to outperform machine learning-based approaches based on Deep Sets \cite{Zaheer2017DeepSets} and \textsc{EdgeConv} \cite{Wang2018DGCNN}, utilized by the widely used Particle Flow Network \cite{Komiske2019EFN} and \textsc{ParticleNet} \cite{QuGouskos2020ParticleNet} algorithms. An ensemble of these set transformer classifiers is trained, one per background, and the output of each is passed to a genetic algorithm to determine cuts on the neural network output(s) that optimize the metric $S/\sqrt{S+B}$, which is used to determine the overall signal significance for each decay channel examined in the analysis. Our findings indicate that the set transformer-based algorithm significantly enhances signal significance over traditional physics-motivated multivariate analyses, even ML methods such as BDTs and MLPs.

In addition, to enable a more direct comparison with both optical $\gamma\gamma$ collider concepts and prospective $e^+e^-$ Higgs factories (such as the ILC and FCC-ee), we apply the techniques used in this analysis chain to perform optical $\gamma\gamma$ and $e^+e^-$ Higgs analyses. We use similar fast detector simulation, luminosity scenarios, and selection criteria, and set transformer-based machine learning workflow, and compute sensitivities for optical $\gamma\gamma \to H$ production and for associated production $e^+e^- \to ZH$, considering the dominant Higgs decay modes. This treatment better isolates genuine physics differences, enabling direct, quantitative comparisons of expected precisions on $\sigma\times\mathrm{Br}$ and key Higgs couplings across colliders.

The rest of this paper is structured as follows.  Section~\ref{sec:XCC} provides an overview of the XCC and current results from existing $\gamma\gamma$ collider concepts. Section~\ref{sec:sampandsims} explains how the Monte
Carlo simulation samples are produced for this study. Event reconstruction and pre-selection criteria are detailed in Sec.~\ref{sec:eventreco} for all channels considered. In Section~\ref{sec:MLAnalysis}, we discuss the motivation and details of our machine learning workflow, and in Section~\ref{sec:res}, the main results are presented. We conclude with a short discussion and outlook on future work in Section~\ref{sec:conc}.

\section{XCC and Related Work}\label{sec:XCC}
The concept of a $\gamma\gamma$ Higgs factory is not new, with previous $ \gamma\gamma $ collider concepts \cite{Ginzburg:1981vm, Ginzburg:1982yr, Telnov:1989sd, Asner:2001vh} dating back to the 1990s. These collider designs relied on the Compton back-scattering of optical-wavelength lasers from multi-GeV electron beams. These optical Compton colliders (OCC) concepts provided access to processes unique to $ \gamma\gamma $ interactions, chief among them the loop-induced direct production of Higgs bosons, which proceeds directly through the $s$-channel without phase space suppression due to co-produced $Z$-boson, and is this feasible at a much lower $\sqrt{s}$. Still, OCCs suffered
from reduced luminosity at the desired energy peak with a broad, asymmetric $\gamma\gamma$ center-of-mass energy spectrum arising from the Compton back-scattering process at $x<4.8$ and large non-linear QED effects. This spectrum 
complicated precision measurements and limited the physics potential of  
traditional OCC schemes with respect to
$e^+e^-$ colliders.

The XCC represents a paradigm shift from previous $\gamma\gamma$ collider concepts by utilizing electron beams and X-ray free-electron laser pulses to generate high-luminosity, near monochromatic $\gamma\gamma$ collisions. For a comprehensive treatment of the XCC, we direct the interested reader to Ref.~\cite{Barklow2023XCC}, but we briefly go over the salient features relevant to this study. The XCC consists of three primary components: a high-gradient electron linac, an X-ray free-electron laser (XFEL) line, a pair of Compton interaction points (IPCs), a main $\gamma\gamma
$ interaction point (IP) where the photon collisions take place. Polarized electron beams are produced with a cryogenic RF photo-injector and accelerated in
distributed-coupling\cite{Bane:2018fzj} 
 C-band linac structures to $62.9$ {GeV} for
 single-Higgs operation. Mid-linac, the beam is interleaved i.e.~every other bunch at $30$ GeV is directed to the XFEL line, passes through a helical undulator to generate circularly polarized $1$ keV 0.7~Joule X-ray pulses, and is focused to 30~nm
 waists with Kirkpatrick–Baez mirrors. The remaining bunches continue through the linac to reach the full $62.8$ GeV.

\begin{figure}
    \centering
    \includegraphics[width=\linewidth]{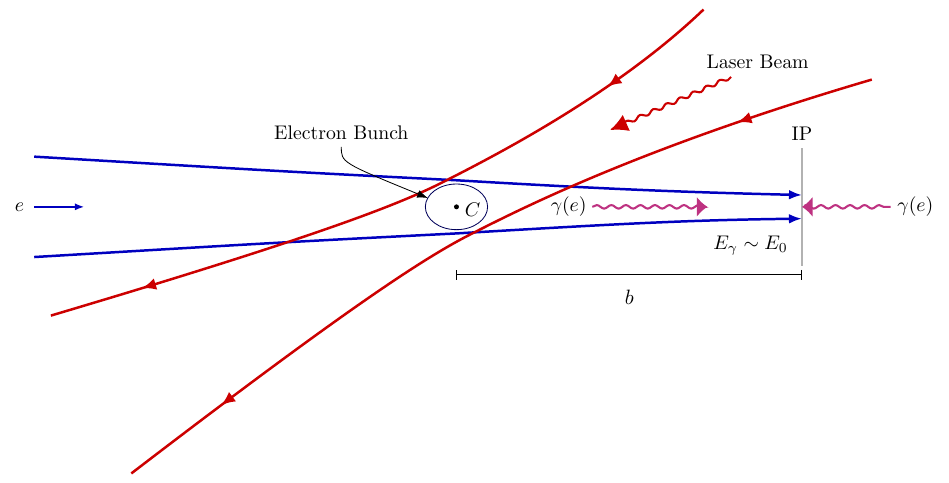}
    \caption{A close-up illustration of the Compton process and associated objects near the IP on the left side of the collider. The right side is identical by symmetry.}
    \label{fig:XCC_IP}
\end{figure}

At the Compton interaction point (IPC), the high-energy electron bunches meet the counter-propagating XFEL pulses, producing inverse-Compton photons that reach the electron beam energy of $62.9~\mathrm{GeV}$.
The scattered photons' energy and polarization are determined by the chosen helicities of the electrons and laser light. To keep the spectrum near-monochromatic and limit nonlinear QED effects, the design operates at low laser strength, with $ \xi^2 \approx 0.1 $. The scattered photons propagate to the $ \gamma\gamma $ interaction point (IP), where two photon beams are brought into head-on collision. The IPC and IP are separated by only $ 60~\mu\mathrm{m} $ to curb transverse spreading from the photons’ angular divergence. Since only about $20\%$ of electrons convert at the IPC, residual $ e^+ e^-$, $ e\gamma $, and $ e^-e^- $ interactions also occur at the IP and are included in the luminosity and background estimates. Fig.~\ref{fig:XCC_IP} shows a close-up of the interaction region.

The XCC adopts a small crossing angle of 2 mrad, with crab cavities rotating the electron beams to achieve effectively head-on collisions. A solenoidal field of 5T shields the detector and confines low-angle background tracks. Downstream extraction is provided by a large-aperture final-focus quadrupole (QD0), and careful masking and beamline layout minimize energy deposition in QD0. This multi-staged design yields a clean, sharply peaked $\gamma\gamma$ spectrum, enabling precision studies at 125 GeV with a path to 280 and eventually even multi-TeV operation (up to about 10 TeV) via wakefield electron acceleration. Independent tuning of the electron beam and the laser system, together with well-characterized interactions at the IPC and IP, provides fine control of the collision environment, establishing a solid basis for the single-Higgs program at $\sqrt{s} = 125$ GeV.

A key parameter characterizing $\gamma\gamma$ collisions is the
$e^--\textrm{laser}$
center-of-mass energy squared in units of the electron mass squared:
 $ x = 4 E_e w_0 / m_e^2 $, where $ E_e $ is the electron beam energy, $ w_0 $  the laser photon energy, and $ m_e $  the electron mass. The Compton edge for the scattered photon energy is given by $ xE_e/(x+1)$. Thus, 
 as $ x $ grows larger, the high-energy portion of the photon spectrum sharpens and the endpoint approaches $ E_e $, which is advantageous since photons near the peak carry the most physics reach. Large $ x $ values, however, trigger linear QED pair creation once $ x > 4.82 $ via $ \gamma \gamma_0 \to e^+ e^- $, with $ \gamma $ the Compton photon and $ \gamma_0 $ the laser photon. To suppress this background and avoid the loss  high energy Compton photons, past optical $ \gamma\gamma $ collider designs operated with $ x < 4.8 $.

\begin{figure}
    \centering
    \begin{minipage}{0.495\textwidth}
        \includegraphics[width=\linewidth]{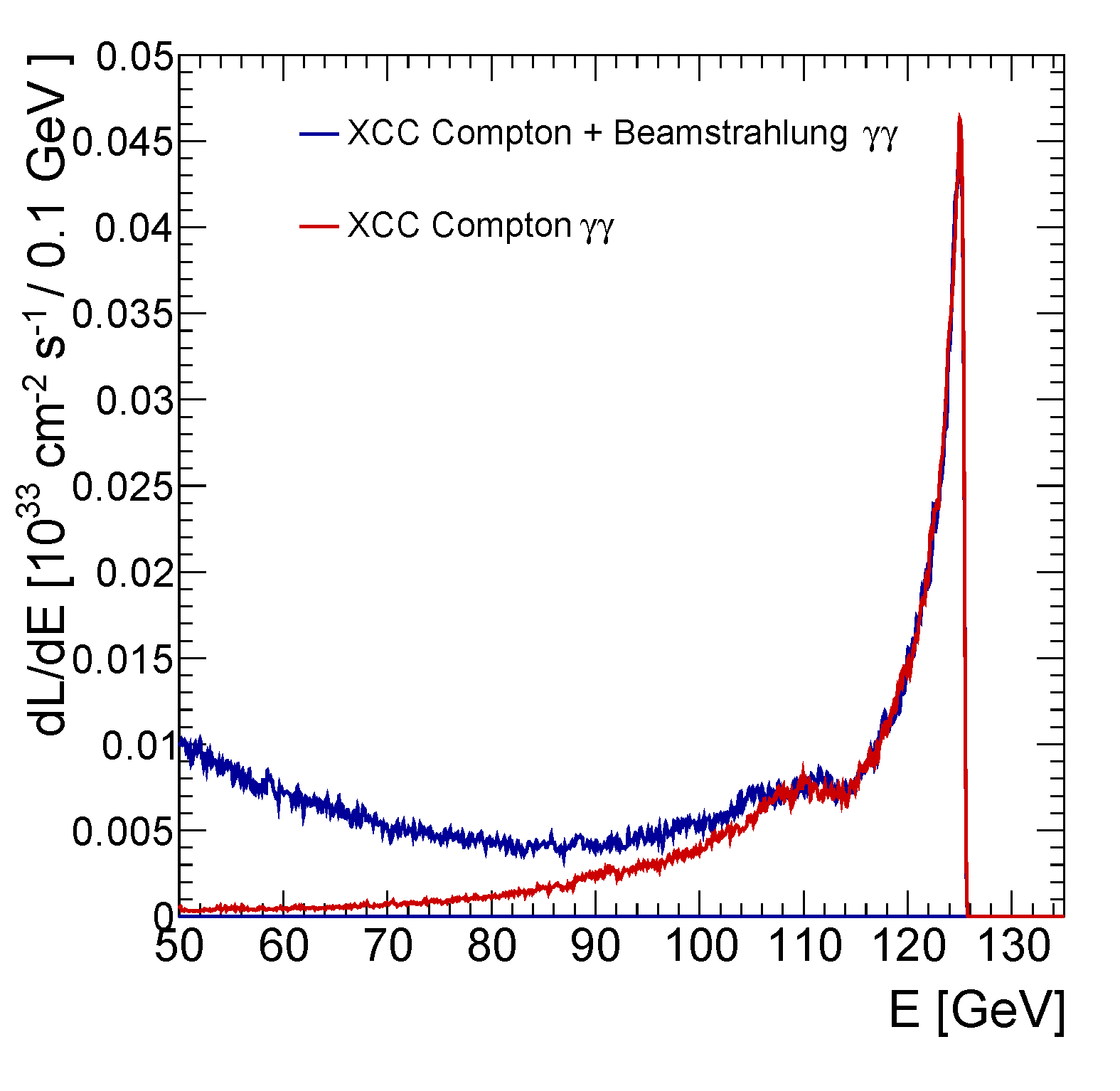}
    \end{minipage}
    \begin{minipage}{0.495\textwidth}
        \includegraphics[width=\linewidth]{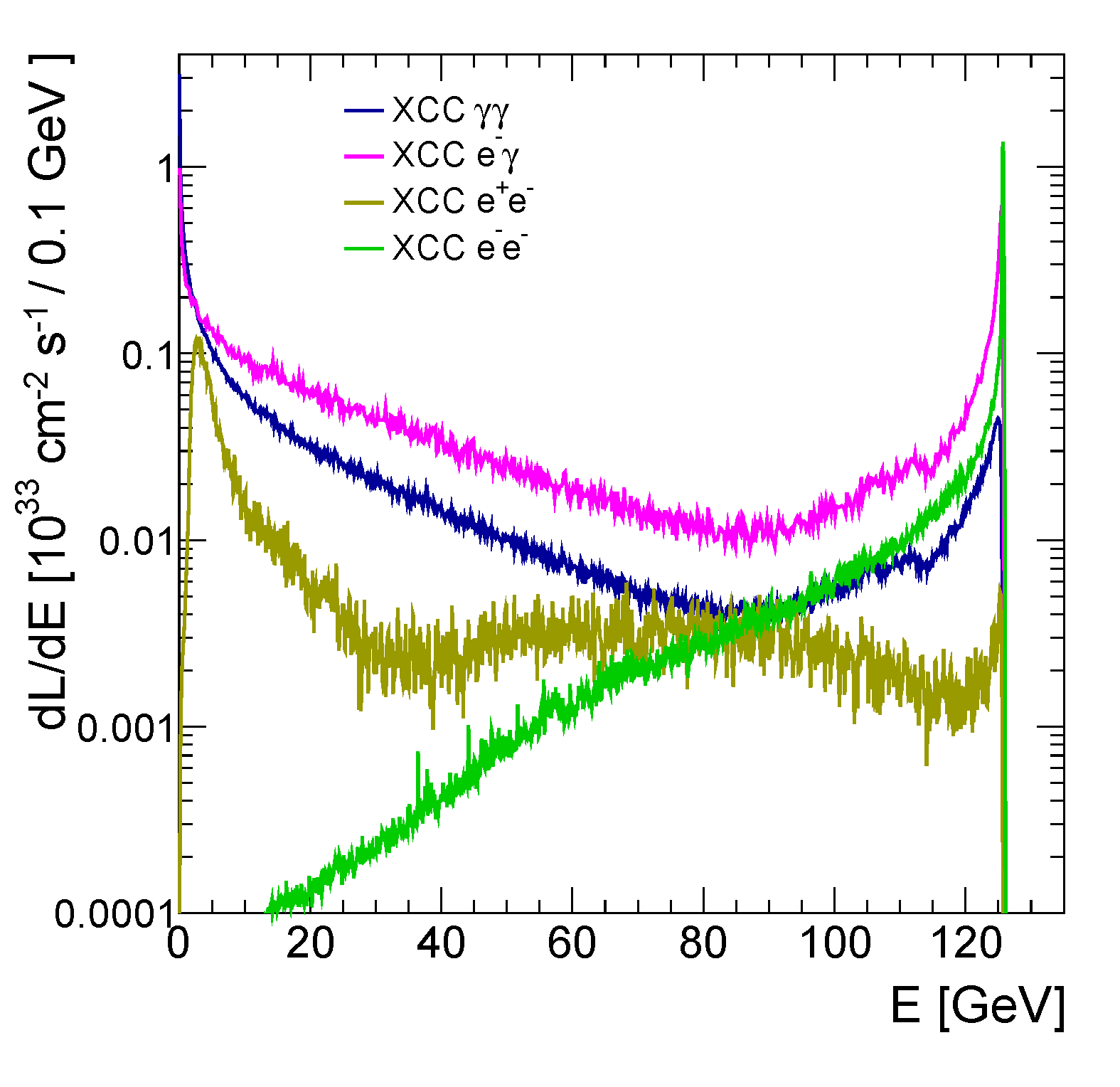}
    \end{minipage}
    \caption{The XCC $\gamma\gamma$ luminosity spectrum with (blue) and without (red) Beamstrahlung photons for the center-of-mass energy range $50\ \mathrm{GeV} < \sqrt{\widehat{s}} < 125\ \mathrm{ GeV}$ (left) and the luminosity spectra for collisions of  $\gamma\gamma$  (blue), $e^-\gamma$ (red) $e^+e^-$ (gold) and $e^-e^-$ (green) for the full center-of-mass energy range $0\ \mathrm{GeV} < \sqrt{\widehat{s}} < 125\ \mathrm{ GeV}$ (right).}
    \label{fig:XCC_lumi}
\end{figure}

With modern x-ray FEL technology, it is possible for $ \gamma\gamma $ colliders to run at very large $ x $ (e.g.~$ x \ge 1000 $), yielding a luminosity spectrum that is sharply peaked near the maximum center-of-mass energy. This concentrated spectrum effectively outweighs losses from linear-QED $ e^+e^- $ production. In addition, choosing electron helicity $ \lambda_e $ and photon circular polarization $ P_c $ with $ \lambda_e P_c > 0 $ 
provides helicity suppressions of the $\gamma_\mathrm{laser}\gamma\rightarrow e^+e^-$ process,
further increasing the peak in the $\gamma\gamma$ luminosity spectrum.  The left-hand side of Fig.~\ref{fig:XCC_lumi} compares the XCC $\gamma\gamma$ luminosity spectra with and without Beamstrahlung photons. The spectra is dominated by Compton photons for $\sqrt{\widehat{s}}> 100\ \mathrm{GeV}$ and by Beamstrahlung photons for $\sqrt{\widehat{s}}< 100\ \mathrm{GeV}$. 
In order to define the luminosity measure most relevant to Higgs physics studies,
the ''integrated $\gamma\gamma$ luminosity'' shall, in the following, refer to the integrated luminosity with $100\ \mathrm{GeV} < \sqrt{\widehat{s}} < 125\ \mathrm{ GeV}$.
The right-hand side of Fig.~\ref{fig:XCC_lumi} compares the luminosity spectra for $\gamma\gamma$, $e\gamma$, $e^+e^-$, and $e^-e^-$  at the XCC over the full center-of-mass energy range $0\ \mathrm{GeV} < \sqrt{\widehat{s}} < 125\ \mathrm{ GeV}$.

Existing OCC $\gamma\gamma\rightarrow H$ simulations conducted with realistic spectra, polarized beams, and fast detector effects find sensitivity to the Higgs couplings underwhelming  compared to their $e^+e^-$ counterparts. In the CLICHE \cite{Asner:2001vh} studies, running at the Higgs resonance with about 200 fb$^{-1}$ per year yields roughly 22,000 Higgs events for $m_H \approx 115$ GeV, enabling statistical precisions of about 2\% for $\sigma(\gamma\gamma \to H)\times\mathrm{Br}(H\to b\bar b)$, 5\% for $H\to W^+W^-$, and near 8\% for $H\to\gamma\gamma$.
TESLA-style \cite{Niezurawski:2002ap} analyses 
project about a $ 2$\% precision on $\Gamma(H\to\gamma\gamma)\times\mathrm{Br}(H\to b\bar b)$, improving to about 1.7\% with neutrino-energy corrections. These studies established that constraints on Higgs $\kappa$ and $\sigma\times \mathrm{Br}$ at an optical $\gamma\gamma$ collider are not competitive with that of electron-positron Higgs factories. Our studies will show that an XFEL $\gamma\gamma$ collider markedly improves the Higgs physics case
with respect to optical $\gamma\gamma$ colliders and can equal or surpass that of $e^+e^-$ colliders


\section{Samples and Simulation}\label{sec:sampandsims}
 Initial beam–beam and beam-laser interactions and non-linear QED effects for $\gamma \gamma$, $e^\pm \gamma$, and $e^\pm e^\mp$ modes are simulated using the \textsc{Cain} \cite{Chen:1994jt} Monte Carlo package.
 This package has been extended in our studies to include the trident process $e^-\gamma_{\ \textrm{laser}}\rightarrow e^-e^+e^-$ in electron laser scattering\cite{barklow2021private} and the proper handling of large-$x$ non-linear QED effects\cite{tauchi2021private}. We generate events with an energy spread extending downward from $\sqrt{s}=125$ GeV and assume an integrated $\gamma\gamma$ luminosity of $\mathcal{L}_\mathrm{int}=500~\textrm{fb}^{-1}$ over a 10-year run, exploiting the sharply peaked luminosity spectrum characteristic of the XCC. 
 Signal and background events are then generated at the parton level, including initial state radiation, using \textsc{Whizard} \cite{Kilian:2007gr, Moretti:2001zz} (v3.1.5) convolved with  $\gamma \gamma$, $e^\pm \gamma$, and $e^\pm e^\mp$ \textsc{Cain} spectra. The \textsc{OpenLoops} \cite{Buccioni:2019sur} (v2.1.4) package is additionally interfaced with \textsc{Whizard} and is used to calculate loop-level processes. To filter background generation in phase space not relevant for this analysis, generator-level cuts of $m_\textrm{hadrons} > 50~\mathrm{GeV}$ ($m_\textrm{visible} > 40~\mathrm{GeV}$) are applied for the hadronic (leptonic and semi-leptonic) backgrounds. These requirements are justified since the number of events failing this cut that would then pass the (more stringent) pre-selections is negligible. This assumption was verified by generating the samples both with and without the cut in earlier iterations of the study. Parton-level events are subsequently interfaced with \textsc{Pythia} \cite{Sjostrand:2006za} (v6.427) for parton showering and hadronization. 
 
 Finally, we model detector response and event reconstruction with \textsc{Delphes}~(v3.5.0) \cite{deFavereau2014Delphes} fast simulation, using an XCC-specific geometry and particle-ID parameterization derived from Ref.~\cite{Behnke:2013lya}. Starting from the SiD configuration developed for the ILC, we produced a tailored XCC card that preserves the baseline tracking and calorimeter settings while adapting acceptance and forward instrumentation to the XCC beam configuration. Object definitions (electrons, muons, jets, and missing transverse momentum) follow the SiD parameterization, including standard smearing, particle propagation, and isolation procedures. The XCC, however, suffers from increased incoherent $e^+e^-$ pair production (IPP) from Bethe–Heitler $\gamma\gamma^* \to e^+e^-$, Breit–Wheeler $\gamma \gamma \to e^+e^-$, and Landau–Lifshitz $\gamma^*\gamma^* \to e^+e^-$ processes compared to its $e^+e^-$ counterparts. While full detector simulation studies are required to accurately determine the occupancy induced by this background and thus the compatibility of existing SiD design, the worst-case end result will be to push the beam pipe and innermost vertex detector to larger radii. Thus, the nominal XCC detector used in this study has the beam pipe moved from $r=12\,\mathrm{mm}$ to $r=15\,\mathrm{mm}$, and the innermost vertex detector from $r=14\,\mathrm{mm}$ to $r=17\,\mathrm{mm}$. \textsc{Cain} simulations demonstrate that the 1.7~cm radius is well above the minimum required as $< 0.01$\% of IPP charged particles  strike a 1.7~cm radius cylindrical inner vertex detector
with angular extent $\abs{\cos\theta}<0.95$, 
which can be compared to 0.1\% at a 250 GeV CoM energy $e^+e^-$ linear collider with a 1.4~cm radius, $|\cos\theta|<0.98$ detector. For this change, we leverage the \textsc{DetectorGeometry} and \textsc{TrackCovariance} modules in \textsc{Delphes}; the former takes as input, a geometric description of the detector's tracking system and the latter, using the geometry, provides an estimate for the track parameters, and associated covariance matrix, for each charged particle. A technical description on the implementation of these modules is provided in Ref.~\cite{Bedeschi:2022rnj}. Moreover, the impact on heavy-flavor tagging as a function of the innermost vertex detector radius is investigated and presented in Appendix~\ref{app:XCCFTAG}.
 
We also incorporate (conservatively)\footnote{See footnote for the $H\to b\overline{b}$ channel in Sec.~\ref{footnote:hbbftag} for details.} improved jet-flavor-tagging performance that reflects recent advances in ML-based taggers as shown in Appendix~\ref{app:XCCFTAG}. Additional details, including the choice of working point efficiencies, mis-tag rates, are provided in later sections. In addition, Particle Flow Objects (PFOs) from the very forward (and backward) regions of the detector, specifically for $|\cos\theta|>0.95$, are not considered in this study. This is due to a large 
flux of  X-ray photons from the IPCs
that end up in these regions,
making instrumentation challenging. We note that this is a very conservative assumption. In practice, dedicated background suppression algorithms are expected to be able to extend the physics object reconstruction to the $0.95 < \abs{\cos\theta} < 0.99$ range, albeit, likely at the cost of some efficiency as a function of increasing $\abs{\cos\theta}$.

Pileup interactions arising from additional $\gamma\gamma \rightarrow \mathrm{hadrons}$ collisions in each bunch crossing in the readout window are also present at the XCC, similar to $e^+e^-$ linear colliders. The total cross section for $\gamma\gamma \rightarrow \mathrm{hadrons}$ varies slowly between $\sqrt{\widehat{s}} = 0.3$ and $400$ GeV, with a mean value of about $0.4~\mu\text{b}$. The expected number of pileup events is obtained by convolving the $\gamma\gamma$, $e^-\gamma$, and $e^-e^-$ luminosity spectra with the total $\gamma\gamma \rightarrow \mathrm{hadrons}$ cross section. The spectra from the $e^-$ beams are also included to account for hadron production from collisions of virtual photons with both virtual and real photons. This procedure yields 
1.1 pileup events per bunch crossing for 
$\sqrt{s} = 125$ GeV.
These pileup levels are much smaller than at the LHC and
similar to that of the ILC operating at $\sqrt{s}=500$~GeV. For this study, pileup events are included using the default \textsc{Delphes} overlay implementation. It is worth mentioning that the impact of pileup is negligible since most of the pileup PFOs are present outside the detector acceptance $|\cos\theta| < 0.95$. Future studies employing full detector simulation with extended coverage up to $|\cos\theta| < 0.99$ that will explicitly investigate pileup subtraction using dedicated machine learning algorithms (such as \textsc{Puppi} \cite{Bertolini:2014bba}, \textsc{Puma} \cite{Maier:2021ymx}, \textsc{Pumml} \cite{Komiske:2017ubm}, and \textsc{DeepSub} \cite{Qureshi:2025ylv}), is forthcoming.

PFOs are subsequently clustered into jets using the $e^+e^-$ inclusive anti-$k_\mathrm{T}$ algorithm \cite{Weinzierl:2010cw} with \textsc{FastJet} \cite{Cacciari:2011ma} (v3.3.4), which uses a sequential recombination scheme based on the metric:
\begin{equation}
\label{eq:yij}
    d_{ij} = 2 \min\left(E_i^{-2}, E_j^{-2}\right) \frac{1 - \cos \theta_{ij}}{1-\cos R},
\end{equation}
where $\theta_{ij}$ is the angular separation between particles $i$ and $j$. For this study, we choose $E_\textrm{min} = 8$ GeV and a relatively large $R=1.5$ to mitigate long low-jet-mass tails arising from hadronic activity left outside of the jets. 

Our signal events consist of $\gamma\gamma \to H$ followed by decay.
The total $\gamma\gamma \to H$ production cross-section is 
3~nb 
at $\sqrt{s} = 125$~GeV assuming a Higgs of mass $m_H=125$~GeV, width $\Gamma_H=4$~MeV, and  monochromatic $\gamma$ beams with 100\% circular polarization oriented with $J_z=0$.  This cross-section cannot be used directly to determine the Higgs event rate as it is impossible to build a $\gamma\gamma$ collider with a beam energy spread much less than 4~MeV.  
To calculate the number of Higgs events one has to convolve the cross-section for $\gamma\gamma\rightarrow H$ with the $\gamma\gamma$ luminosity spectrum.  This process yields
1.1 million $\gamma\gamma \to H$ events over a 10-year period for the
XCC design in Ref.~\cite{Barklow2023XCC}.  

All standard model backgrounds were considered for this study, grouped into 5 hadronic and 13 leptonic / semi-leptonic categories.
These backgrounds were generated using an identical simulation methodology as the signal. Table \ref{tab:hadbacks1} contains the expected number of events for each background. 

\begin{table}
\centering
    \begin{tabular}{ ll   }
\hline
 \textbf{Background Process} & \textbf{Expected Yield}  \\
 \hline 
$\gamma \gamma \to qq\overline{q}\overline{q}$ & 381,829  \\
$\gamma\gamma \to q\overline{q}$ & 513,755,395\\
$\gamma e \to q\overline{q}e $ & 74,887,283\\
$\gamma e \to q\overline{q}\nu $ & 626,064\\
 $e^+e^- \to q\overline{q}$ & 459,001,600  \\ 
\hline 
$\gamma\gamma \to \ell \ell \overline{\ell} \overline{\ell} $ & 225,100,651\\
$\gamma\gamma \to \ell \ell \overline{q} \overline{q} $ & 9,771,761\\
$\gamma\gamma \to \ell \ell $ &                 2,135,184,138\\
$\gamma\gamma \to \gamma\gamma$ & 94,130\\
$\gamma\gamma \to \ell \overline{\nu} q\overline{q}$ & 61,708\\
$ \gamma e \to \ell \ell \ell $ & 1,189,141,510\\
$\gamma e \to \gamma e$ & 2,566,604,060 \\
$\gamma e\to e \ell \ell$ & 2,409,926,020\\
$ e^+ e^- \to \ell \overline{\ell} \ell \overline{\ell} $ & 456,869\\ 
$e^+e^- \to \ell \overline{\ell} q\overline{q} $ & 865,248\\
$e^-e^- \to \ell \ell \ell \overline{\ell}$ & 5,880,017\\ 
$e^+e^- \to \ell \overline{\ell }$       &      2,245,722,200\\
$e^-e^- \to e^-e^-$ & 27,825,925,000\\
\hline
\end{tabular}
\caption{List of hadronic (top) and (semi-)leptonic (bottom) background processes included in the analysis and their expected yields for an integrated 
$\gamma\gamma$ luminosity of $\mathcal{L}_\mathrm{int} = 500~\text{fb}^{-1}$ corresponding to a 10-year run-time at $\sqrt{s} = 125$~GeV.\label{tab:hadbacks1}}
\end{table}

\section{Event Reconstruction and Preselection Criteria}\label{sec:eventreco}
Before being passed to the machine learning workflow, the simulated signal and background events are 
filtered to isolate signal-like topologies while suppressing the large backgrounds present in $\gamma\gamma$ collisions. We refer to these filters as ``pre-selections'' with the particular pre-selection criteria contingent on the channel being considered, and are described in the remainder of this section. For any channel, background processes contributing to $\ll 1\%$ of the total background are not listed in this section, but are still considered as inputs for the machine learning algorithm in the later stages of the analysis. 

\subsection{\texorpdfstring{\boldmath $H\to b\overline{b}$}{}}

\begin{figure}
    \centering
    \begin{minipage}{0.495\textwidth}
        \includegraphics[width=\linewidth]{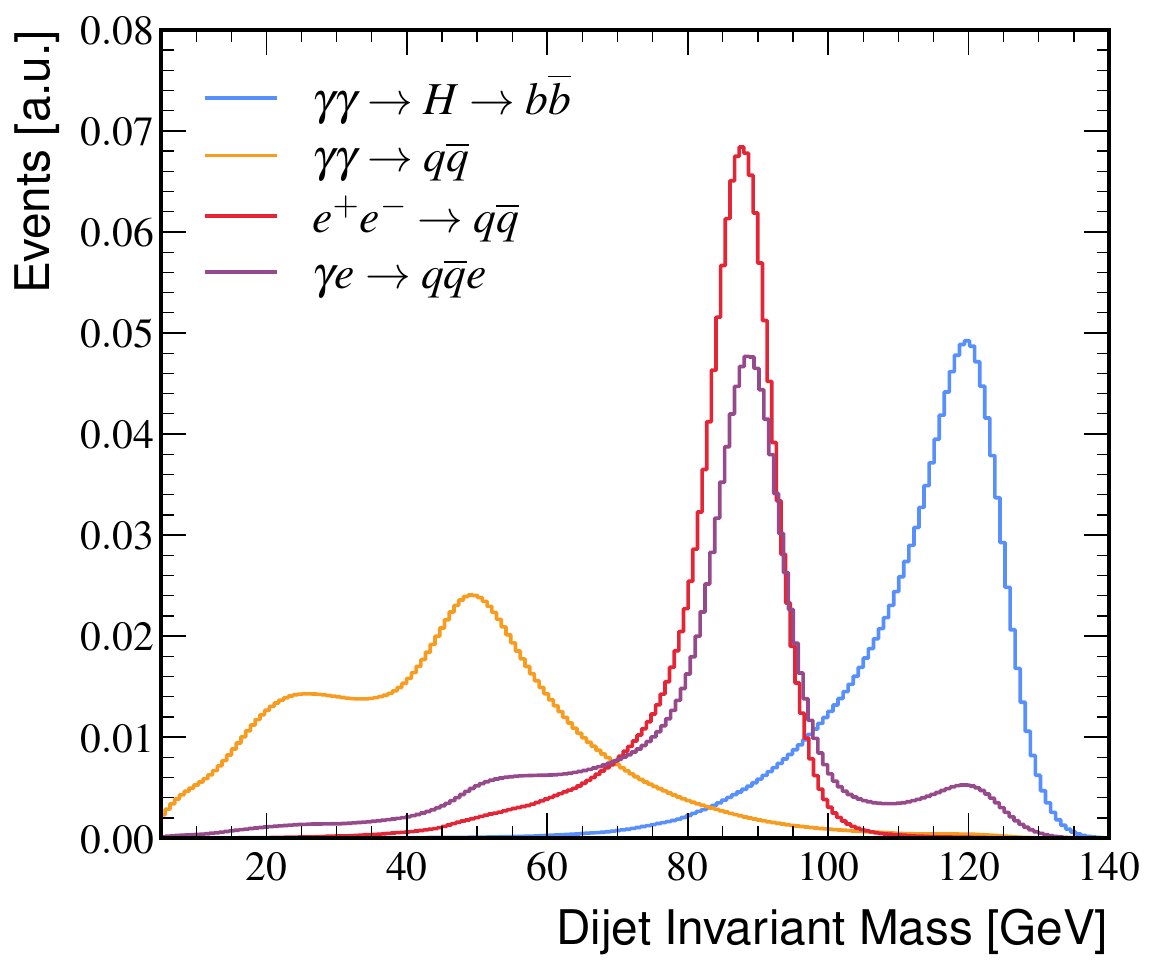}
    \end{minipage}
    \begin{minipage}{0.495\textwidth}
        \includegraphics[width=\linewidth]{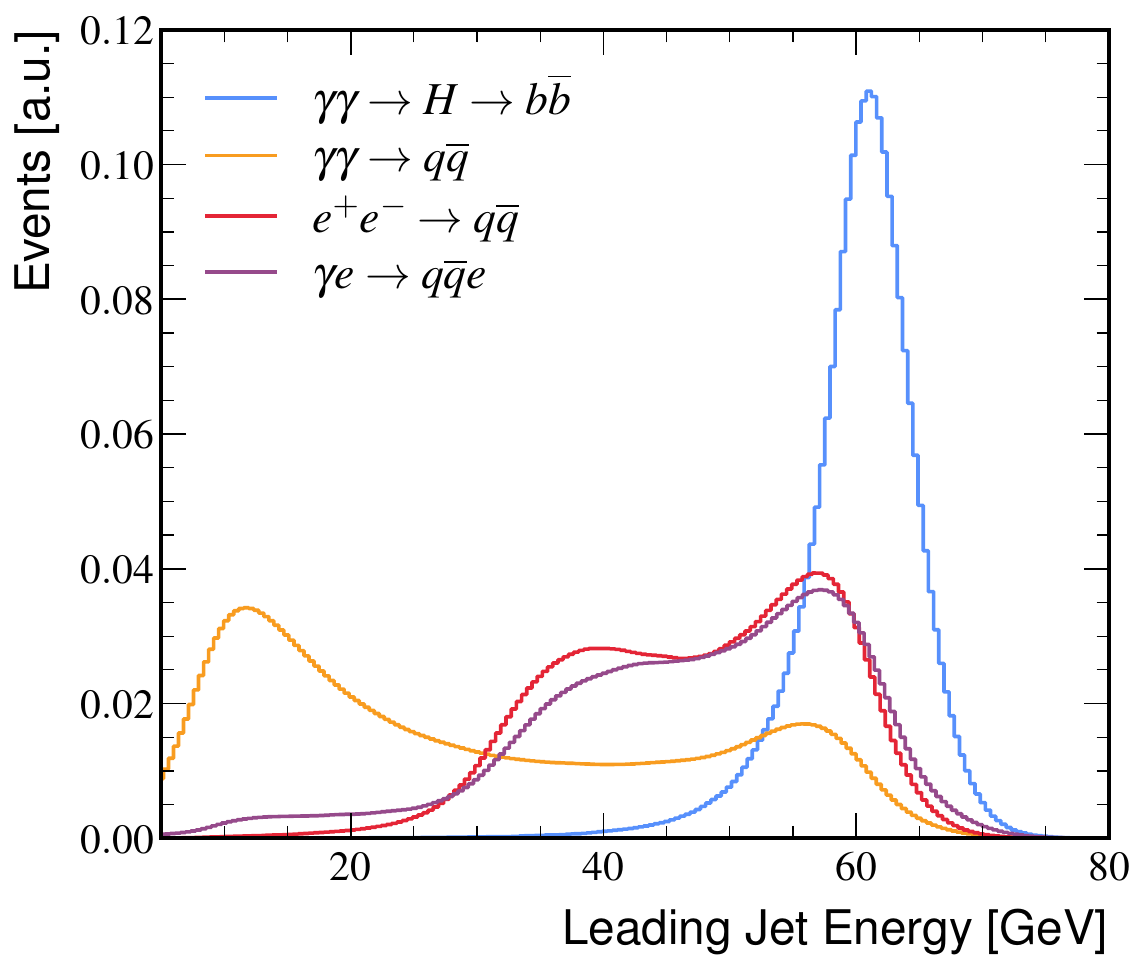}
    \end{minipage}
    \begin{minipage}{0.495\textwidth}
        \includegraphics[width=\linewidth]{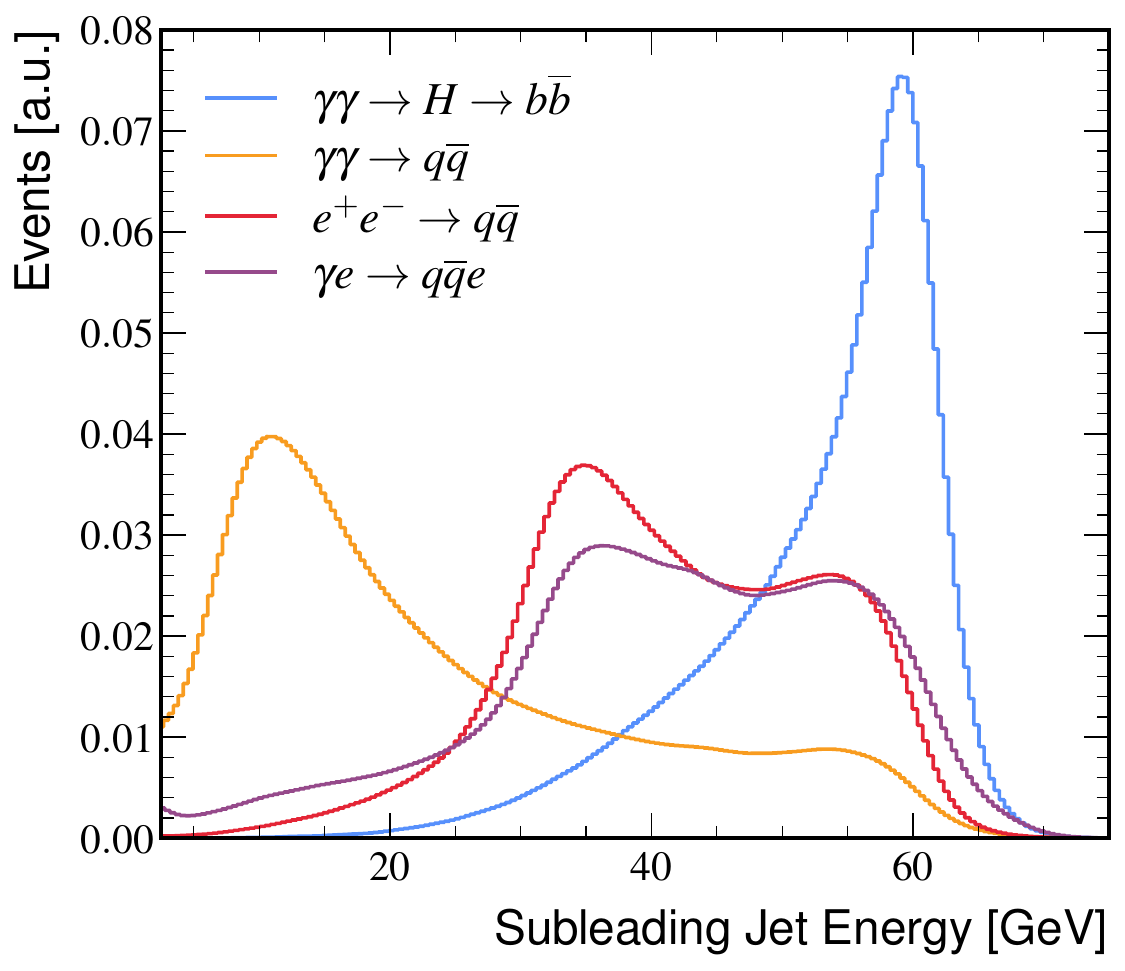}
    \end{minipage}
    \begin{minipage}{0.495\textwidth}
        \includegraphics[width=\linewidth]{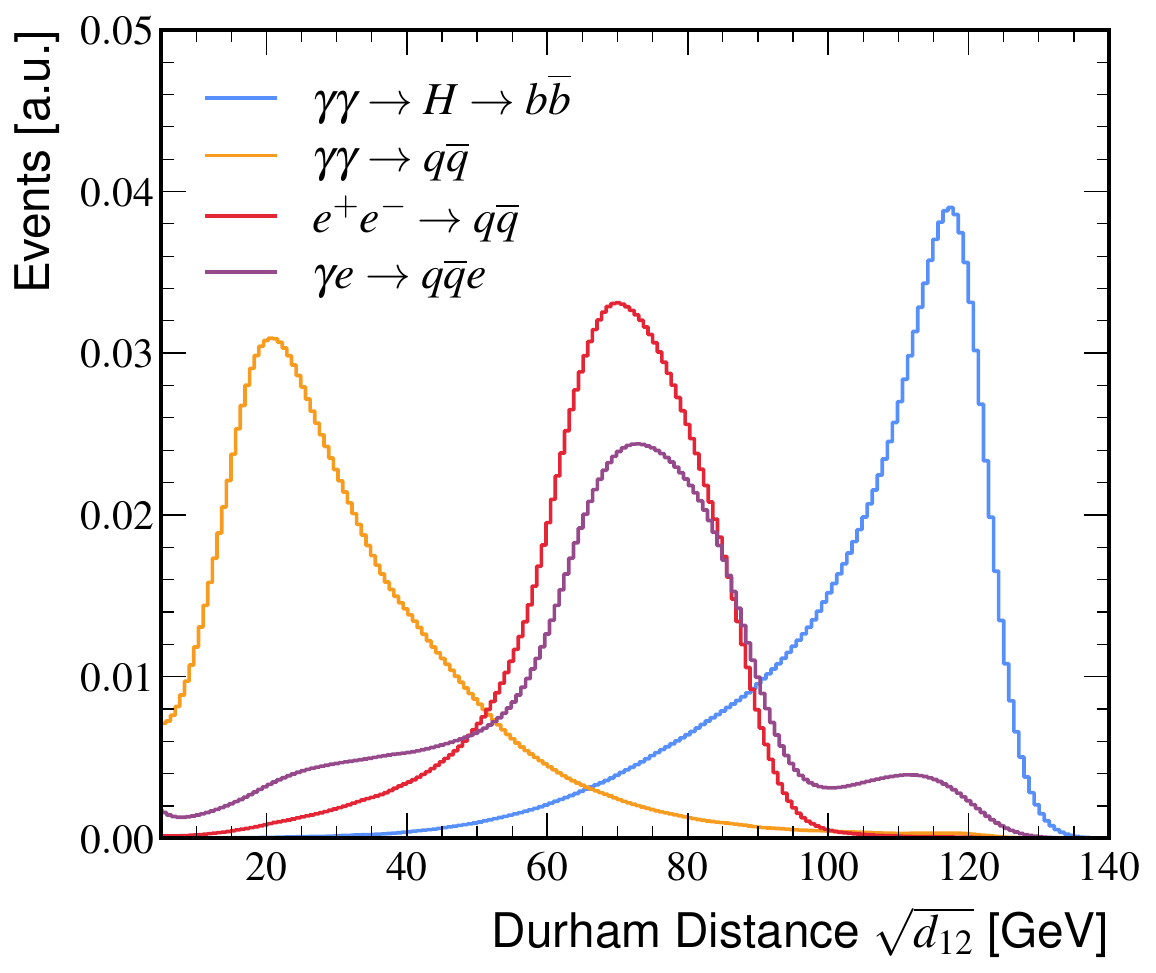}
    \end{minipage}
    \caption{Distributions of the dijet invariant mass (top left), leading jet energy (top right), subleading jet energy (bottom left), and the Durham distance between the jets (bottom right) for the $H\to b\overline{b}$ signal and dominant backgrounds prior to any selection cuts. The distributions are normalized such that the area under the curve is unity.}
    \label{fig:Hbb_plots}
\end{figure}
The $H\to b\overline b$ channel is critical for any Higgs analysis since it often provides the highest sensitivity due to the large $H\to b\overline{b}$ branching fraction ($\sim58\%$). For physics object reconstruction, we first require two $b$-tagged jets with $E>8$ GeV to ensure well-reconstructed, central objects and to suppress soft/forward QCD fragments. Events containing an isolated, well-identified lepton with $E>10$ GeV are vetoed to remove leptonic and semileptonic backgrounds. To illustrate the impact of our preselection filters, Fig.~\ref{fig:Hbb_plots} shows a few representative kinematic distributions (normalized to unity) for this channel before any further cuts are applied. The top-left plot shows the reconstructed dijet invariant mass. The soft $\gamma\gamma \to q\overline{q}$ continuum dijet mass spectrum arises from point-like $t$ and $u$-channel production. On the other hand, jet candidates from the $e^+e^-\to q\overline{q}$ and $\gamma e\to q\overline{q}e$ backgrounds originate from $Z/\gamma^*$ decays, and therefore, have a peak around the $m_Z\approx 91$ GeV resonance. Similarly, the top-right and top-left plots show the leading (highest-energy) and subleading (second-highest energy) jet energy distributions. Finally, the bottom-left plots shows the Durham distance $\sqrt{d_{12}}$ between the leading and subleading jets, defined as:
\begin{equation}
    d_{12} = 2\min\left\{E_1^2, E_2^2\right\}\left(1-\cos \theta_{12}\right).
\end{equation}
For $b$-tagging, we conservatively\label{footnote:hbbftag}\footnote{The actual performance of our flavor tagging algorithm, even after pushing the inner vertex detector to $r=17\,\mathrm{mm}$ in \textsc{Delphes}, is substantially better than the working point used here. In particular, for an 85\% $b$-jet efficiency, we observe mistag rates of $<1\%$ for $c$-jets and $<0.5\%$ for light quark jets. While full simulation studies are needed to fully understand the impact of the IPP background, we assume this aggressive degradation (comparable with LHC numbers) as a worst-case scenario. The same is assumed for all flavor tagging working points used in this study.} assume a ``Medium'' working point of the \textsc{ParticleNet} algorithm (see Appendix~\ref{app:XCCFTAG}) which gives an 85\% $b$-jet efficiency with a $5\%$ mistag rate for charm $1\%$ mistag rate for light quark and gluon $=\{u,d,s,g\}$ jets. The choice of the $b$-tagging working point is determined through an optimization procedure which maximizes signal significance. Charm and light-flavor contamination is further reduced by the two-tag requirement at a working point optimized for strong $c$ and light rejection; squaring the per-jet mistag efficiency provides substantial suppression of $\gamma\gamma\to q\bar q$ and $\gamma\gamma/e^+e^-\to q\bar q$ for $q\neq b$. Finally, we require $E_{\mathrm T}^{\rm miss}<18$ GeV.


\begin{table}
\centering
    \begin{tabular}{ l l l}
\hline
 \textbf{Process} & \textbf{Before Preselection} & \textbf{After Preselection}  \\ \hline
 $\gamma\gamma \to H\to b \overline{b}$ & 635,800 & 353,392 \\
 \hline 
$\gamma \gamma \to qq\overline{q}\overline{q}$ & 381,829 & 2,367 \\
$\gamma\gamma \to q\overline{q}$ & 513,755,395& 11,719\\
$\gamma e \to q\overline{q}e $ & 74,887,283& 67,792\\
 $e^+e^- \to q\overline{q}$ & 459,001,600& 595,378 \\ 
 \hline
\end{tabular}

\caption{Number of events before and after the $H\to b\overline{b}$ pre-selection filters for signal and backgrounds. \label{tab:hbb}}
\end{table}

\subsection{\texorpdfstring{\boldmath $H\to c\overline{c}$}{}}
\begin{figure}
    \centering
    \begin{minipage}{0.495\textwidth}
        \includegraphics[width=\linewidth]{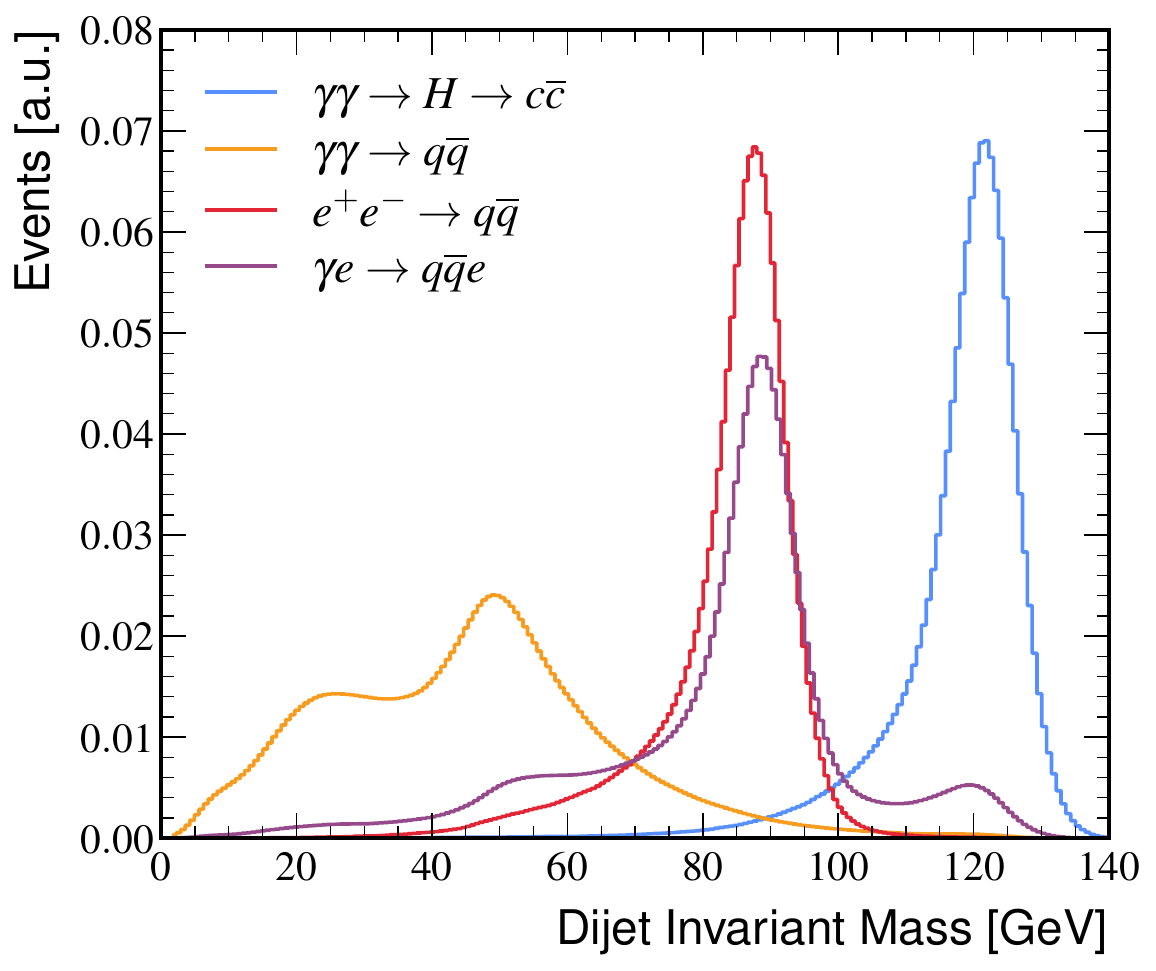}
    \end{minipage}
    \begin{minipage}{0.495\textwidth}
        \includegraphics[width=\linewidth]{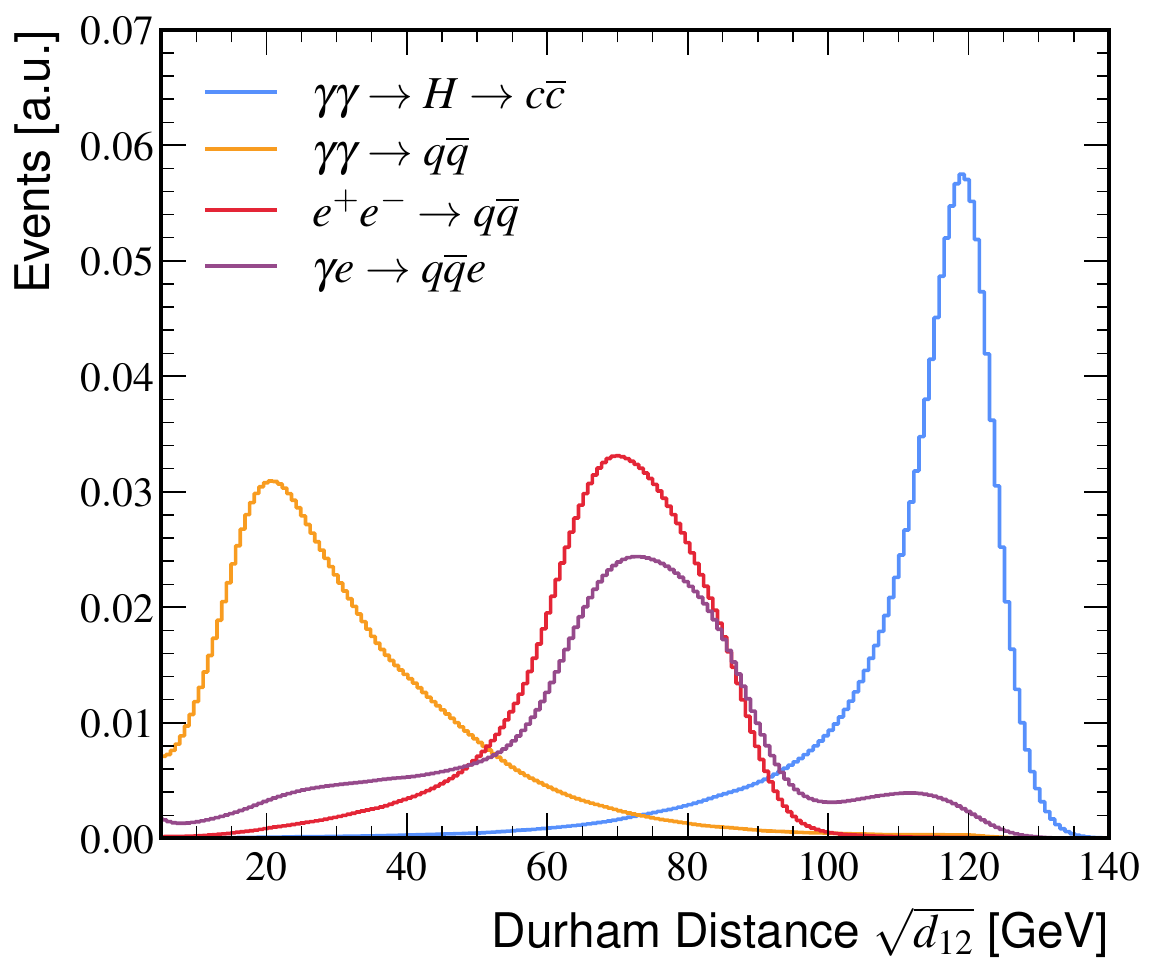}
    \end{minipage}
    \caption{Distributions of the dijet invariant mass (left), leading jet energy and the Durham distance between the jets (bottom right) for the $H\to c\overline{c}$ signal and dominant backgrounds prior to any selection cuts. The distributions are normalized such that the area under the curve is unity.}
    \label{fig:Hcc_plots}
\end{figure}
The $H\to c\overline{c}$ channel has traditionally been challenging at $\gamma\gamma$ colliders due to large $\gamma\gamma \to c\overline{c}$ background. This stems from the charm quark carrying a $ 2e/3 $ charge compared to $e/3$ for the bottom quark, which leads to a factor of 16 (23) larger $\gamma\gamma \to c\overline{c}$ background before (after) QCD corrections. However, much of this background can be effectively removed by cuts on event kinematic distributions. Similar to the $b\overline{b}$ channel, we require two jets with $E>8$ GeV, and veto events with isolated leptons with $E>10$ GeV. To mitigate multi-prong topologies and jets with soft, wide angle radiation, we impose a minimum Durham distance $\sqrt{d_{12}}>85$ GeV between the two jets, which disfavors soft QCD splittings and configurations resembling merged $W/Z$ decays, without affecting the back-to-back in $\phi$ topology of the signal dijets. The dijet invariant mass is then required to satisfy $m(jj)>100$ GeV, efficiently removing the steeply falling low-mass $\gamma\gamma\to c\overline{c}$ continuum and events near the $Z$ peak ($m({jj})\approx90$ GeV) while retaining the shoulder toward the Higgs resonance. As before, we require low missing transverse energy, $E_{\mathrm T}^{\rm miss}<18$ GeV, which suppresses channels with prompt neutrinos and semileptonic heavy-flavor decays yet has minimal impact on fully hadronic $H\to c\bar c$ events.   The effect of the $m(jj)$ and $\sqrt{d_{12}}$ cuts can be inferred from the left and right plots in Fig.~\ref{fig:Hcc_plots}.

\begin{table}[h]
\centering
    \begin{tabular}{ l l l}
\hline
 \textbf{Process} & \textbf{Before Preselection} & \textbf{After Preselection}  \\ \hline
 $\gamma\gamma \to H\to c \overline{c}$ & 33,000 & 22,203 \\
 \hline 
$\gamma \gamma \to qq\overline{q}\overline{q}$ & 381,829 & 3,597 \\
$\gamma\gamma \to q\overline{q}$ & 513,755,395& 48,909\\
$\gamma e \to q\overline{q}e $ & 74,887,283& 103,804\\
 $e^+e^- \to q\overline{q}$ & 459,001,600& 660,037 \\ 
 \hline
\end{tabular}
\caption{Number of events before and after the $H\to c\overline{c}$ pre-selection filters for signal and backgrounds. \label{tab:hcc}}
\end{table}

We select events with two $c$-tagged jets and suppress heavy-flavor look-alikes with an explicit $b$-tag veto; any event containing a jet passing the $b$-tag working point is discarded to reduce $H\to b\bar b$ and generic $b$-rich backgrounds. For charm jet identification, our strategy uses a ``Loose''  configuration of the \textsc{ParticleNet} algorithm, which gives a $90\%$ charm efficiency with a $5\%$ mistag rate for other quark jets. Overall, the strategy isolates a $c\bar c$ resonance, yielding a near $70\%$ efficiency for $H\to c\bar c$, while significantly suppressing backgrounds as summarized in Table \ref{tab:hcc}.

\subsection{\texorpdfstring{\boldmath $H\to g{g}$}{}}
\begin{figure}
    \centering
    \begin{minipage}{0.495\textwidth}
        \includegraphics[width=\linewidth]{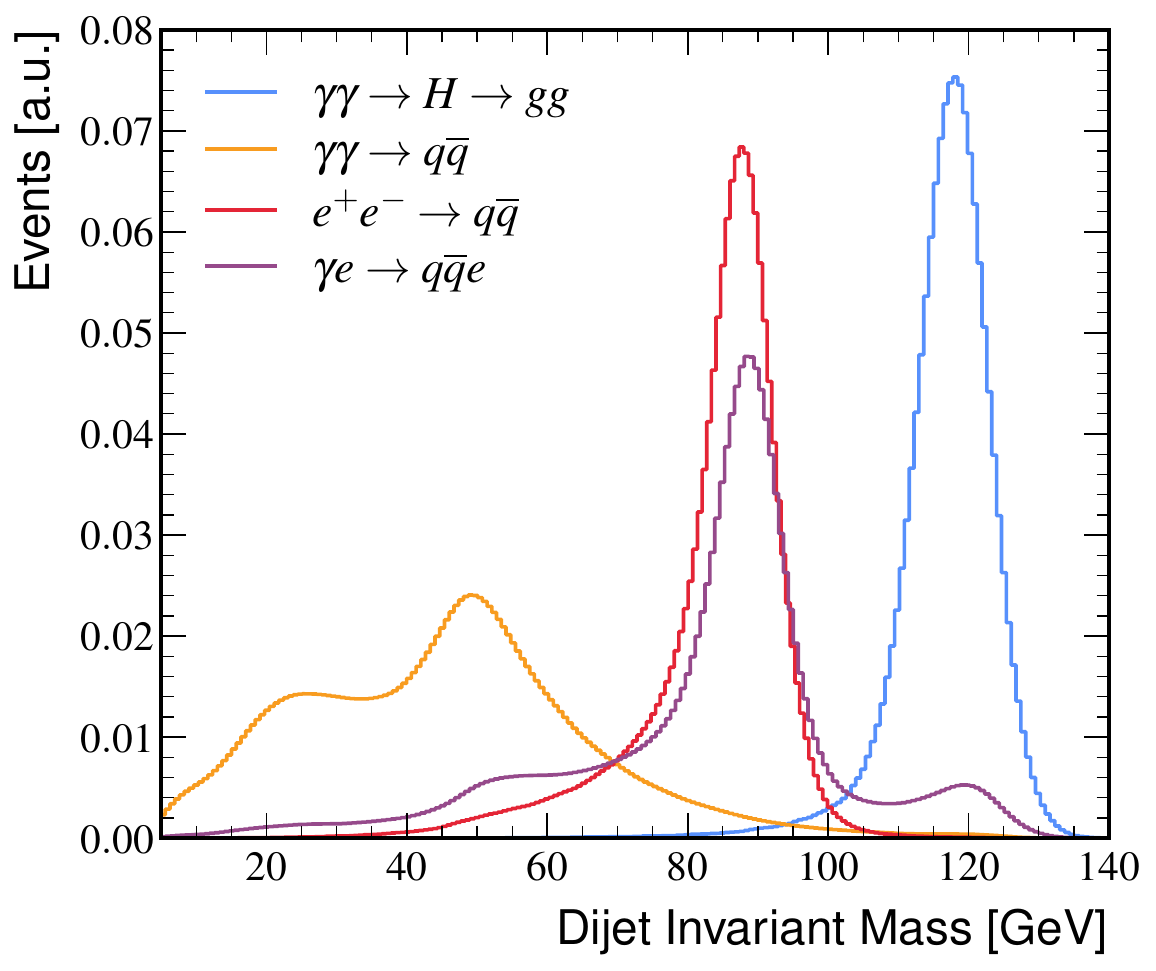}
    \end{minipage}
    \begin{minipage}{0.495\textwidth}
        \includegraphics[width=\linewidth]{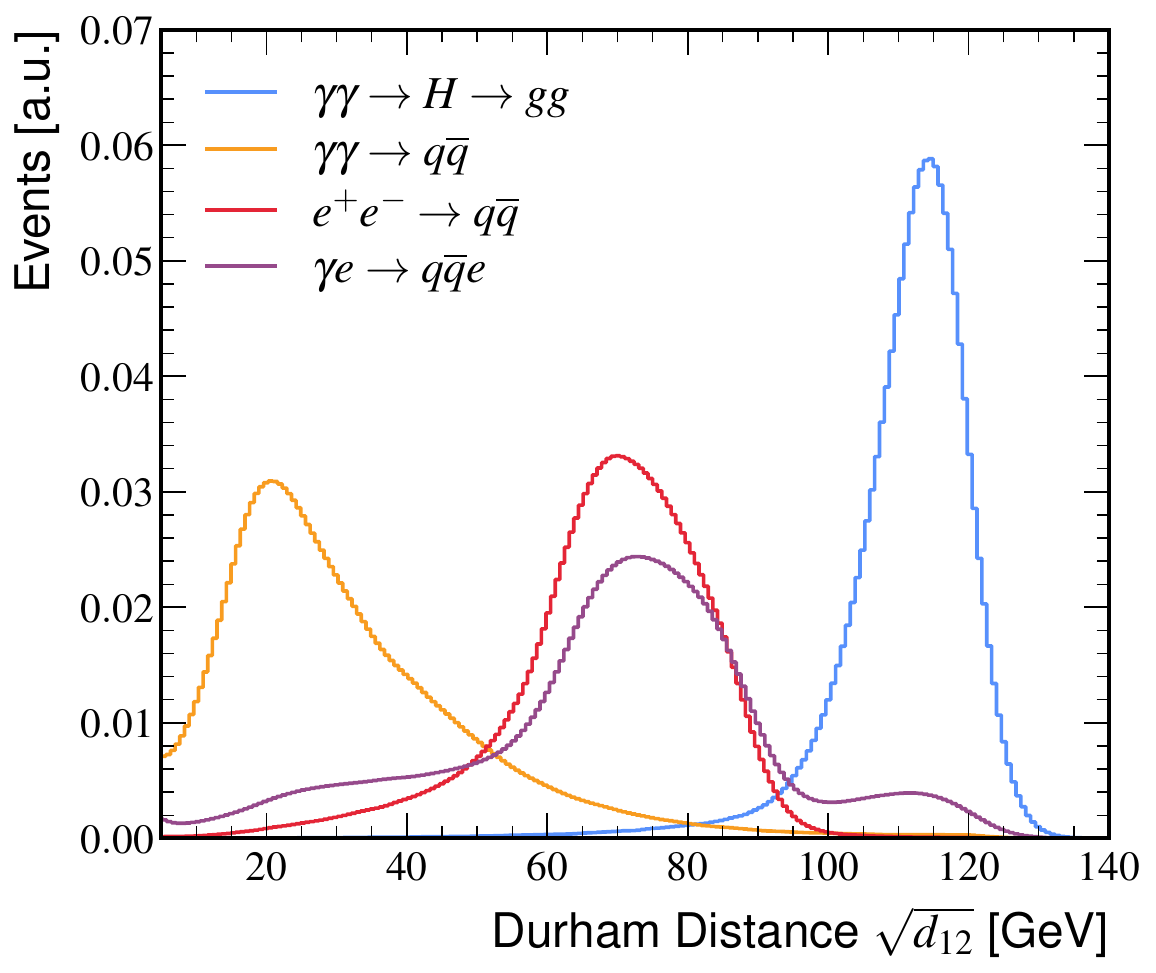}
    \end{minipage}
    \caption{Distributions of the dijet invariant mass (left), leading jet energy and the Durham distance between the jets (bottom right) for the $H\to gg$ signal and dominant backgrounds prior to any selection cuts. The distributions are normalized such that the area under the curve is unity.}
    \label{fig:Hgg_plots}
\end{figure}

\begin{table}[h]
\centering
    \begin{tabular}{ l l l}
\hline
 \textbf{Process} & \textbf{Before Preselection} & \textbf{After Preselection}  \\ \hline
 $\gamma\gamma \to H\to gg$ & 94,602 &59,833\\
 \hline 
 $\gamma\gamma \to H\to b \overline{b}$ & 635,800 & 9,347\\
$\gamma\gamma \to H\to W W^* \to qq\overline{q}\overline{q}$ & 166,664 & 31,463 \\
$\gamma\gamma \to H\to ZZ^*\to qq\overline{q}\overline{q}$ & 20,552& 5,386 \\
 \hline
$\gamma \gamma \to qq\overline{q}\overline{q}$ & 381,829 & 15,094 \\
$\gamma\gamma \to q\overline{q}$ & 513,755,395& 50,802\\
$\gamma e \to q\overline{q}e $ & 74,887,283& 198,899\\
 $e^+e^- \to q\overline{q}$ & 459,001,600& 881,392 \\ 
 \hline
\end{tabular}
\caption{Number of events before and after the $H\to gg$ pre-selection filters for signal and backgrounds. \label{tab:hgg}}
\end{table}

For the $H\to gg$ channel we select a dijet topology consistent with two gluon-initiated jets while rejecting heavy-flavor and electroweak backgrounds. Events must contain two or three jets with $E>8$ GeV; the third jet accommodates soft, wide-angle gluon ISR/FSR without discarding good signal. We veto isolated leptons with relative isolation $< 0.15$ and $E>15$ GeV as before to suppress channels with prompt leptons from $\gamma e$ and electroweak processes. To favor gluon-like topologies, we require a charged-track multiplicity $N_{\rm track} > 25$ at the event level, exploiting the higher particle multiplicity of gluon jets relative to quark jets, owing to their larger Casimir color factor. Heavy-flavor contamination is removed with explicit $(b/c)$-tag vetoes: any event containing a jet passing the $(b/c)$-tag at the ``Medium'' working point of the \textsc{ParticleNet} algorithm is rejected. To suppress multi-jet and boosted vector-boson-fusion–like configurations, we impose $\sqrt{d_{12}}>80$ GeV between the two leading jets. Missing transverse energy is required to be modest, $E_T^{\rm miss}<18$ GeV, reducing final states with prompt neutrinos or semi-leptonic heavy-flavor decays. Finally, we select events with a leading dijet invariant mass $m(jj)>95$ GeV.  The effect of the $m(jj)$ and $\sqrt{d_{12}}$ cuts can be inferred from the left and right plots in Fig.~\ref{fig:Hgg_plots}.  Collectively, these criteria yield $\approx 70\%$ $H\to gg$ efficiency, with controlled quark and heavy-flavor backgrounds and virtually no sensitivity to leptonic or semi-leptonic channels, as shown in Table \ref{tab:hgg}. Notably, however, the preselection is inclusive enough that other Higgs decays leak in and must be considered as background. In particular, hadronic decay modes of $H\to WW^*$ and $H\to ZZ^*$ frequently produce two or three-jet topologies with low $E_{\mathrm T}^{\rm miss}$, no isolated leptons, and pass the $\sqrt{d_{12}}$ and mass requirements; their jets can also be mis-tagged or untagged in ways that mimic the target final state. Consequently, these channels survive at appreciable rates, and must be treated as irreducible. However, as will be seen in later sections, these backgrounds are effectively removed using machine learning.

\subsection{\texorpdfstring{\boldmath $H\to s\overline{s}$}{}}
\begin{figure}
    \centering
    \begin{minipage}{0.495\textwidth}
        \includegraphics[width=\linewidth]{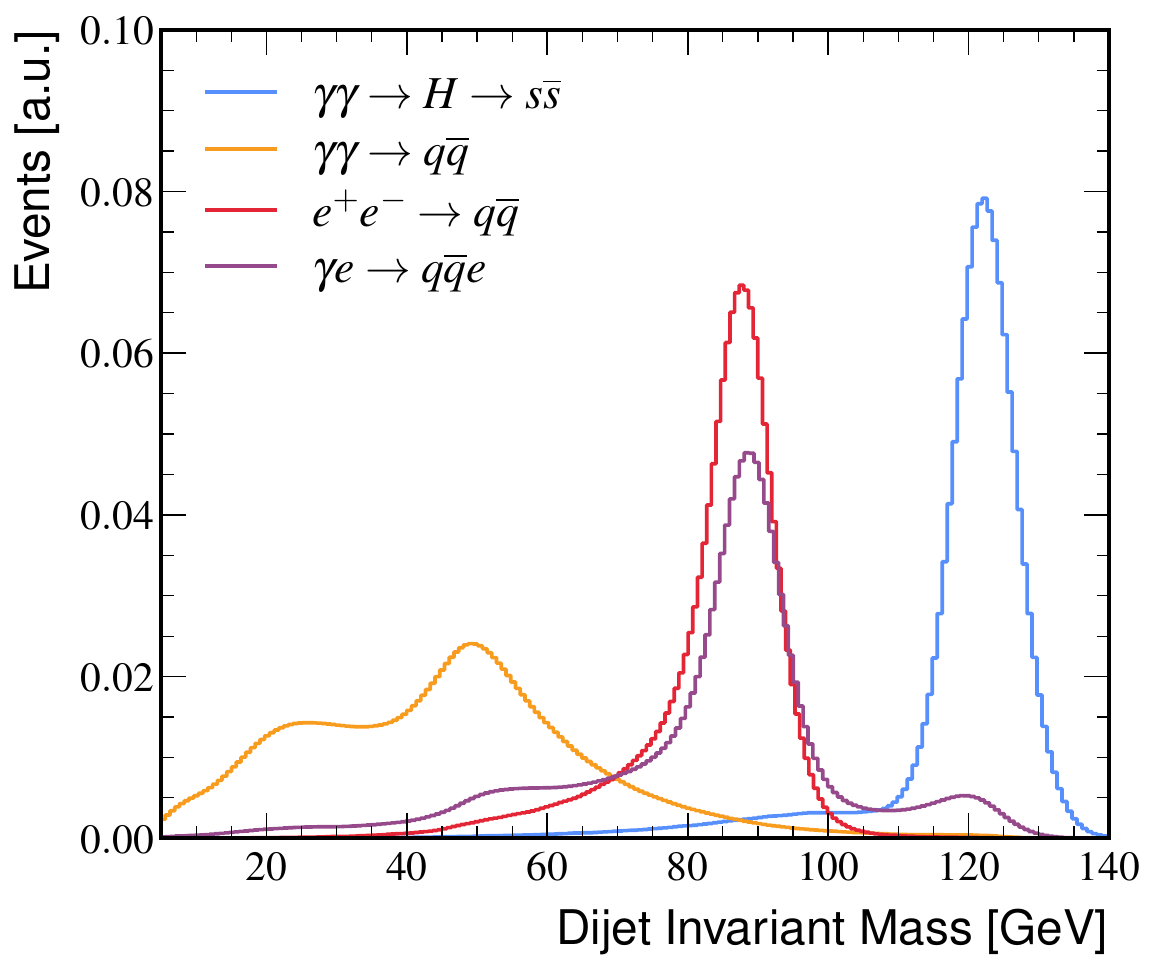}
    \end{minipage}
    \begin{minipage}{0.495\textwidth}
        \includegraphics[width=\linewidth]{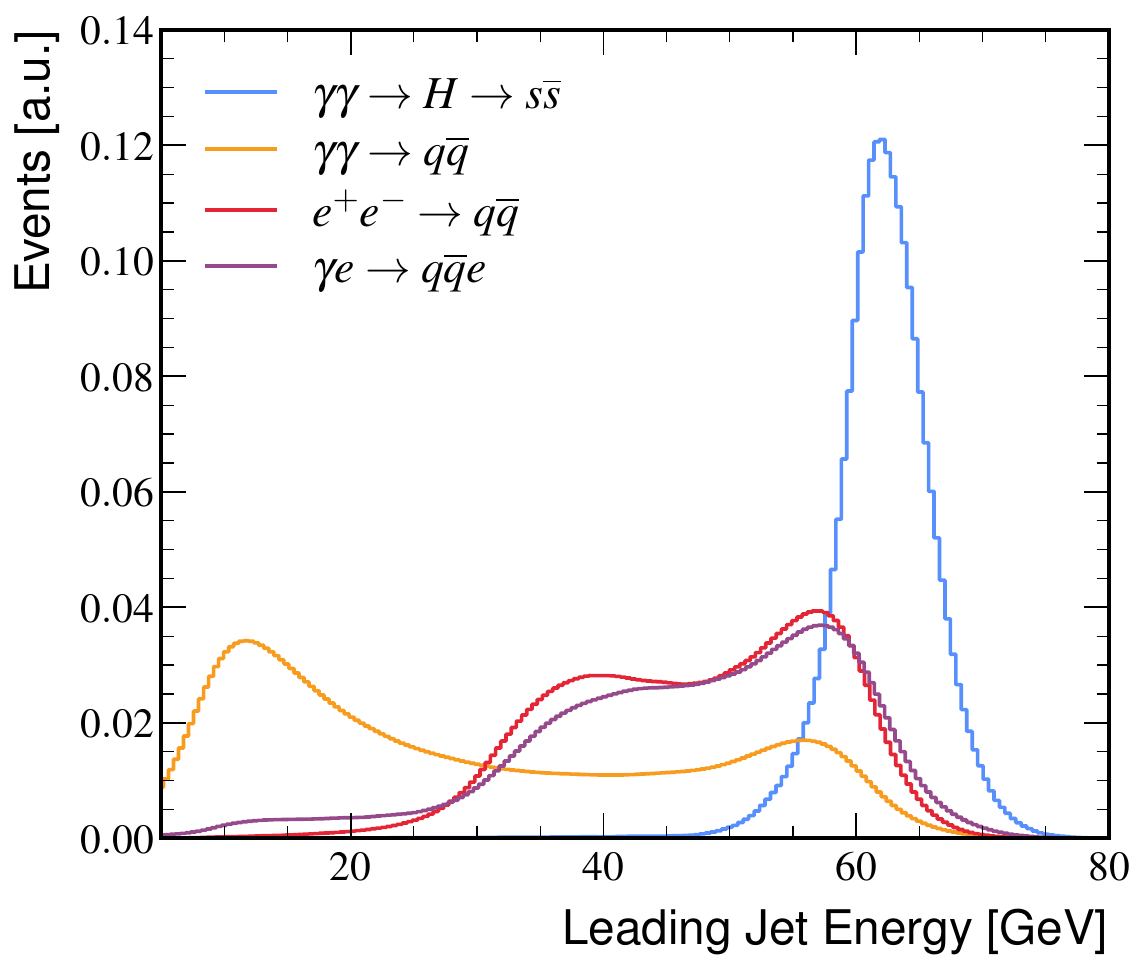}
    \end{minipage}
    \begin{minipage}{0.495\textwidth}
        \includegraphics[width=\linewidth]{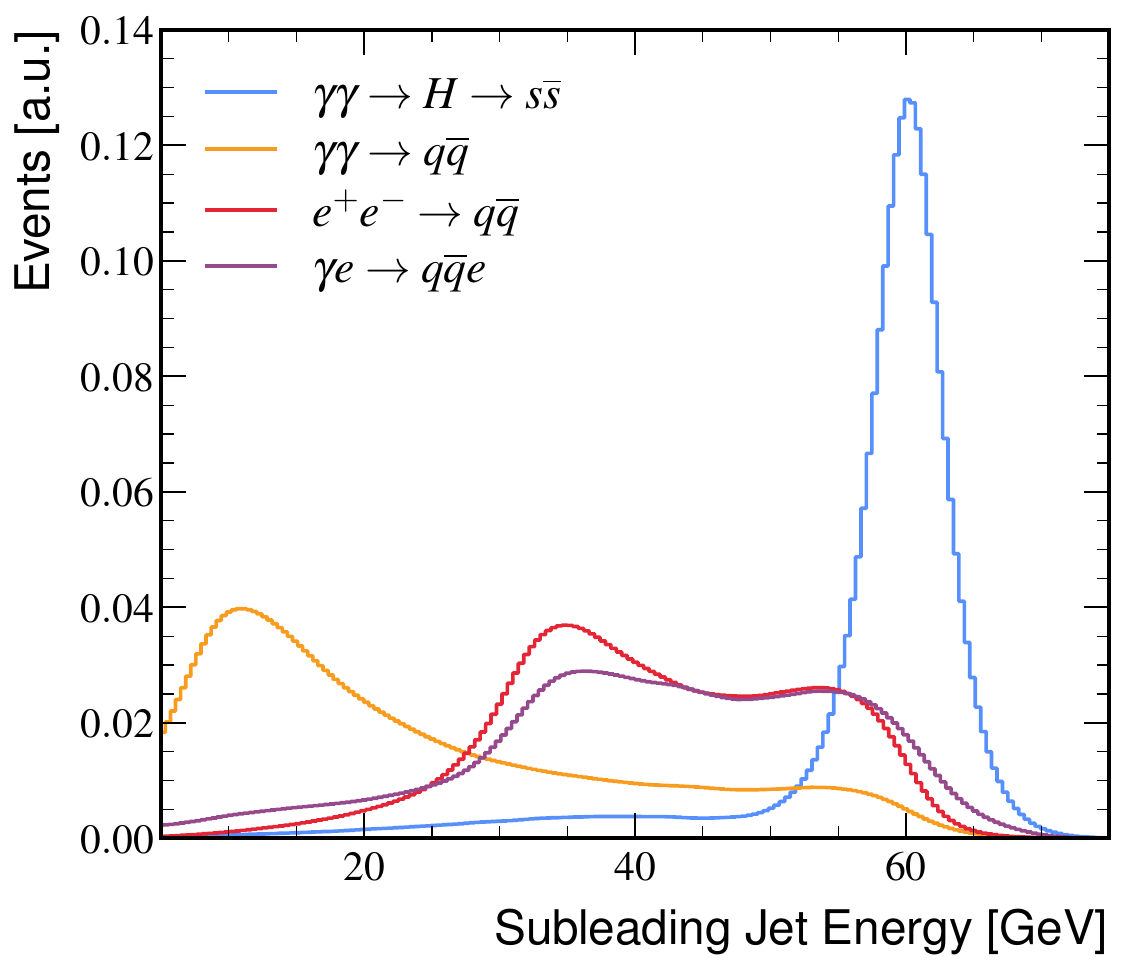}
    \end{minipage}
    \begin{minipage}{0.495\textwidth}
        \includegraphics[width=\linewidth]{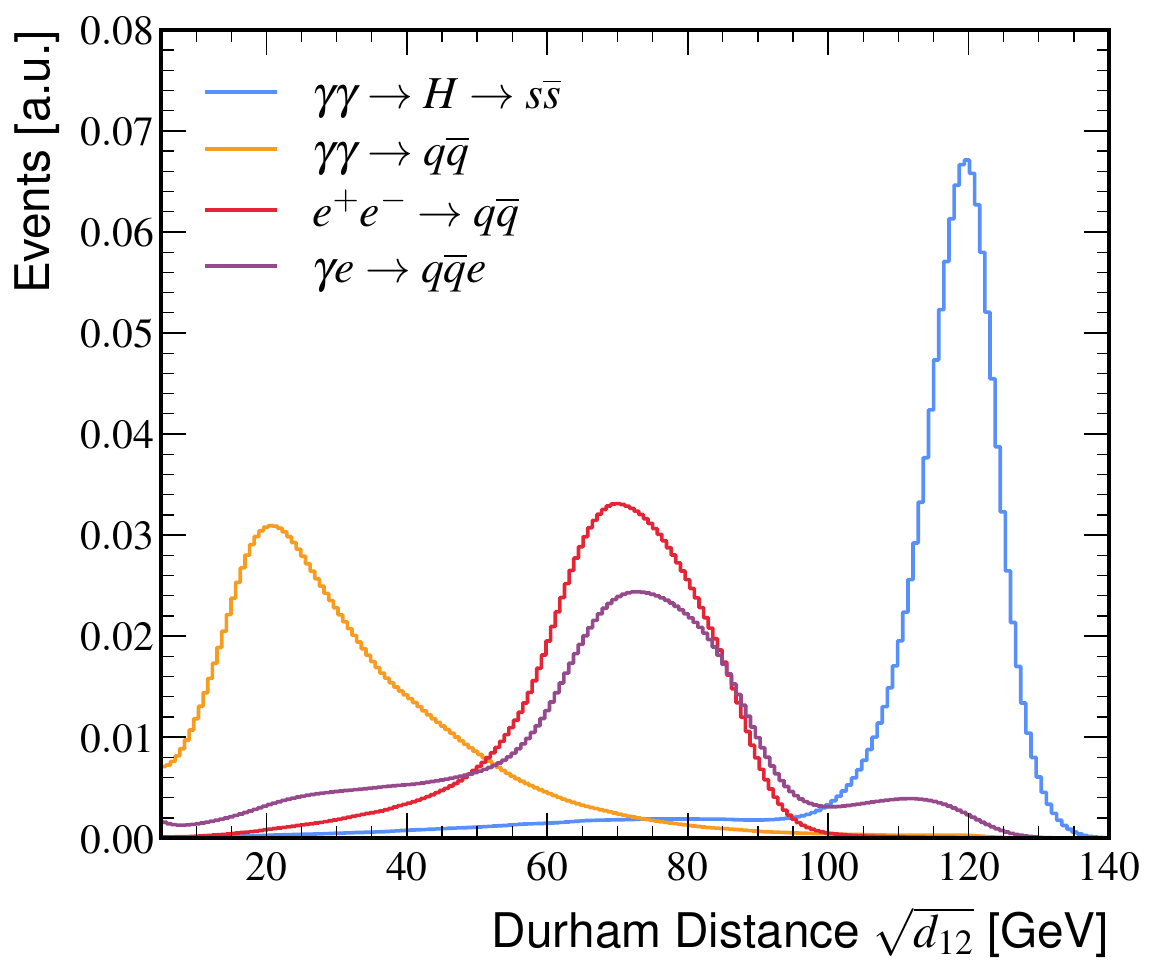}
    \end{minipage}
    \caption{Distributions of the dijet invariant mass (top left), leading jet energy (top right), subleading jet energy (bottom left), and the Durham distance between the jets (bottom right) for the $H\to s\overline{s}$ signal and dominant backgrounds prior to any selection cuts. The distributions are normalized such that the area under the curve is unity.}
    \label{fig:ss_plots}
\end{figure}

\begin{table}
\centering
    \begin{tabular}{ l l l}
\hline
 \textbf{Process} & \textbf{Before Preselection} & \textbf{After Preselection}  \\ \hline
 $\gamma\gamma \to H\to s \overline{s}$ & 440 & 238 \\
 \hline 
$\gamma \gamma \to qq\overline{q}\overline{q}$ & 381,829 & 6,671\\
$\gamma\gamma \to q\overline{q}$ & 513,755,395& 17,235\\
$\gamma e \to q\overline{q}e $ & 74,887,283& 94,236\\
 $e^+e^- \to q\overline{q}$ & 459,001,600& 32,388 \\ 
 \hline
\end{tabular}
\caption{Number of events before and after the $H\to s\overline{s}$ pre-selection filters for signal and backgrounds. \label{tab:hss}}
\end{table}

The $H\to s\overline{s}$ channel has traditionally been considered prohibitively challenging due to the minuscule $H\to s\overline{s}$ branching fraction at $\mathcal{O}\left(10^{-4}\right)$. Further, signal events compete with massive hadronic backgrounds. A $\gamma\gamma$ collider, however, provides a unique opportunity to probe $H\to s\overline{s} $ decays since the main $\gamma\gamma \to s\overline{s}$ background process is highly suppressed due to the strange quark charge of $e/3$. Thus, we select a dijet topology with dedicated strange jet tagging, following a ``Medium'' working point of the \textsc{ParticleNet} algorithm, which yields an 82.5\% efficiency for $s$-jets, and mistag rates of $5\%$, $2.5\%$, and $50\%$ for  $c$, $b$, and light $= \{u,d, g\}$ jets respectively. This working point was chosen after a scan over multiple working points, selecting one which maximizes the sensitivity metric i.e. $S/\sqrt{S+B}$. Events must contain at least two strange tagged jets with $E>8$ GeV and no isolated leptons. To further disfavour multi-prong or boosted configurations we require a minimum $\sqrt{d_{12}}=95$ GeV between the two leading jets, and we keep only low–missing-energy topologies with $E_{\mathrm T}^{\rm miss}<18$ GeV. Finally, we require a leading dijet invariant mass $m(jj) > 105$ GeV, similar to the other channels, which trims the low-mass continuum and the $Z$-peak region while retaining the Higgs resonance. The effect of the $m(jj)$ and $\sqrt{d_{12}}$ cuts can be inferred from the top-left and bottom-right plots in Fig.~\ref{fig:ss_plots}, which show these distributions. Overall, the pre-selections result in about a $60\%$ efficiency for the signal, while reducing the backgrounds by 3-4 orders of magnitude as summarized in Table \ref{tab:hss}.

\subsection{\texorpdfstring{\boldmath $H\to VV^* \to qq\overline{qq}$}{}}
\begin{figure}
    \centering
    \begin{minipage}{0.495\textwidth}
        \includegraphics[width=\linewidth]{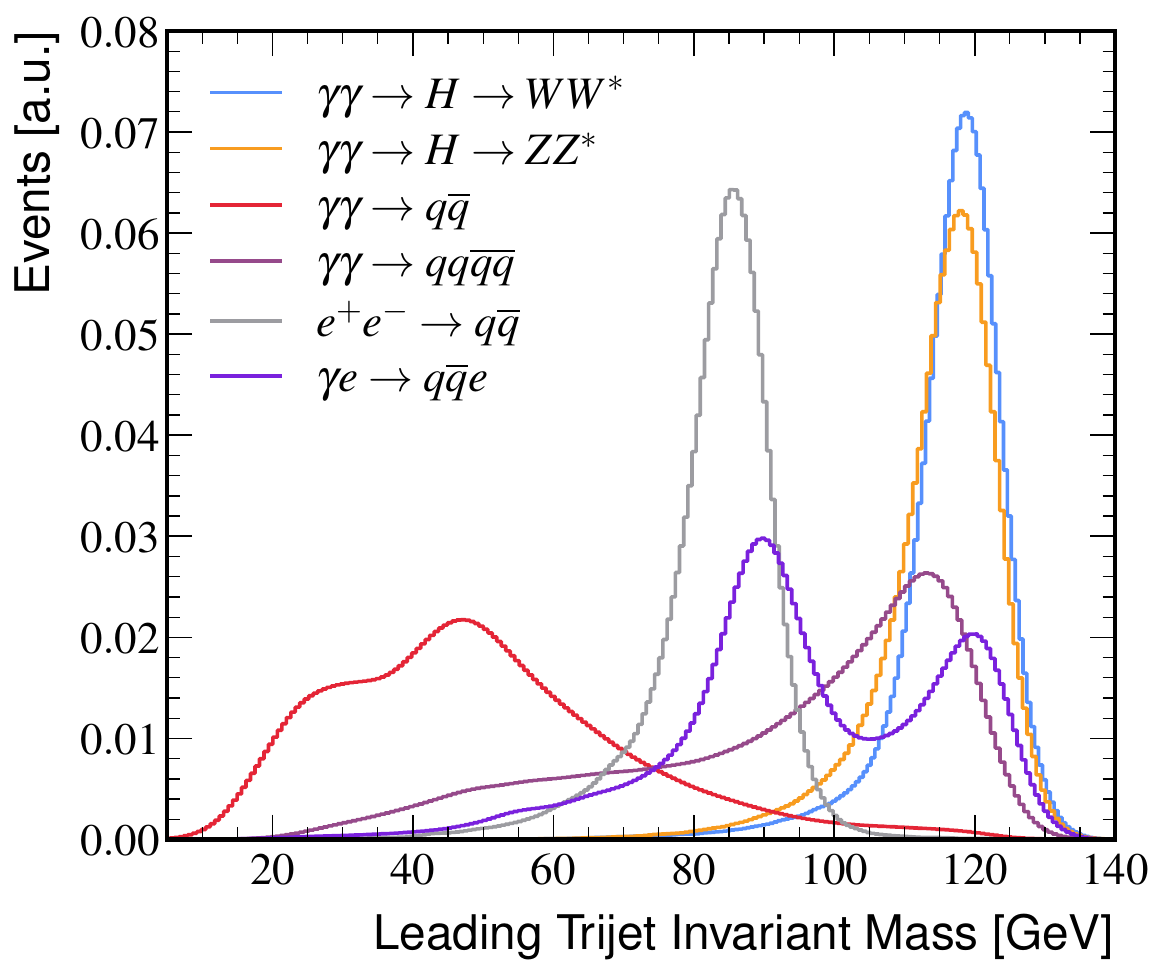}
    \end{minipage}
    \begin{minipage}{0.495\textwidth}
        \includegraphics[width=\linewidth]{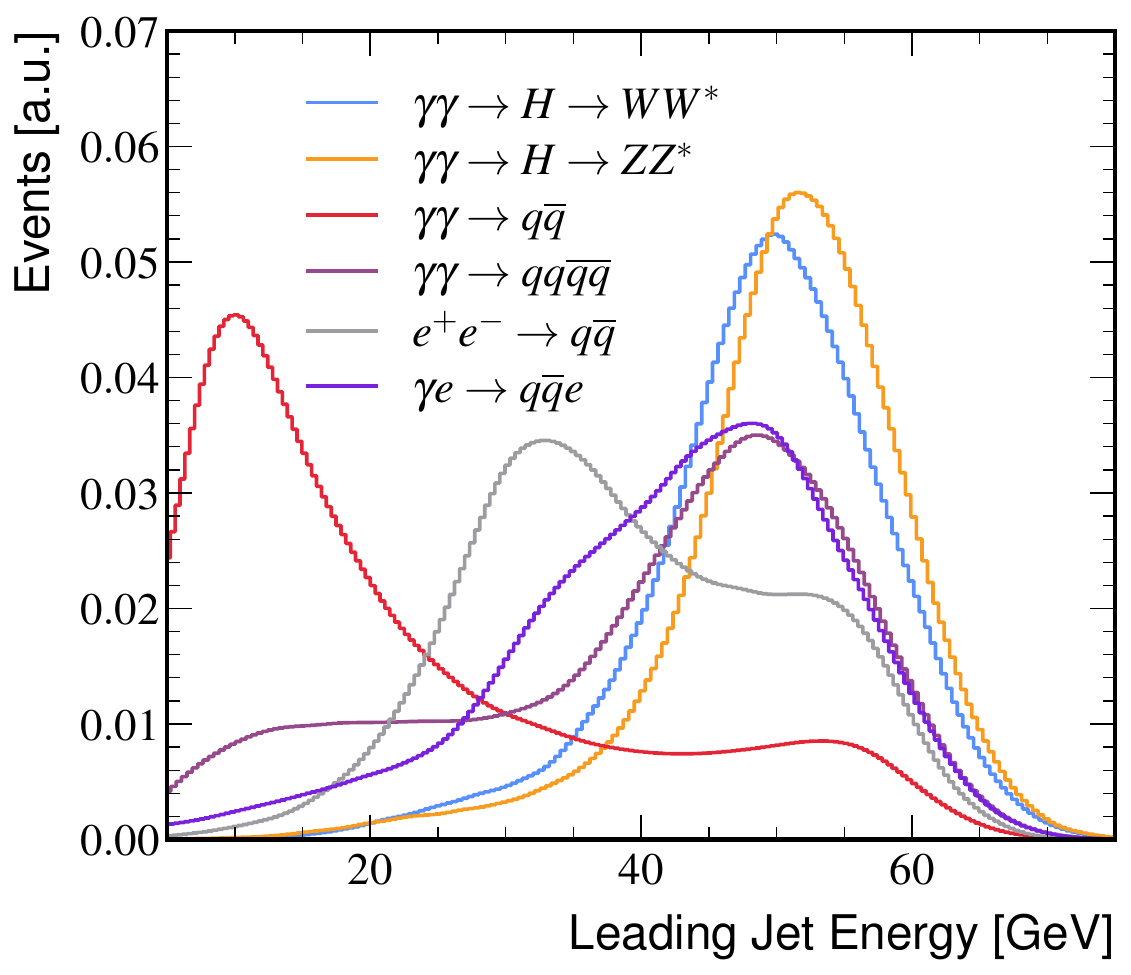}
    \end{minipage}
    \caption{Distributions of the trijet invariant mass (left), leading jet energy and the Durham distance between the jets (right) for the $H\to VV^*\to qq\overline{qq}$ signal and dominant backgrounds prior to any selection cuts. The distributions are normalized such that the area under the curve is unity.}
    \label{fig:HVVqqqq_plots}
\end{figure}

\begin{table}
\centering
    \begin{tabular}{ l l l}
\hline
 \textbf{Process} & \textbf{Before Preselection} & \textbf{After Preselection}  \\ \hline
 $\gamma\gamma \to H\to ZZ^*\to qq\overline{qq}$ & 20,552 & 5,995 \\
 $\gamma\gamma \to H\to WW^* \to qq\overline{qq}$ & 166,664 & 41,889 \\
 \hline 
  $\gamma\gamma \to H\to b\overline{b}$ & 635,800 & 52,602 \\
 $\gamma\gamma \to H\to c\overline{c}$ & 33,000 & 3,190 \\
  $\gamma\gamma \to H\to gg$ & 94,602 & 15,126 \\
 \hline 
$\gamma \gamma \to qq\overline{q}\overline{q}$ & 381,829 & 16,794 \\
$\gamma\gamma \to q\overline{q}$ & 513,755,395& 29,598\\
$\gamma e \to q\overline{q}e $ & 74,887,283& 51,241\\
 $e^+e^- \to q\overline{q}$ & 459,001,600& 203,655 \\ 
 \hline
\end{tabular}
\caption{Number of events before and after the hadronic $H\to VV^*$ pre-selection filters for signal and backgrounds. \label{tab:hvv}}
\end{table}

For the fully hadronic $H\to VV^* \to qq\overline{qq}$, where $V\in\{Z, W\}$ bosons,
channel, we require events to contain three or four jets with $E>8$ GeV. Further, we veto isolated leptons with $E>15$ GeV to remove semi-leptonic backgrounds. We find that including three jets in addition to four retains genuine four-quark topologies, with tolerance for occasional jet merging due to boost. We further exploit the higher particle multiplicity of hadronic $VV^{*}$ decays by having an event-level track count $N_{\mathrm{tracks}}>35$, which efficiently rejects two-jet continuum $\gamma\gamma \to q\overline{q}$. Prompt neutrino channels are reduced with a low-MET requirement with $E_{\mathrm T}^{\rm miss}<18$ GeV. Finally, we only keep events that pass a leading tri-jet mass $m(jjj)>100$ GeV, which removes backgrounds inconsistent with $H\to VV^*$ topology as shown in the left panel of Fig.~\ref{fig:HVVqqqq_plots}. We forego traditional jet pairing algorithms that minimize a $\chi^2$ to the $W/Z$ hypotheses as well as consistency windows around the nominal $W/Z$ masses, as these methods limit the pre-selection signal efficiency. Instead, we use the machine learning algorithm to perform $W/Z$ discrimination as we find it to be more signal efficient. Table \ref{tab:hvv} lists the signal and dominant background counts before and after the pre-selections.

\subsection{\texorpdfstring{\boldmath $H\to \mu^+ \mu^-$}{}}

\begin{figure}
    \centering
    \begin{minipage}{0.495\textwidth}
        \includegraphics[width=\linewidth]{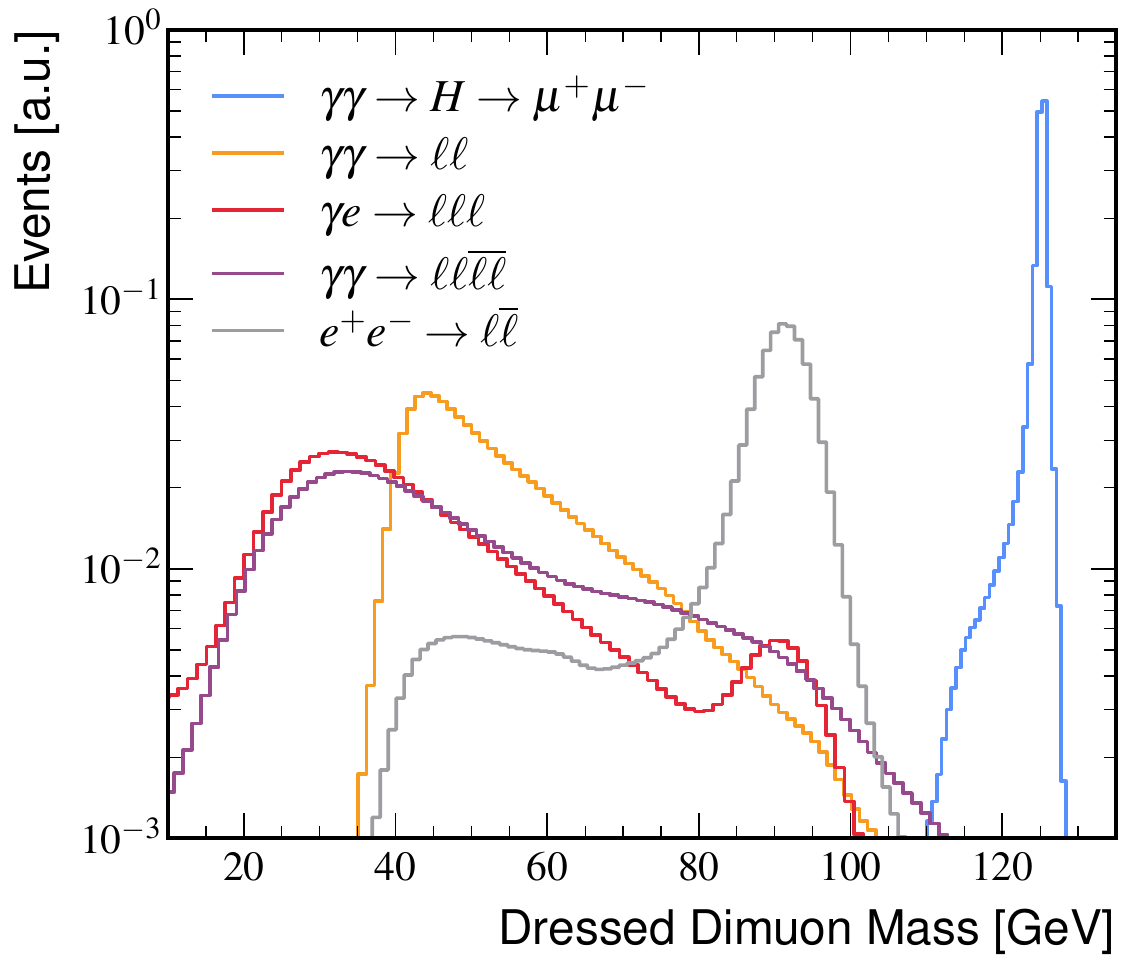}
    \end{minipage}
    \begin{minipage}{0.495\textwidth}
        \includegraphics[width=\linewidth]{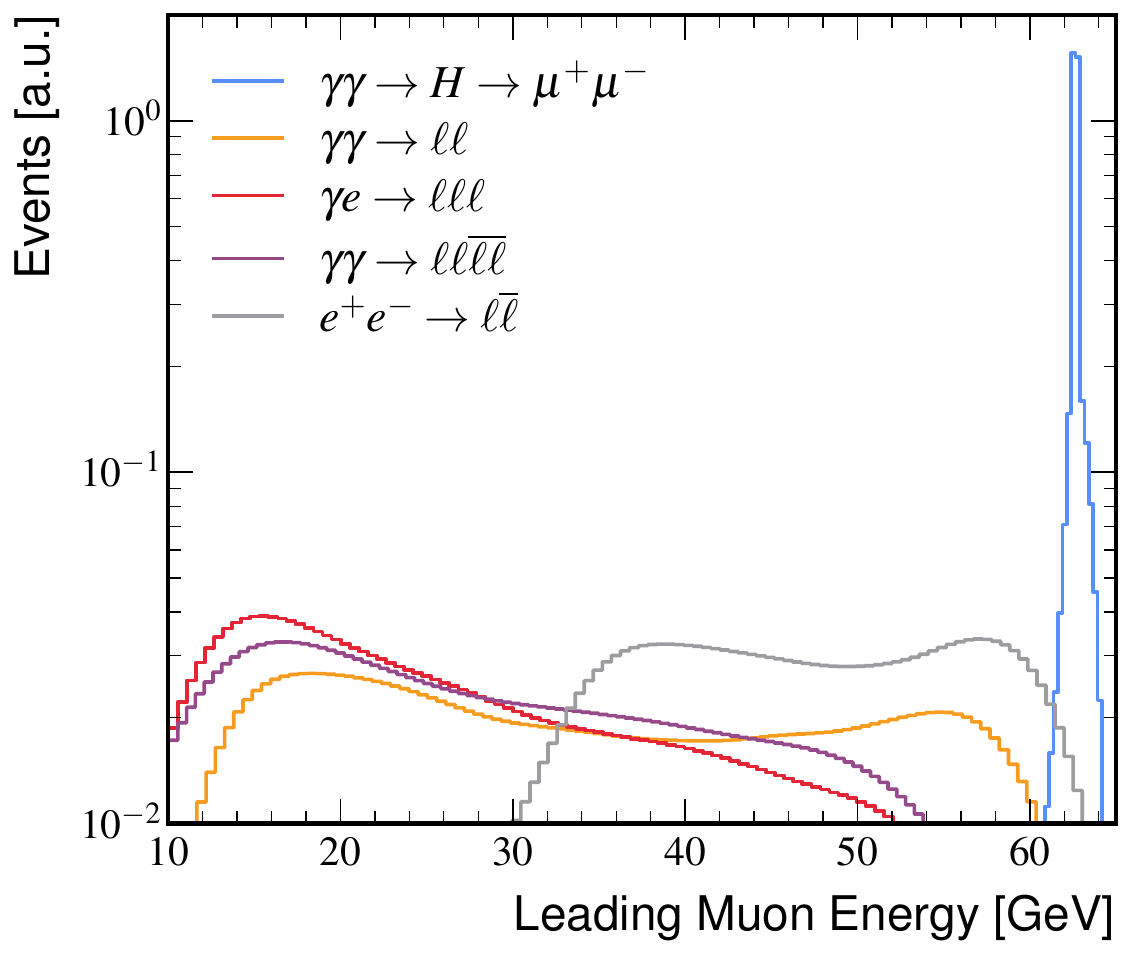}
    \end{minipage}
    \caption{Distributions of the dressed dimuon invariant mass (left) and leading muon energy (right) for the $H\to \mu^+\mu^-$ signal and dominant backgrounds prior to any selection cuts. The distributions are normalized such that the area under the curve is unity.}
    \label{fig:Hmumu_plots}
\end{figure}

\begin{table}
\centering
    \begin{tabular}{ l l l}
\hline
 \textbf{Process} & \textbf{Before Preselection} & \textbf{After Preselection}  \\ \hline
 $\gamma\gamma \to H\to \mu^+\mu^-$ & 238 & 178 \\ \hline
$\gamma \gamma \to \ell \ell \overline{\ell \ell}$ & 225,100,651 & 1,901 \\
$\gamma\gamma \to \ell \overline{\ell}$ & 2,135,184,138& 724,095\\
$\gamma e \to \ell \ell \ell  $ & 1,189,141,510& 11,297\\
 $e^+e^- \to \ell\overline{\ell}$ & 2,245,722,200 & 103,303\\ 
 \hline
\end{tabular}
\caption{Number of events before and after the $H\to \mu^+\mu^-$ pre-selection filters for signal and backgrounds. \label{tab:hmumu}}
\end{table}

The $H\to \mu^+\mu^-$ channel offers an exceptionally clean, low-activity final state, and thus we exploit the excellent muon momentum resolution to isolate a narrow resonance near the Higgs mass. Events are required to contain exactly two oppositely charged, well-identified muons with $E>15$ GeV and relative isolation $<0.15$, which suppresses heavy-flavor decays and jets faking muons. We veto additional electrons or hadronic $\tau$ candidates to remove electroweak processes with extra leptons and $\gamma e$ topologies, and apply a jet veto (excluding the lepton candidates) to mitigate any hadronic activity that can leak into the dimuon category. Because the signal is expected to be fully visible, we further require low missing transverse energy, $E_{\mathrm T}^{\rm miss}<5$ GeV, to reject channels with neutrinos. The dimuon invariant mass is then restricted to $ 115<m(\mu^+\mu^-)<130$ GeV, eliminating the dominant $Z$ peak at $m_Z\approx 90$ GeV. Following standard practice in previous $H\to \mu^+\mu^-$ studies \cite{CMS:2021gxc}, muons are ``dressed'' with nearby FSR photons if present within $\Delta R < 0.1$ to sharpen the mass peak. The distributions of the dressed dimuon invariant mass, which peaks sharply around the Higgs mass, and the leading muon energy are shown in Fig.~\ref{fig:Hmumu_plots}. All in all, these pre-selection filters have a signal efficiency of nearly $80\%$ while eliminating nearly all sources of background save those listed in Table \ref{tab:hmumu}. 

\subsection{\texorpdfstring{\boldmath $H\to \tau^+ \tau^-$}{}}

\begin{figure}
    \centering
    \begin{minipage}{0.495\textwidth}
        \includegraphics[width=\linewidth]{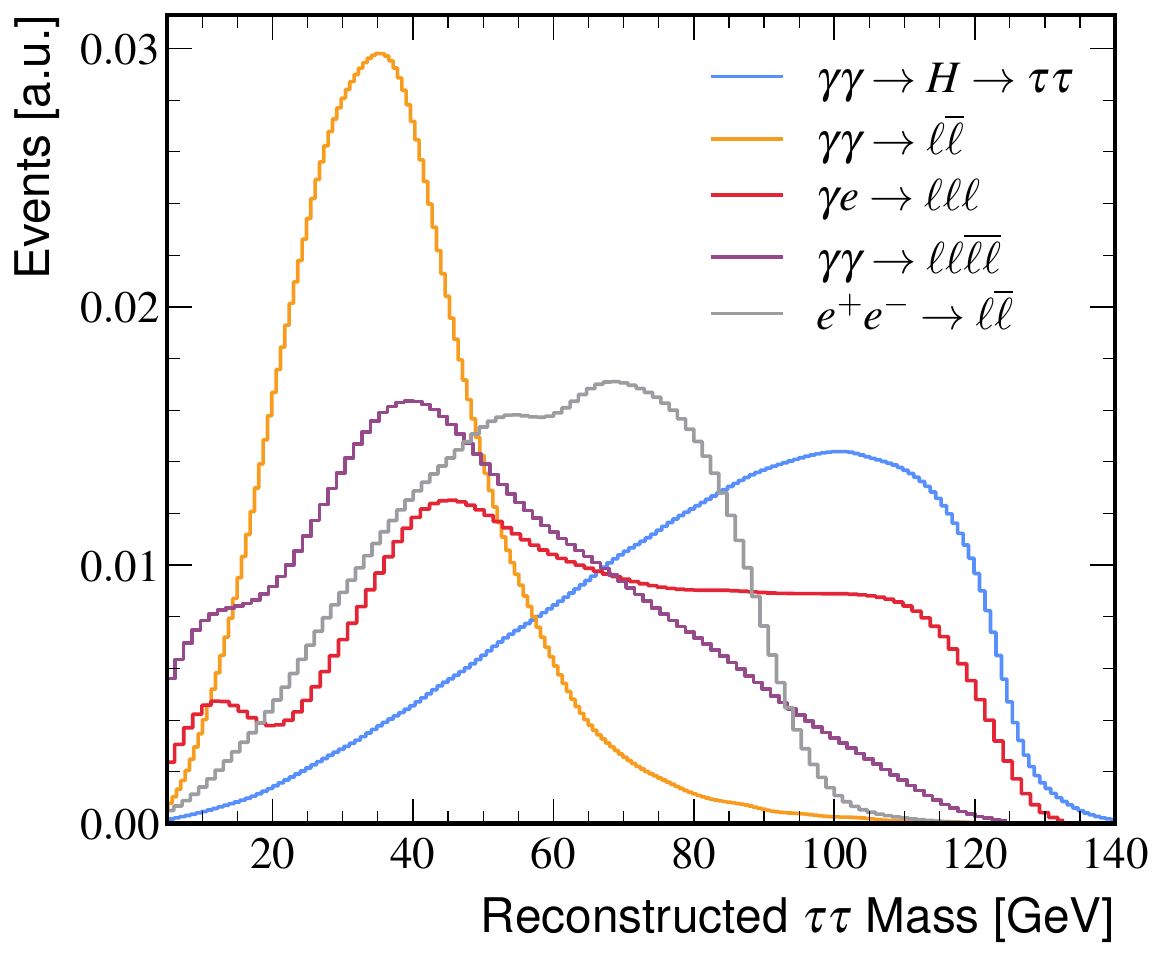}
    \end{minipage}
    \begin{minipage}{0.495\textwidth}
        \includegraphics[width=\linewidth]{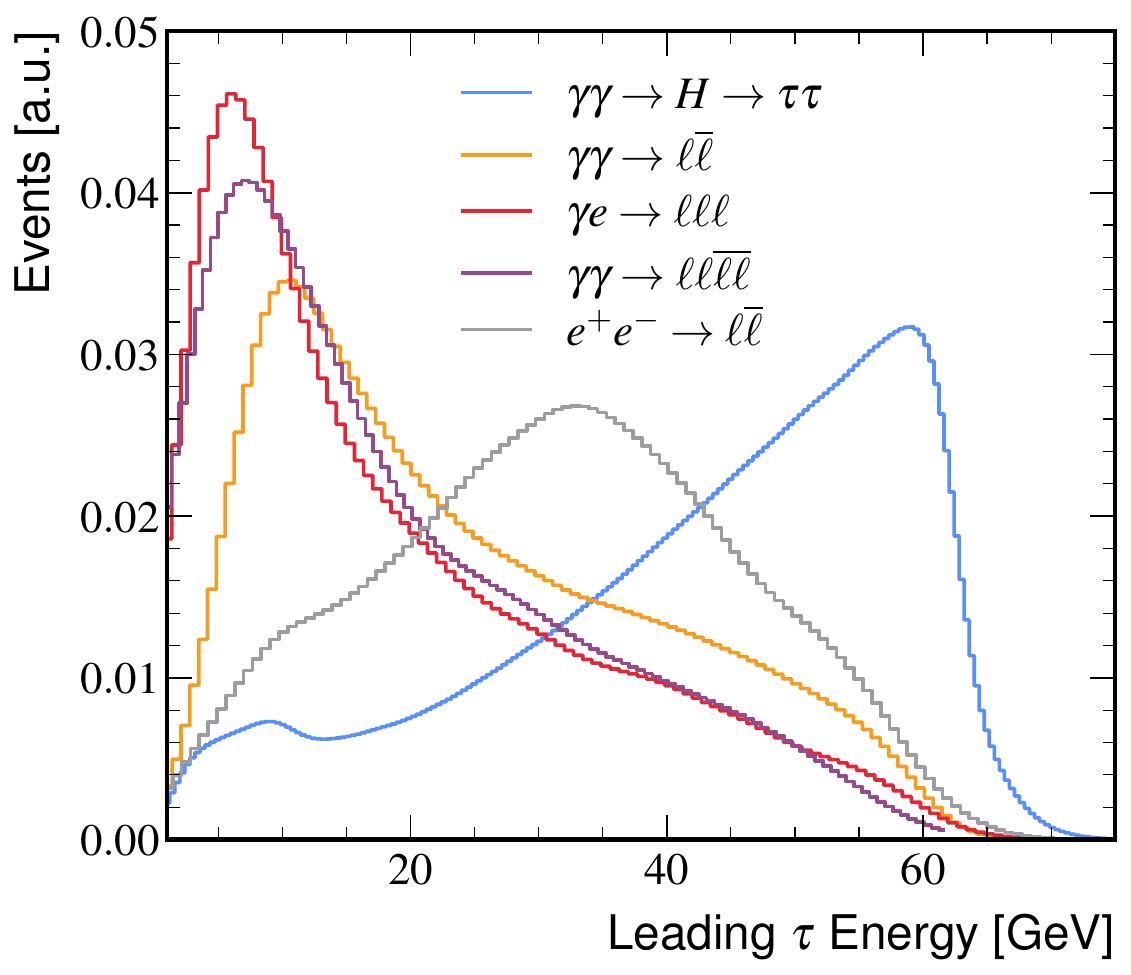}
    \end{minipage}
    \caption{Distributions of the reconstructed ditau invariant mass (left) and leading tau energy (right) for the $H\to \tau\tau$ signal and dominant backgrounds prior to any selection cuts. The distributions are normalized such that the area under the curve is unity.}
    \label{fig:Htautau_plots}
\end{figure}

For the reconstruction of hadronically decaying $\tau$ leptons, denoted $\tau_h$, several methods \cite{Muennich:1443551, Lange:2023gbe, Suehara:2024qoc, Jeans:2018anq, Kawada:2015wea} have been devised for searches at the ILC, FCC-ee, and other future colliders. Common strategies include first clustering narrow, low-multiplicity jets and classifying their decay mode (i.e.~1-prong or 3-prong) using track multiplicity, secondary-vertex fits, impact-parameter significance, and $\pi^0/\gamma$ reconstruction from ECAL clusters. Candidate taus are then isolated with particle-flow sums in angular cones and vetted with kinematic/shape variables (visible mass, transverse mass, jet width) to suppress light-flavor and gluon jets as well as electrons and muons with associated FSR. Unsurprisingly, state-of-the-art taggers utilize all available phase space information i.e.~kinematic, calorimetric and tracking variables in tandem using modern machine-learning methods such as graph neural netwroks/transformers. The best performing architectures achieve about a $90\%$ $\tau$-identification efficiency with $<1\%$ mistag rates. For this study, we adopt slightly more conservative approach and follow the default \textsc{Delphes} working point of the IDEA, SiD, and ILD $\tau$ taggers that yield an 85\% $\tau$ signal efficiency and a $1\%$ mistag rate, applied uniformly across channels and energies.

In keeping with previous $H\to \tau^+\tau^-$ analyses \cite{CMS:2021gxc}, all possible $\tau^+ \tau^-$ decay modes are considered for this channel, except for those with two muons or two electrons. We utilize this light di-lepton veto because of the low branching fraction ratio and large background contributions associated with these modes. The di-tau mass $m(\tau\tau)$ is reconstructed as the mass of all visible objects combined with the missing momentum $p_\mathrm{T}^\mathrm{miss}$. The left panel of Fig.~\ref{fig:Htautau_plots} shows the reconstructed $\tau\tau$ mass for the signal and dominant backgrounds and the right panel shows the leading $\tau$ energy. We select events that have a reconstructed $\tau\tau$ mass $>60$ GeV. The signal and background counts before and after preselection are shown in Table~\ref{tab:h_tautau}.

\begin{table}
\centering
\begin{tabular}{ l l l}
\hline
 \textbf{Process} & \textbf{Before Preselection} & \textbf{After Preselection}  \\ \hline
 $\gamma\gamma \to H\to \tau^+ \tau^-$ & 69,080 & 41,504 \\ \hline
$\gamma \gamma \to \ell \ell \overline{\ell \ell}$ & 225,100,651 & 77,726 \\
$\gamma\gamma \to \ell \overline{\ell} \nu \overline{\nu}$ &24,655 &2,695 \\
$\gamma\gamma \to \ell \overline{\ell}$ & 2,135,184,138& 1,831,361\\
$\gamma e \to \ell \ell \ell  $ & 1,189,141,510& 3,553,369\\
 $e^+e^- \to \ell\overline{\ell}$ & 2,245,722,200 & 4,833,863 \\ 
 \hline
\end{tabular}
\caption{Number of events before and after the $H\to \tau^+ \tau^-$ pre-selection filters for signal and backgrounds. \label{tab:h_tautau}}
\end{table}

\subsection{\texorpdfstring{\boldmath $H\to WW^* \to \ell \overline{\nu} q\overline{q}$}{}}
\begin{figure}
    \centering
    \begin{minipage}{0.495\textwidth}
        \includegraphics[width=\linewidth]{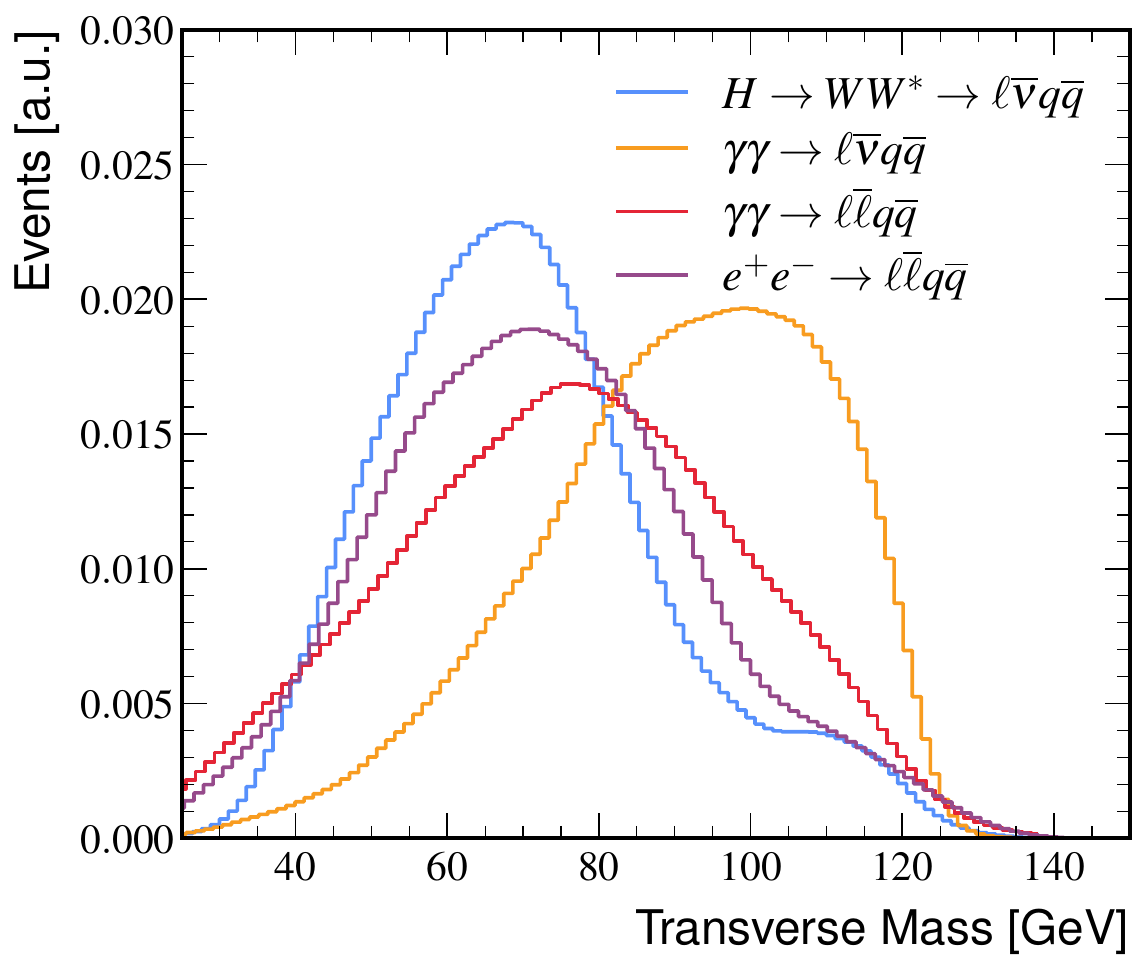}
    \end{minipage}
    \begin{minipage}{0.495\textwidth}
        \includegraphics[width=\linewidth]{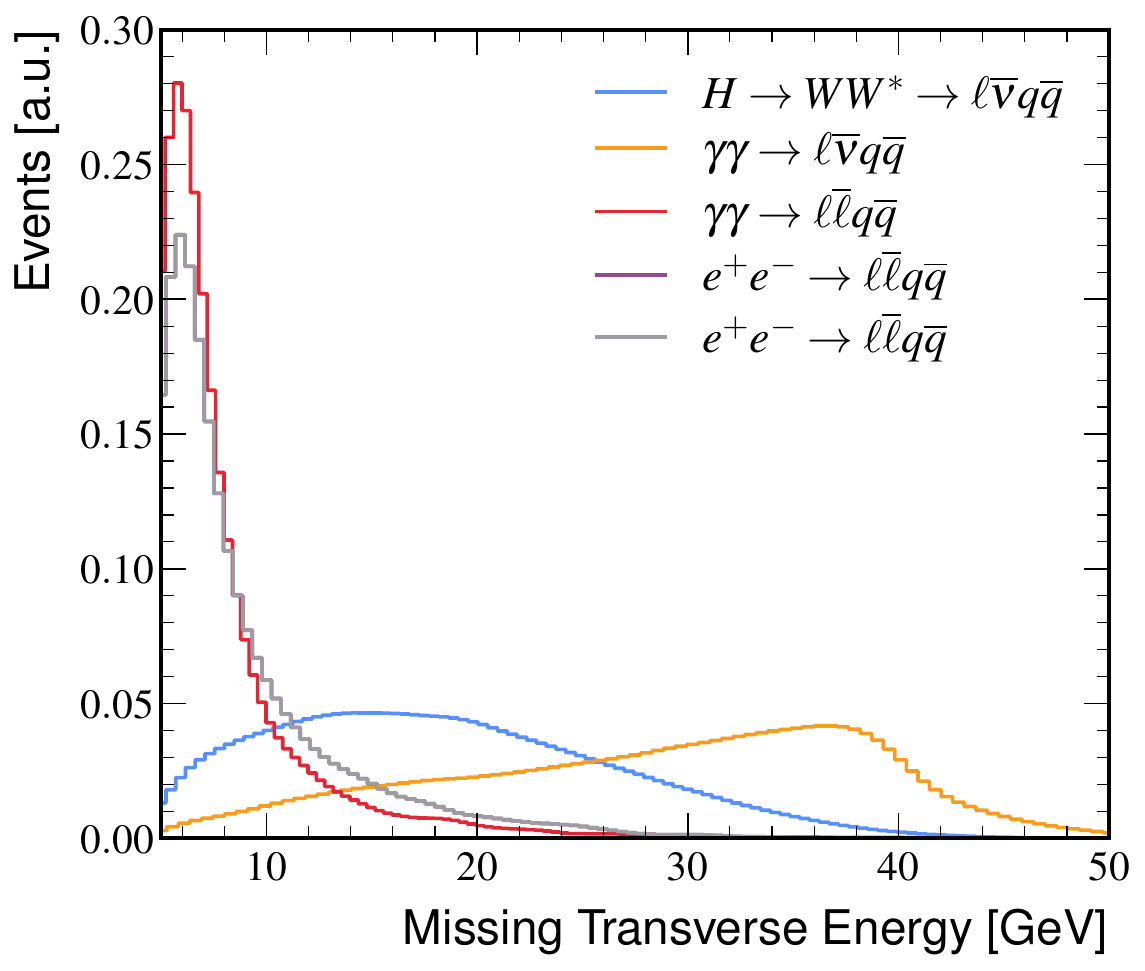}
    \end{minipage}
    \caption{Distributions of the reconstructed transverse mass (left) and missing transverse energy (right) for the $H\to WW^* \to \ell \overline{\nu} q\overline{q}$ signal and dominant backgrounds prior to any selection cuts. The distributions are normalized such that the area under the curve is unity.}
    \label{fig:Hww_2qllnu_plots}
\end{figure}

\begin{table}
\centering
\begin{tabular}{ l l l }
\hline
\textbf{Process} & \textbf{Before Preselection} & \textbf{After Preselection} \\ \hline
$\gamma\gamma \to H\to WW^* \to \ell \overline{\nu }q\overline{q}$ & 51,767 & 26,262 \\ \hline
$\gamma\gamma \to \ell\overline{\nu} q\overline{q}$ & 61,708 & 14,510 \\
$\gamma\gamma \to \ell^+\ell^-q\overline{q}$ & 9,771,761 & 6,623 \\
$e^+e^- \to \ell^+\ell^- q\overline{q}$ & 190,407 & 409\\
\hline
\end{tabular}
\caption{Number of events before and after the $H\to WW^* \to \ell \overline{\nu }q\overline{q}$ pre-selection filters for signal and backgrounds. \label{tab:hwwlnuqq}}
\end{table}
The semi-leptonic $H\to WW^{*} \to \ell \overline{\nu} q\overline{q}$ channel is characterized by a (often merged) dijet topology with modest missing momentum and one lepton. As such, we reconstruct events requiring either one or two jets with $E>8$ GeV, an isolated lepton or $\tau$-tagged jet with $E>10$ GeV and $E_\mathrm{T}^\mathrm{miss}>10$ GeV. Vetos are applied for additional jets and leptons. Furthermore, to mitigate backgrounds resulting form spurious missing energy, we require a maximum transverse mass requirement of $m_\mathrm{T} < 80$ GeV. Fig.~\ref{fig:Hww_2qllnu_plots} shows the distributions of the transverse mass and missing transverse energy after physics object selection but before any filters or cuts are applied on the distributions.

\subsection{\texorpdfstring{\boldmath $H\to \gamma\gamma$}{}}

\begin{table}
\centering
\begin{tabular}{ l l l l }
\hline
 \textbf{Process} & \textbf{Before Preselection} & \textbf{w.~DRO} & \textbf{w/o~DRO}  \\ \hline
 $\gamma\gamma \to H\to \gamma\gamma$& 2,519 & 2,227 & 1,914 \\ \hline
$\gamma \gamma \to \gamma\gamma$  & 69,080 & 911 & 914\\
 \hline
\end{tabular}
\caption{Number of events before and after the $H\to \gamma\gamma$ pre-selection filters with and without a dual-readout calorimeter for signal and background. \label{tab:h_AA}}
\end{table}

\begin{figure}
    \centering
    \begin{minipage}{0.495\textwidth}
        \includegraphics[width=\linewidth]{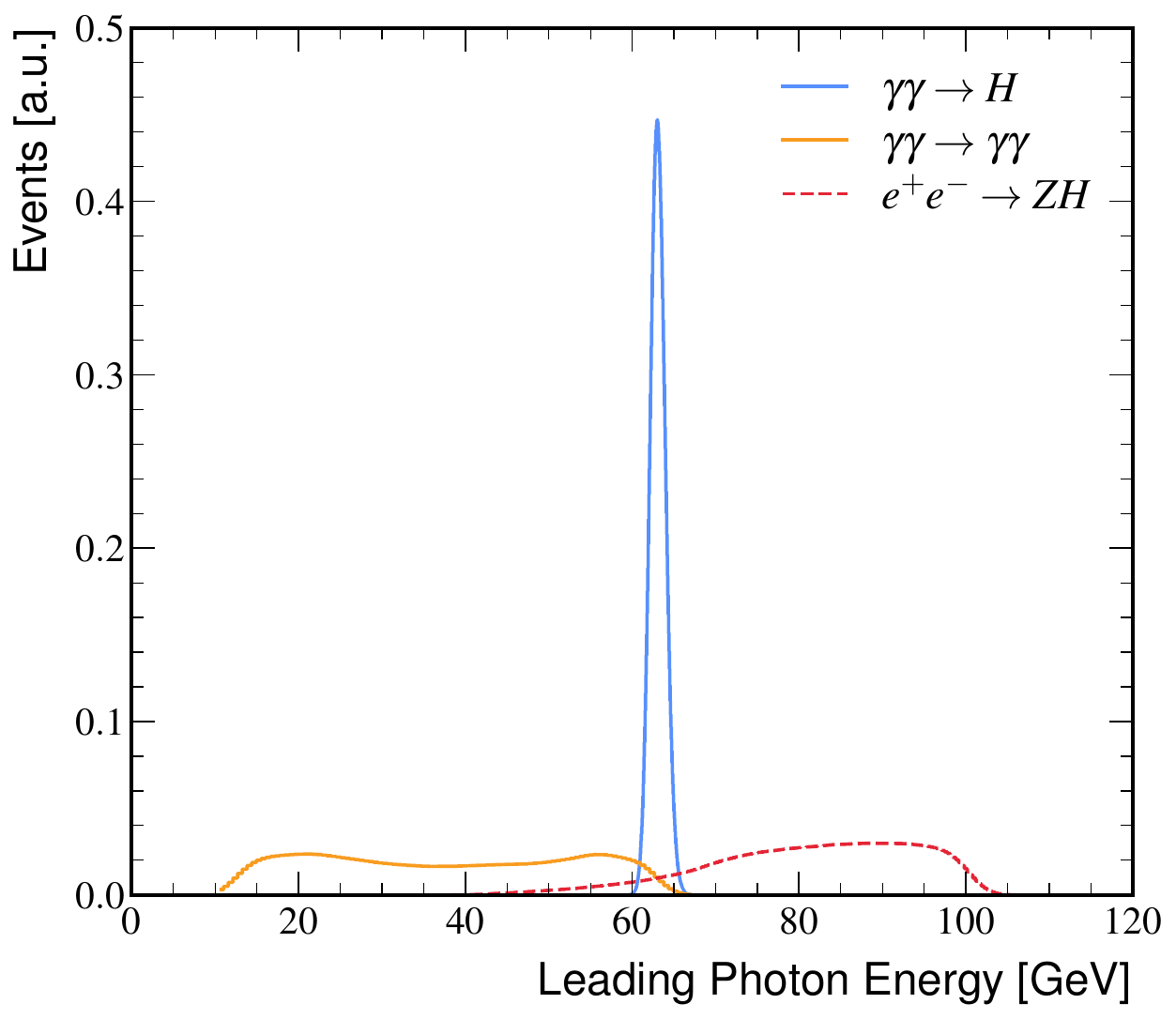}
    \end{minipage}
    \begin{minipage}{0.495\textwidth}
        \includegraphics[width=\linewidth]{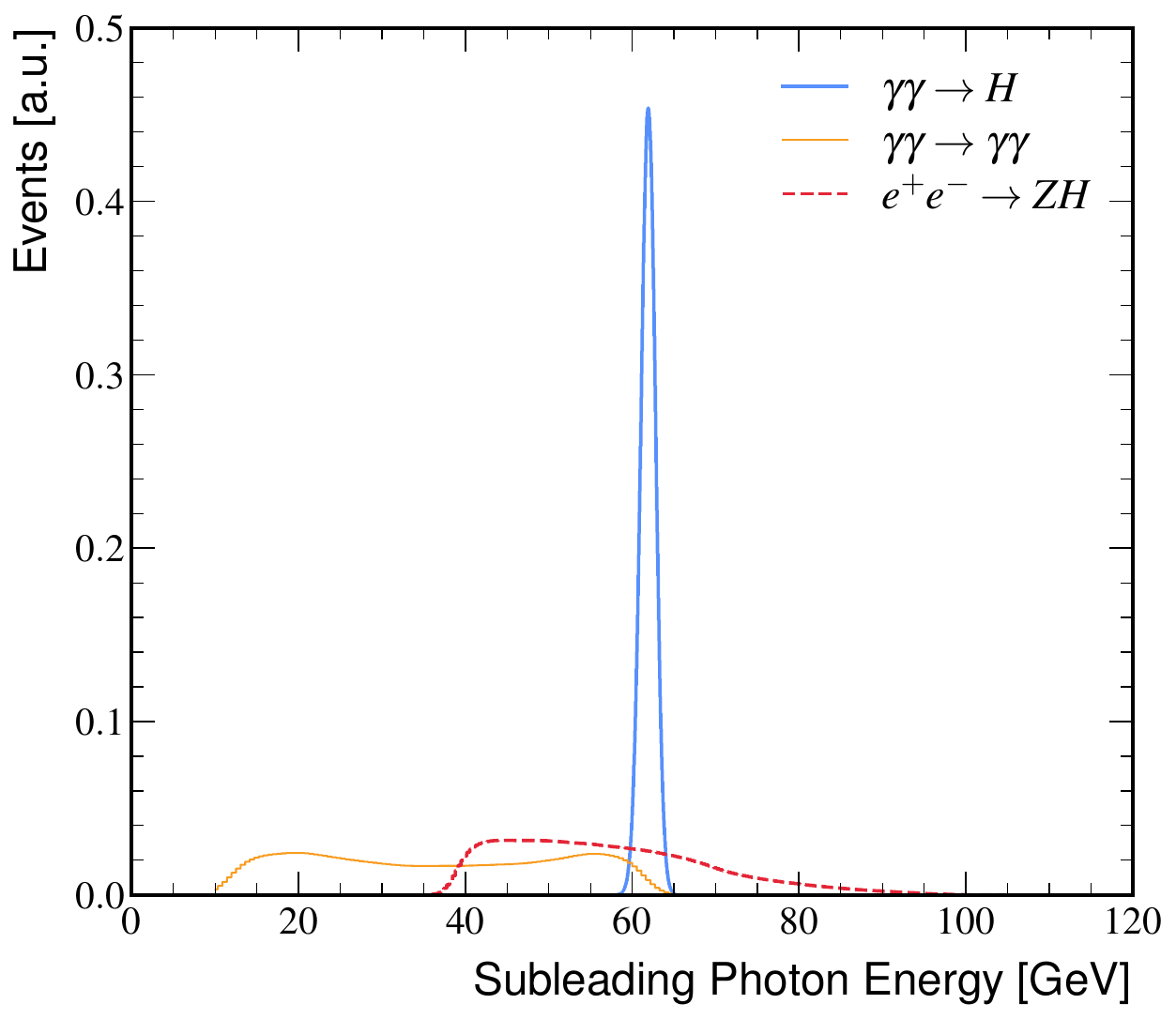}
    \end{minipage}
    \begin{minipage}{0.495\textwidth}
        \includegraphics[width=\linewidth]{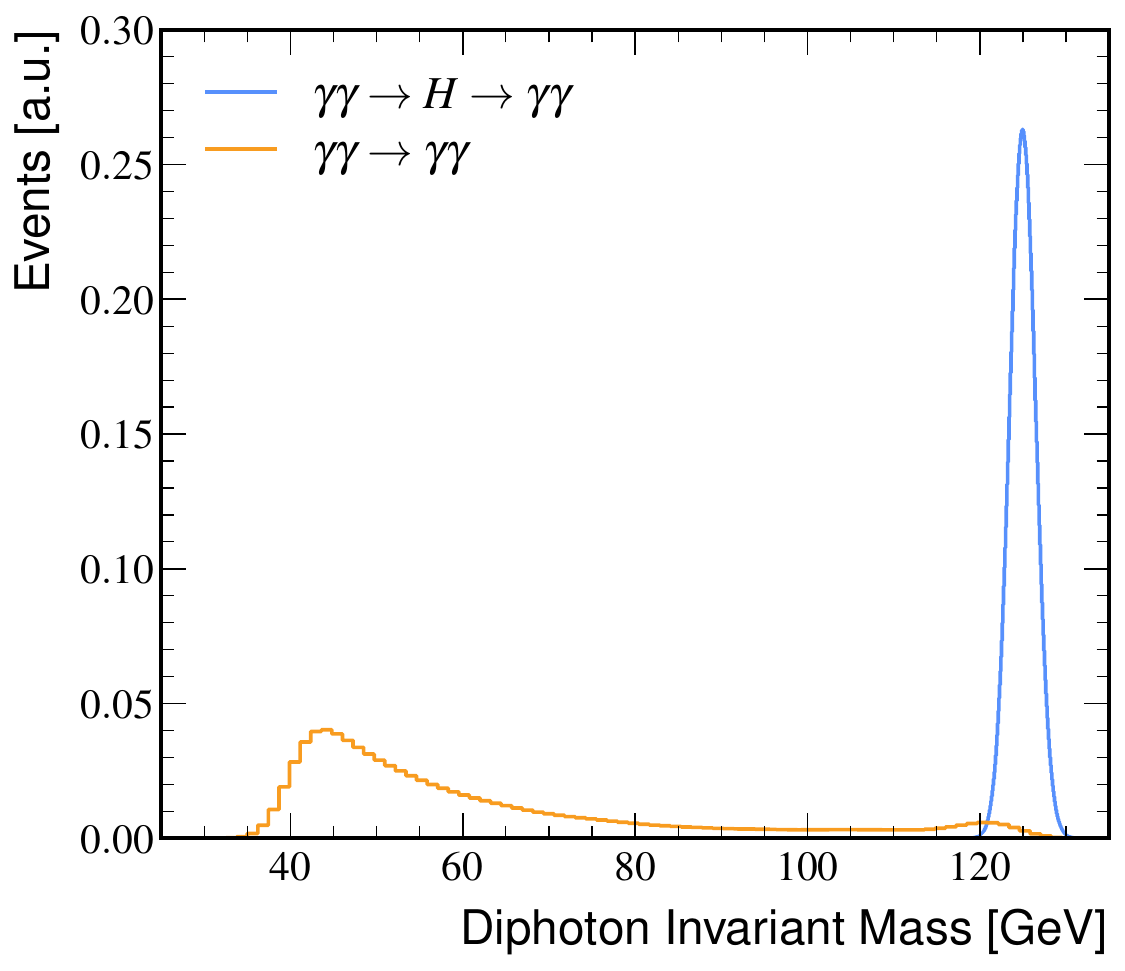}
    \end{minipage}
    \begin{minipage}{0.495\textwidth}
        \includegraphics[width=\linewidth]{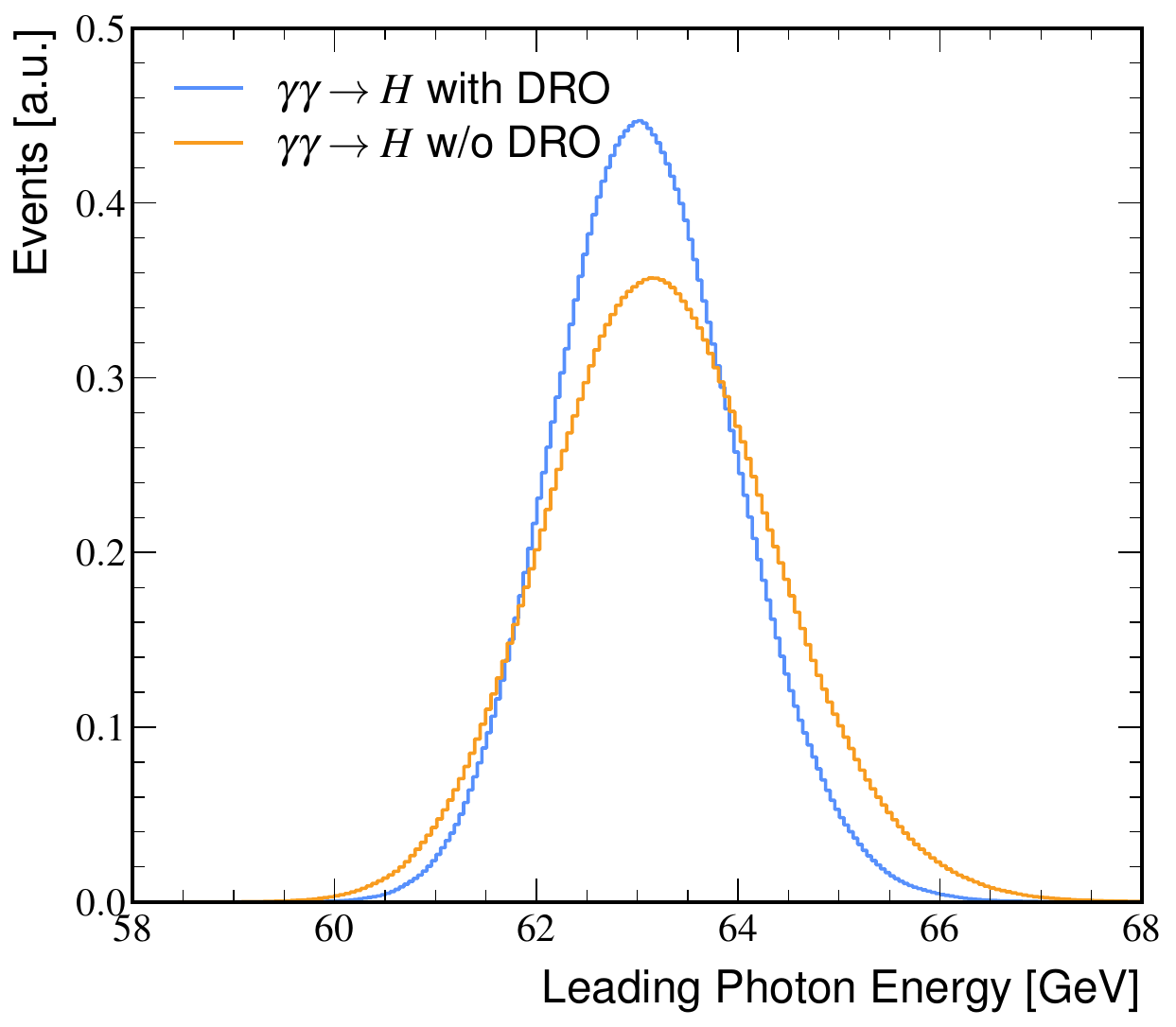}
    \end{minipage}
    \caption{Distributions of the leading photon energy (top left), subleading photon energy (top right) for the $H\to \gamma\gamma$ signal and the dominant background $\gamma\gamma \to \gamma\gamma$ prior to any selection cuts. We also overlay the $e^+e^-\to ZH, (Z\to \nu \nu), (H\to \gamma\gamma)$ for a comparison of the photon energies. The distributions are normalized such that the area under the curve is unity.}
    \label{fig:HAA_plots}
\end{figure}

The $H\to\gamma\gamma$ channel enjoys a unique advantage at a $\gamma\gamma$ collider such as the XCC, relative to $e^+e^-$ Higgs factories, due to resonant $s$-channel Higgs production. In particular, the Higgs decay photons are produced symmetrically and yield a monochromatic energy spectrum, modulo detector resolution, peaked around $m_H/2$, which can be easily distinguished from background processes. The top panels of Fig.~\ref{fig:HAA_plots} illustrate this by showing the leading and subleading photon energy distributions for both signal and background. This figure also shows the corresponding distributions for $e^+e^-\to ZH$ with $Z\to\nu\nu$ and $H\to\gamma\gamma$, highlighting the much broader spread in the latter case. The photon energy spectrum directly motivates the use of a high–resolution calorimeter, such as the Hybrid Dual-Readout (DRO) technology envisioned for the IDEA FCC-ee detector, which we implement in our \textsc{Delphes} setup. We refer the interested reader to Ref.~\cite{IDEAStudyGroup:2025gbt} for more detail, but by separately measuring the scintillation and Cerenkov components of the shower, DRO calorimetry substantially sharpens the reconstructed photon energy and thus further enhances the already narrow $H\to\gamma\gamma$ peak. This improvement is illustrated in the bottom right panel of Fig.~\ref{fig:HAA_plots}, where the leading photon energy distribution for $\gamma\gamma\to H$ with Dual-Readout is visibly narrower and more sharply peaked around $m_H/2$ than in the case without DRO. For this study, we present analyses both with and without DRO calorimetry.

Our pre-selection exploits this narrow photon energy spectrum by requiring two isolated photons, with the leading photon energy constrained to lie between 61 and 65 GeV and the subleading photon energy between 60 and 64 GeV. We further impose a diphoton invariant-mass requirement within the Higgs resonance window, $119 < m(\gamma\gamma) < 128$ GeV. A veto is applied to events containing an isolated lepton or jet with $E>8$ GeV. These criteria effectively suppress the $\gamma\gamma \to \gamma\gamma$ background as well as backgrounds originating from $e^+e^-$ and $e^-e^-$ pairs from photon conversions in the detector material, as well as events in which jets or other objects are misidentified as photons (fakes), while retaining high efficiency for genuine $H\to\gamma\gamma$ decays. The only significant background surviving the cuts is $\gamma\gamma\to \gamma\gamma$ as summarized in Tab.~\ref{tab:h_AA}.

\subsection{\texorpdfstring{\boldmath $H\to \gamma Z (\to \ell \overline{\ell})$}{}}

\begin{table}
\centering
\begin{tabular}{ l l l}
\hline
 \textbf{Process} & \textbf{Before Preselection} & \textbf{After Preselection}  \\ \hline
 $\gamma\gamma \to H\to \gamma Z(\to \ell\ell)$ & 165 & 117\\ \hline
$\gamma \gamma \to \gamma Z$ & 1,642 & 28\\
$\gamma\gamma \to \ell \ell$ & 2,135,184,138 & 4,272\\
$\gamma e \to \ell \ell \ell$ & 1,189,141,510 & 946 \\
$\gamma e \to \ell \ell $ & 2,409,926,020   & 4,820  \\
$e^-e^- \to e^-e^-  $ & 27,825,925,000    & 256,518  \\
 \hline
\end{tabular}
\caption{Number of events before and after the $H\to \gamma Z(\to \ell\ell)$ pre-selection filters for signal and background. \label{tab:h_AZ_ll}}
\end{table}

Similar to the $H\to\gamma\gamma$ channel, the $s$-channel production yields a photon energy distribution for $H\to\gamma Z$ that is sharply peaked around $(m_H^2-m_W^2)/2m_H \approx 29$ GeV. Combined with the fully-leptonic decays of the $Z$ boson, precise reconstruction of the Higgs mass is possible by considering the invariant mass of the photon di-lepton system. Our preselection is thus motivated by the aforementioned features of the fully-leptonic $H\to Z \gamma$ topology. For event selection, we require exactly two light leptons with $E>30$ (20) GeV for the leading (subleading) lepton. In addition, events must contain exactly one isolated photon with $24~\mathrm{GeV} < E< 36$ GeV. Final states with hadronically decaying $\tau_h$ are not considered. Further, we require the di-lepton mass to be around the $Z$-boson peak at $80 ~\mathrm{GeV} < m(\ell \ell) < 105$ GeV. In addition, we require virtually no missing energy with $E_T^\mathrm{miss}<2.5$ GeV. Finally, we require the invariant mass of the photon di-lepton system to be sharply peaked around the Higgs mass at $118 ~\mathrm{GeV} < m(\gamma\ell \ell) < 125$ GeV. The signal and background counts after the filters are summarized in Table \ref{tab:h_AZ_ll}.

\subsection{\texorpdfstring{\boldmath $H\to \gamma Z (\to q\overline{q})$}{}}

\begin{table}
\centering
\begin{tabular}{ l l l}
\hline
 \textbf{Process} & \textbf{Before Preselection} & \textbf{After Preselection}  \\ \hline
 $\gamma\gamma \to H\to \gamma Z(\to qq)$ & 1,540 & 703\\ \hline
$\gamma \gamma \to \gamma Z$ & 1,642 & 108\\
$e^-e^- \to e^-e^-  $ & 27,825,925,000    & 19,279  \\
$\gamma^*\gamma^* \to q \overline{q}$ & 128,469,182 & 5,287 \\
$\gamma\gamma \to \ell \ell$ & 2,409,926,020 & 2,029
 \\ \hline
\end{tabular}
\caption{
Number of events before and after the $H\to \gamma Z(\to qq)$ pre-selection filters for signal and background.\label{tab:h_AZ_qq}}
\end{table}

Similar to the fully leptonic channel, the nearly monochromatic photon in $H\to Z\gamma$ is independent of the subsequent $Z$ decay. For the hadronic mode $Z\to q\bar q$, this clean photon signature can be combined with the dijet system to reconstruct the Higgs mass from the invariant mass of the $\gamma jj$ system. Hence, we require events to contain exactly one or two jets, accounting for merged scenarios, with the leading (subleading) jet satisfying $E > 25~(15)~\mathrm{GeV}$, and exactly one isolated photon with $24~\mathrm{GeV} < E < 36~\mathrm{GeV}$. Events with isolated light leptons or hadronically decaying $\tau_h$ are vetoed. To select $Z$ candidates, we demand that the dijet invariant mass lie near the $Z$-boson resonance, $70~\mathrm{GeV} < m(jj) < 110~\mathrm{GeV}$. Finally, we require the invariant mass of the photon–dijet system to be consistent with Higgs production, $95~\mathrm{GeV} < m(\gamma jj) < 115~\mathrm{GeV}$. Table \ref{tab:h_AZ_qq} lists the signal and background counts before and after the preselections.

\subsection{\texorpdfstring{\boldmath $H\to \gamma Z (\to \nu\overline{\nu})$}{}}

\begin{table}
\centering
\begin{tabular}{ l l l}
\hline
 \textbf{Process} & \textbf{Before Preselection} & \textbf{After Preselection}  \\ \hline
 $\gamma\gamma \to H\to \gamma Z(\to \nu \nu)$ & 318 & 201\\ \hline
$\gamma \gamma \to \gamma Z$ & 1,642 & 89\\
$\gamma e \to \ell \ell \ell$ & 1,189,141,510 & 21,998 \\
$\gamma e \to e\ell \ell $ & 2,409,926,020   & 69,888  \\
$e^-e^- \to e^-e^-  $ & 27,825,925,000    & 667,822  \\
 \hline
\end{tabular}
\caption{Number of events before and after the $H\to \gamma Z(\to \nu \nu)$ pre-selection filters for signal and background. \label{tab:h_AZ_vv}}
\end{table}
 For the $H\to \gamma Z (\to \nu\overline{\nu})$ channel, we target a single photon with missing energy topology. The event selection requires a single isolated photon with transverse energy between 24 and 36 GeV, similar to the other $H\to \gamma Z$ channels. In addition, we require large missing transverse energy $E_T^\mathrm{miss}$ from the $Z$ boson decaying to neutrinos. Finally, we veto any jets or leptons in the final state. 

\subsection{\texorpdfstring{\boldmath $H\to VV^* \to \ell \ell \nu \nu$}{}}
\begin{figure}
    \centering
    \begin{minipage}{0.495\textwidth}
        \includegraphics[width=\linewidth]{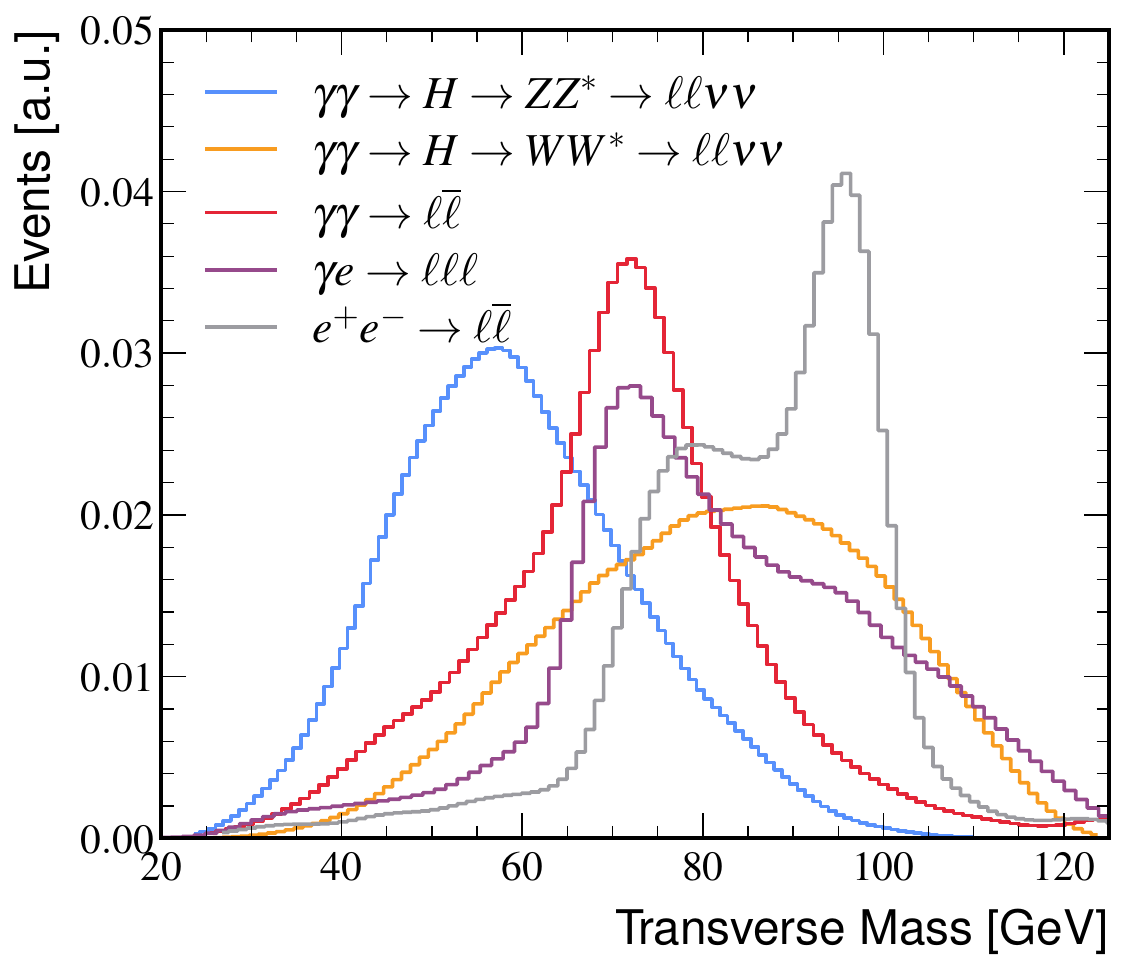}
    \end{minipage}
    \begin{minipage}{0.495\textwidth}
        \includegraphics[width=\linewidth]{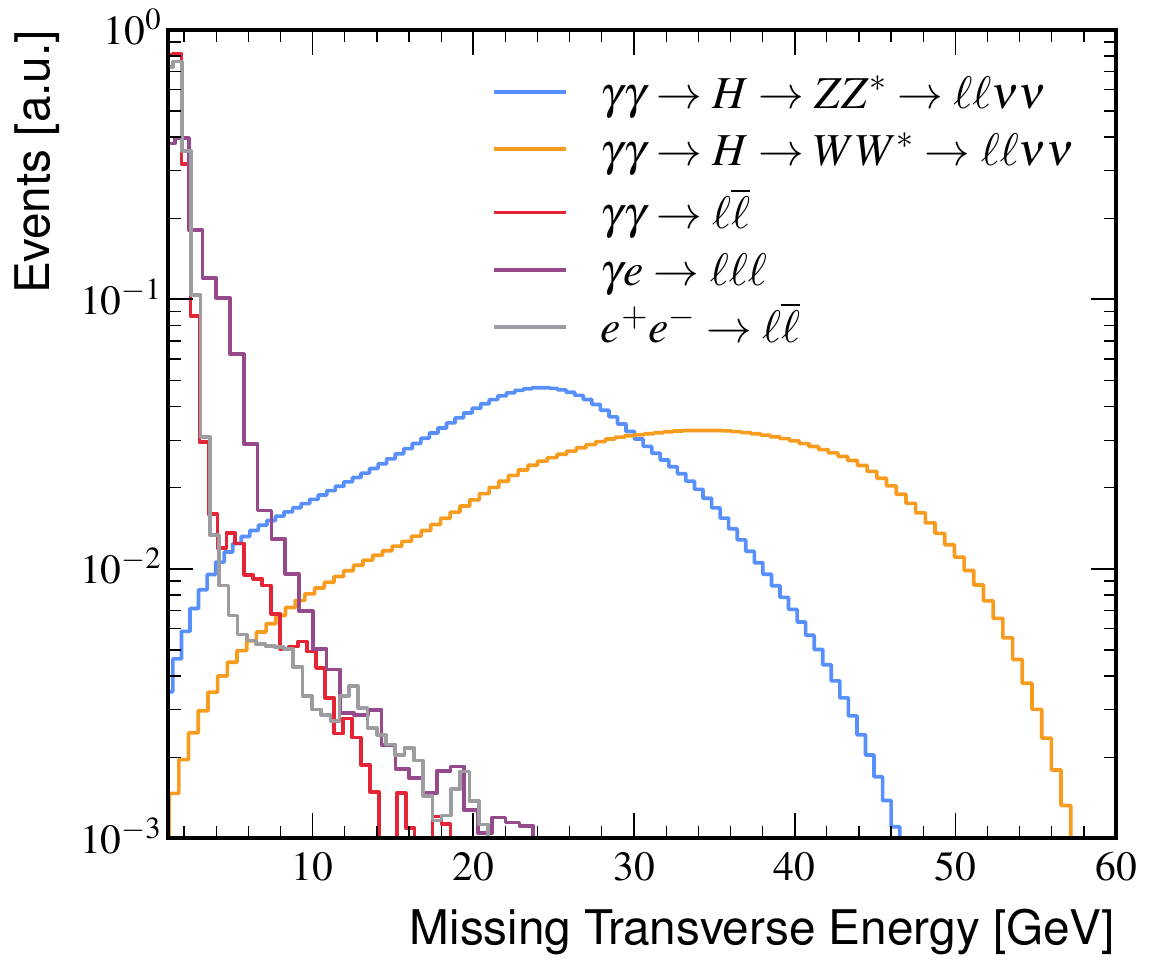}
    \end{minipage}
    \caption{Distributions of the reconstructed transverse mass (left) and missing transverse energy (right) for the $H\to WW^* \to \ell\ell \nu \nu$ signal and dominant backgrounds prior to any selection cuts. The distributions are normalized such that the area under the curve is unity.}
    \label{fig:Hww_2l2nu_plots}
\end{figure}

\begin{table}
\centering
    \begin{tabular}{ l l l}
\hline
 \textbf{Process} & \textbf{Before Preselection} & \textbf{After Preselection}  \\ \hline
 $\gamma\gamma \to H\to WW^* \to \ell \ell \nu \nu$ & 24,655 & 10,397 \\ \hline
$\gamma \gamma \to \ell \ell \overline{\ell \ell}$ & 225,100,651 & 7,209 \\
$\gamma\gamma \to \ell \overline{\ell}$ & 2,135,184,138& 337,484\\
$\gamma e \to \ell \ell \ell  $ & 1,189,141,510& 435,820\\
 $e^+e^- \to \ell\overline{\ell}$ & 2,245,722,200 & 992,609 \\ 
 \hline
\end{tabular}
\caption{Number of events before and after the $H\to WW^* \to \ell \ell \nu \nu$ pre-selection filters for signal and backgrounds. \label{tab:hwwllvv}}
\end{table}

\begin{table}
\centering
    \begin{tabular}{ l l l}
\hline
 \textbf{Process} & \textbf{Before Preselection} & \textbf{After Preselection}  \\ \hline
 $\gamma\gamma \to H\to ZZ^* \to \ell \ell \nu \nu$ & 594 & 269 \\ \hline
$\gamma \gamma \to \ell \ell \overline{\ell \ell}$ & 225,100,651 & 82,211 \\
$\gamma\gamma \to \ell \overline{\ell}$ & 2,135,184,138& 12,318,152\\
$\gamma e \to \ell \ell \ell  $ & 1,189,141,510& 771,967\\
 $e^+e^- \to \ell\overline{\ell}$ & 2,245,722,200 & 3,024,988 \\ 
 \hline
\end{tabular}
\caption{Number of events before and after the $H\to ZZ^* \to \ \nu\overline{\nu}\ell \overline{\ell}$ pre-selection filters for signal and backgrounds. \label{tab:hzzllvv}}
\end{table}

For the fully leptonic $H\to VV^{*} \to\ell\ell\nu\nu$ channels, we target a di-lepton with missing momentum topology with minimal hadronic activity. Jets are still reconstructed with the $e^{+}e^{-}$ inclusive $k_\mathrm{T}$ algorithm with $E_{\min}=8$ GeV, and $R=1.5$, but only to define a jet veto and control radiation. Selected events must contain exactly two isolated electrons, muons, or $\tau$-tagged jets with $E>15$ GeV and relative isolation $<0.15$ in case of the light leptons. A jet veto subsequently removes all hadronic activity. Because two neutrinos are expected, we require significant missing transverse energy, $E_{\mathrm T}^{\rm miss}>25$ GeV, which effectively rejects $Z/\gamma^{*}\to \ell\ell $ and other fully visible processes. Finally, we impose a transverse-mass requirement $m_\mathrm{T}(\vec p_{\rm miss},\ell\ell)>85$ GeV for the $WW^*$ channel and $m_\mathrm{T}<65$ GeV for the $ZZ^*$ channel; this exploits the broad $m_\mathrm{T}$ spectrum of $WW^*$ decays with neutrinos and further separates signal from $Z\to \ell \ell$ with spurious MET and from low-mass electroweak backgrounds. The effect of the  $m_\mathrm{T}$ and $E_{\mathrm T}^{\rm miss}$  cuts on $WW^{*}\to \ell\ell\nu\nu$ can be inferred from the left and right plots in Fig.~\ref{fig:Hww_2l2nu_plots}.  After these selections, the dominant backgrounds are continuum $WW^*$ and $ZZ^{*}\to \ell\ell\nu\nu$ as summarized in Tables \ref{tab:hwwllvv} and \ref{tab:hzzllvv}.

\subsection{\texorpdfstring{\boldmath $H\to ZZ^* \to q\overline{q} \nu \overline{\nu}$}{}}

\begin{figure}
    \centering
    \begin{minipage}{0.495\textwidth}
        \includegraphics[width=\linewidth]{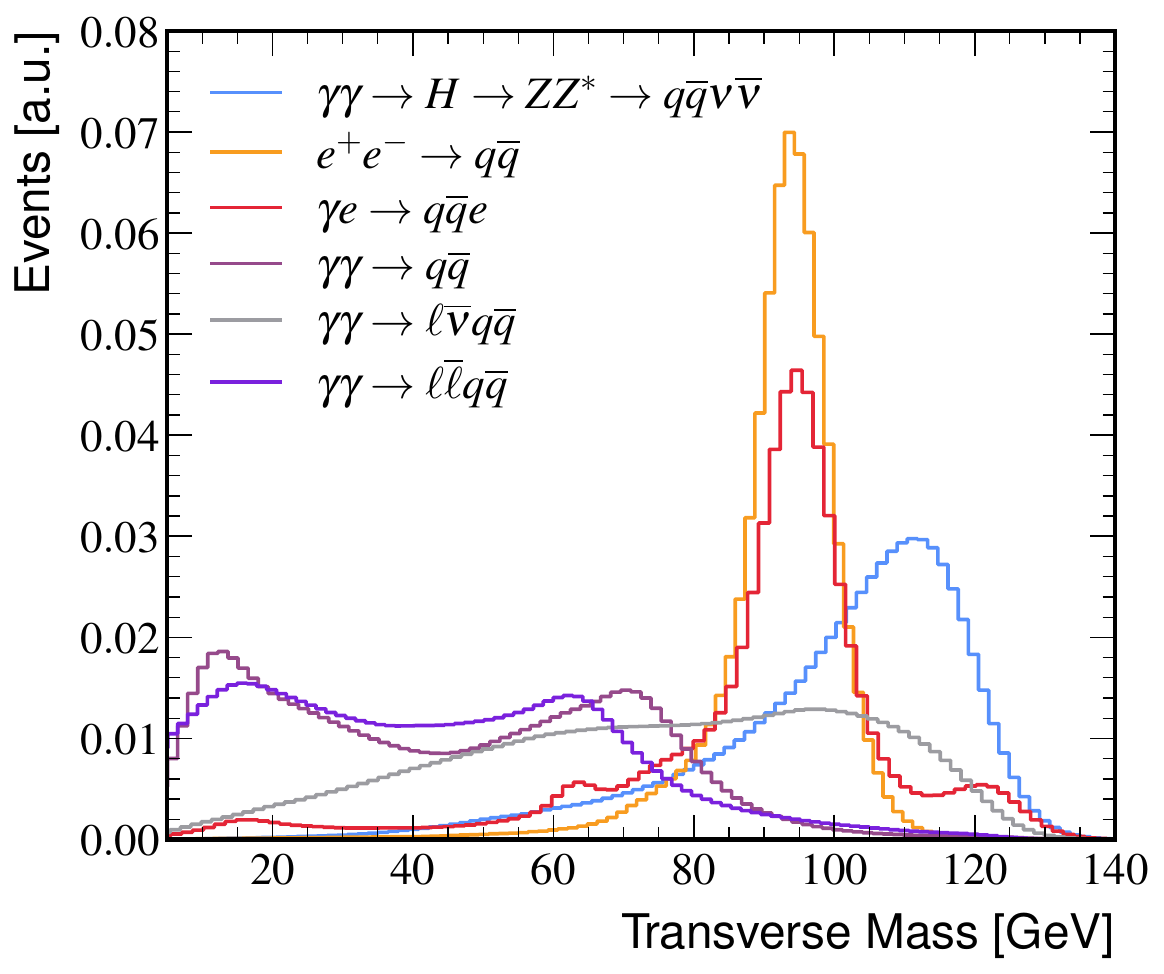}
    \end{minipage}
    \begin{minipage}{0.495\textwidth}
        \includegraphics[width=\linewidth]{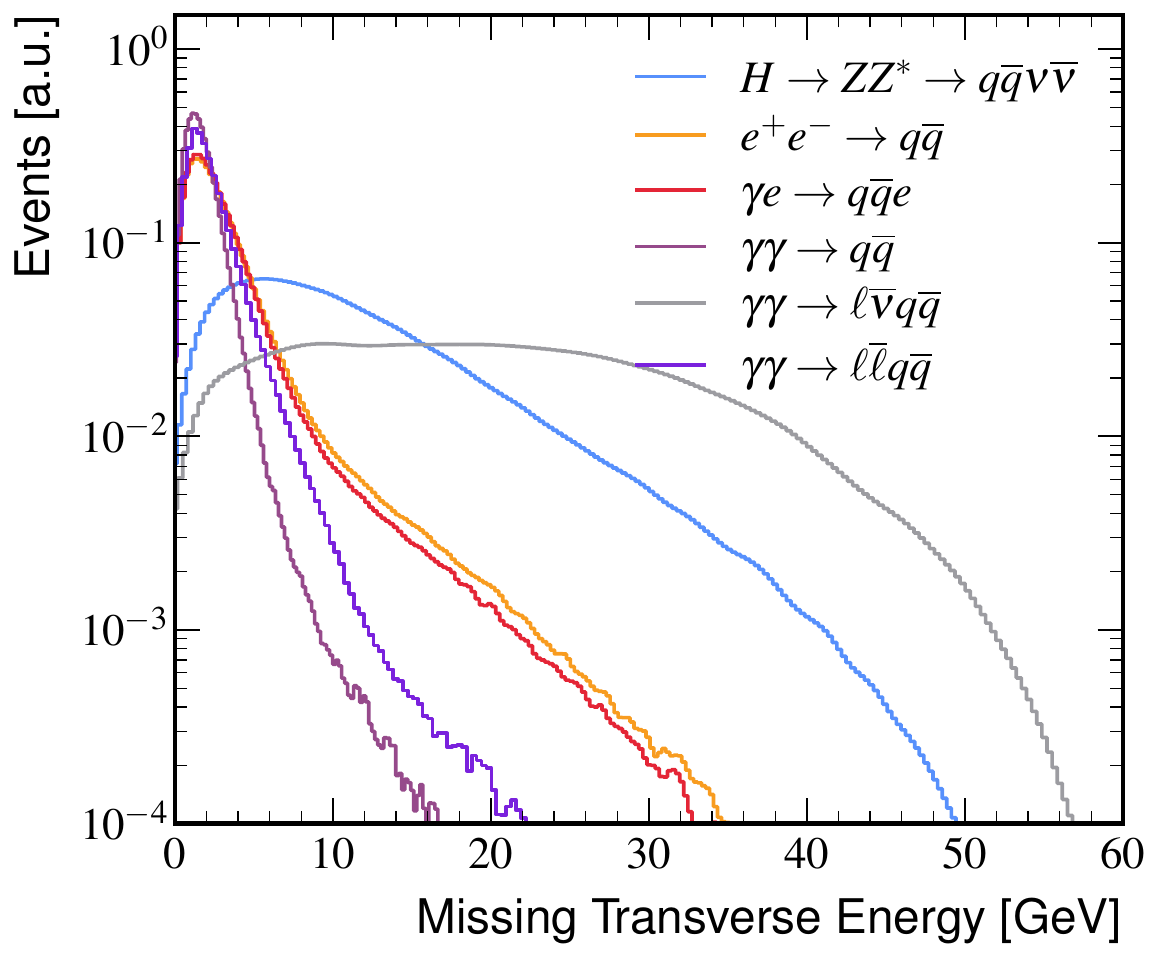}
    \end{minipage}
    \caption{Distributions of the reconstructed transverse mass (left) and missing transverse energy (right) for the $H\to ZZ^* \to q\overline{q} \nu \overline{\nu}$ signal and dominant backgrounds prior to any selection cuts. The distributions are normalized such that the area under the curve is unity.}
    \label{fig:Hzz_2q2nu_plots}
\end{figure}

For the semi leptonic $H\to ZZ^* \to q\overline{q} \nu \overline{\nu}$ channel, we target a dijet with missing momentum topology. Selected events must contain either one or two jets resulting from the $ZZ^*$ decays, and a veto is applied for events with electrons, muons, or $\tau$-tagged jets with $E>15$ GeV and relative isolation $<0.15$ for the light leptons. Since we expect two neutrinos and thus genuine missing energy, we require significant missing transverse energy, $E_{\mathrm T}^{\rm miss}>7.5$ GeV, which effectively rejects $Z/\gamma^{*}\to \ell\ell $ and other fully visible processes. Finally, we impose a transverse-mass requirement $m_\mathrm{T}(\vec p_{\rm miss},\ell\ell)>100$ GeV. The transverse mass and missing energy distributions are shown in Fig.~\ref{fig:Hzz_2q2nu_plots} for the $H\to ZZ^* \to q\overline{q} \nu \overline{\nu}$ signal and domaninant backgrounds and the signal and background counts after the preselections are shown in Table \ref{tab:hzzvvqq}.

\begin{table}
\centering
\begin{tabular}{ l l l }
\hline
\textbf{Process} & \textbf{Before Preselection} & \textbf{After Preselection} \\ \hline

$\gamma\gamma \to H \to ZZ \to q\overline{q}\nu\bar\nu$ & 4,158 & 2,159 \\ \hline

$e^+e^- \to q\overline{q}$ & 459,001,600 & 128,747 \\
$\gamma\gamma \to \ell\ell q\overline{q}$ & 9,771,761 & 26,512 \\
$\gamma e \to q\overline{q}e$ & 74,887,283 & 18,688 \\
$\gamma\gamma \to \ell\overline{\nu}q\overline{q}$ & 61,708 & 12,159 \\
$\gamma e \to q\overline{q}\nu$ & 626,064 & 6,222 \\
$\gamma\gamma \to q\overline{q}$ & 513,755,395 & 5,383 \\
$e^+e^- \to \ell\overline{\ell}q\overline{q}$ & 865,248 & 2565 \\
\hline
\end{tabular}
\caption{Number of events before and after the $H\to ZZ^* \to q\overline{q}\nu \overline{\nu}$ pre-selection filters for signal and backgrounds. \label{tab:hzzvvqq}}
\end{table}

\subsection{\texorpdfstring{\boldmath $H\to ZZ^* \to q\overline{q} \ell \overline{\ell}$}{}}
The combined leptonic-hadronic decays of the $Z$-boson pair from the Higgs decay lead to a dijet plus di-lepton topology. Therefore, for event selection, we require one or two reconstructed jets and two isolated muons (electrons or muons) or two $\tau$-tagged jets with $E>8$ GeV. Similar to the other fully-visible topologies, we require $E_\mathrm{T}<15$ GeV to reject processes with prompt neutrinos. In addition we require the Durham distance between the dijet (or jet, in the case of a single jet) and di-lepton objects to be larger than $\sqrt{d}_{12}>80$ GeV and require the invariant mass of all objects $m_{\mathrm{visible}}>105$ GeV to remove topologies incompatible with a Higgs decay. These filters have a signal efficiency of $\approx 60\%$ while reducing the dominant backgrounds by 2-3 orders of magnitude as summarized in Table~\ref{tab:hzzllqq}.

\begin{table}
\centering
\begin{tabular}{ l l l }
\hline
\textbf{Process} & \textbf{Before Preselection} & \textbf{After Preselection} \\ \hline
$\gamma\gamma \to  H\to ZZ^* \to q\overline{q}\ell \overline{\ell}$ & 2,079 & 938 \\ \hline
$\gamma\gamma \to \ell \overline{ \ell} q\overline{q}$ & 9,771,761 & 77,914 \\
$e^+e^- \to \ell\overline{\ell}q\overline{q}$ & 865,248 & 5,771 \\
$\gamma e \to q\bar q e$                                      & 74,887,283   & 7,526\\
\hline
\end{tabular}
\caption{Number of events before and after the $H\to ZZ^* \to q\overline{q}\ell \overline{\ell}$ pre-selection filters for signal and backgrounds. \label{tab:hzzllqq}}
\end{table}


\subsection{\texorpdfstring{\boldmath $H\to ZZ^* \to \ell \ell \overline{\ell\ell}$}{}}
The $H\to ZZ^*\to \ell \ell \ell \ell$ channel, similar to the $H\to \mu^+\mu^-$ channel, is relatively clean final state, and we seek to utilize the excellent muon and electron resolutions to isolate a narrow $4\ell$ resonance near the Higgs mass. Events are required to contain exactly four oppositely charged, well-identified leptons (including $\tau$-tagged jets) with $E>8$ GeV and relative isolation $<0.15$. We veto any jets (excluding the lepton candidates) to mitigate any hadronic activity. Because the majority of the signal is expected to be fully visible, with the only exception being the leptonic $\tau$ decays, we further require low missing transverse energy, $E_{\mathrm T}^{\rm miss}<5$ GeV, to reject channels with neutrinos. The 4-lepton invariant mass is then restricted to the Higgs resonance i.e.~$ 105<m(\ell \ell \overline{\ell \ell})<130$ GeV. As before, final state electrons and muons are ``dressed'' with nearby FSR photons within a $\Delta R < 0.1$ radius to sharpen the mass peak. These pre-selection filters have a signal efficiency of over $70\%$ while eliminating nearly all sources of background with the exception of the irreducible $4\ell$ backgrounds listed in Table \ref{tab:h_zz_4l}.

\begin{table}
\centering
\begin{tabular}{ l l l }
\hline
\textbf{Process} & \textbf{Before Preselection} & \textbf{After Preselection} \\ \hline
$\gamma\gamma \to H \to ZZ \to \ell\ell\overline{\ell\ell}$ & 297 & 206 \\ \hline
$\gamma\gamma \to \ell\ell\overline{\ell\ell}$ & 225,100,651 & 376,653 \\
$e^+e^- \to \ell\ell\overline{\ell\ell}$ & 456,869 & 31,821 \\
\hline
\end{tabular}
\caption{Number of events before and after the $H\to ZZ^* \to \ell\ell\overline{\ell\ell}$ pre-selection filters for signal and backgrounds. \label{tab:h_zz_4l}}
\end{table}

\section{Analysis with Machine Learning}\label{sec:MLAnalysis}

For this study, the analysis of signal and background events is performed using state-of-the-art machine learning event classifiers paired with a genetic algorithm for multi-dimensional cut optimization. Machine learning approaches offer significant advantages compared to traditional event classification techniques. Unlike conventional multivariate methods, machine learning models can simultaneously consider and encode the entirety of the high-dimensional event space and thus enact sophisticated selection criteria that exploit correlations across high-dimensional features. This makes them well-suited in high-energy physics where event topologies are inherently complex and high-dimensional.
Recently, significant effort has been dedicated to developing point-cloud-based machine learning architectures to process unordered sets of particles with varying cardinalities while incorporating geometric invariances. These models have found widespread adoption in jet classification and LHC searches, with demonstrated advantages over traditional approaches. For this study, we employ a set transformer architecture, a state-of-the-art deep learning model specifically designed for permutation-invariant classification of unordered sets and point clouds. The effectiveness of machine learning models such as boosted decision trees, multi-layer perceptrons, and transformer architectures has been validated in numerous experimental and phenomenological studies. Point-cloud strategies based on deep sets and edge convolutions have been shown to outperform traditional methods like BDTs and MLPs for collider event classification. Set transformers in particular have demonstrated superior performance across a range of tasks, substantially motivating their adoption in our analysis. Our findings indicate that this architecture provides significant improvements in sensitivity compared to earlier BDT-based approaches.

\begin{table}
\centering
\label{tab:infeats}
    \begin{tabular}{ c c c  }
\hline
\textbf{Category} & \textbf{Variable} &\quad \textbf{Definition} \quad \\
\hline
\multirow{4}{*}{Kinematics}

  & $\theta$ & The polar angle of the particle \\
& $\phi$ & The azimuthal angle of the particle \\
& $p_{T}$ & The transverse momentum of the particle \\
& $E$ & The energy of the particle \\
\hline
\multirow{6}{*}{Particle ID}
  & $q$ & The electric charge of the particle \\
& $e$ & Boolean, 1 if the particle is identified as an electron \\
& $\mu$ & Boolean, 1 if the particle is identified as a muon \\
& $\gamma$ & Boolean, 1 if the particle is identified as a photon\\
& $h^\pm$ & Boolean, 1 if the particle is identified as a charged hadron \\
& $h^0$ & Boolean, 1 if the particle is identified as a neutral hadron \\
\hline
\multirow{4}{*}{Trajectory}
  & $d_0$ & Transverse impact parameter of the track \\
& $ d_z$ & Longitudinal impact parameter of the track \\
& $\sigma_{d_0}$ & Uncertainty in the transverse impact parameter \\
& $\sigma_{d_z}$ & Uncertainty in the longitudinal impact parameter \\ \hline
\end{tabular}
\caption{The set of per-particle input features used to train the set transformer model.}
\end{table}

The set transformer accepts as input an unordered set (in our case, of final-state particle flow objects), $X = \{x_i\}_{i=1}^{n}$ with $x_i \in \mathbb{R}^{d_{\text{in}}}$, where $d_{\text{in}}$ is the per-element input feature dimension and $n$ is the variable number of particles in each event. We use the full gamut of particle flow information, including kinematics, particle identification, and trajectory displacement, as input features. The complete list of the 15 features for each particle is summarized in Table~\ref{tab:infeats}. Since $p_\mathrm{T}$ and $E$ typically have long-tail distributions, they are log-transformed before being passed to the network. The architecture implements a permutation-invariant classifier through an encoder–decoder framework specifically designed to handle variable-cardinality sets. The encoder comprises two induced self-attention layers, each of width $d_{\text{hidden}}$. Rather than applying full self-attention between all pairs of input elements, which would incur $\mathcal{O}(n^2)$ computational complexity, the model introduces $m$ learnable inducing vectors $I \in \mathbb{R}^{m \times d_{\text{hidden}}}$ that serve as a compressed summary of the input set. This reduces computational complexity to $\mathcal{O}(mn)$ for fixed $m$, which is particularly important for our application since individual $\gamma\gamma$ collision events can contain $\mathcal{O}(100)$–$\mathcal{O}(1000)$ particles, rendering full self-attention prohibitively expensive.
The encoder operates via a cross-attention mechanism between the inducing vectors and every element in the input set. This cross-attention allows the model to encode both pairwise interactions and higher-order correlations between particles. Multi-headed attention is employed to stack multiple attention blocks, with each block's output preserving information about pairwise and higher-order interactions among set elements, thereby enabling the model to learn complex combinatorial structures. After the encoder maps the input data $X \in \mathbb{R}^{n \times d_{\text{in}}}$ to learned feature representations $Z \in \mathbb{R}^{n \times d_{\text{hidden}}}$, a decoder aggregates these distributed representations into a single global summary vector. The aggregation employs a self-attention-based pooling mechanism to generate a fixed-dimensional representation independent of set cardinality. This aggregated representation is subsequently passed through a multi-layer perceptron feed-forward network to produce the final binary classification output (signal vs. background). Model architecture configuration and training were selected through a mild, manual trial-and-error hyperparameter tuning procedure. The hidden dimension is set to $d_{\text{hidden}} = 256$, the number of learnable inducing points to $m = 32$, and the number of attention heads to $n_h = 8$. Two layers of induced self-attention blocks are employed in both the encoder and decoder stages. The resulting model contains approximately 1.78 million trainable parameters and requires roughly 10 milliseconds per forward pass with a batch size of 256 on a single NVIDIA A100 GPU. The \textsc{AdamW} optimizer is used with an initial learning rate $\alpha = 10^{-4}$, combined with a cosine annealing learning rate scheduler in conjunction with warm restarts every 10 epochs for stable convergence. Binary cross-entropy loss with logits is used for numerical stability in the classification task, comparing model predictions against binary labels (0 for background, 1 for signal). Data are split into training, validation, and test subsets using a 75–10–15 stratified split. Events are processed in batches of 256 during training. 

\begin{figure}
    \centering
    \begin{minipage}{0.495\textwidth}
        \includegraphics[width=\linewidth]{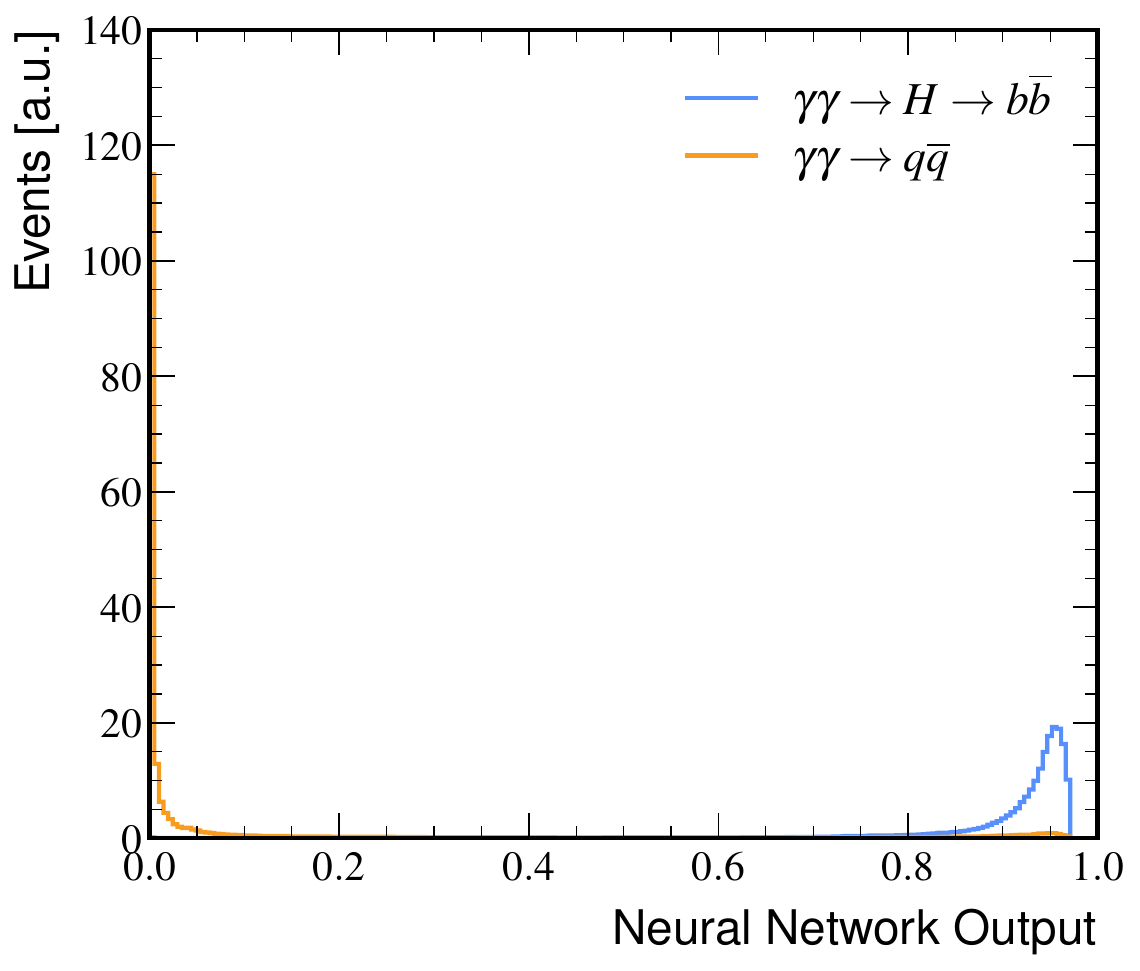}
    \end{minipage}
    \begin{minipage}{0.495\textwidth}
        \includegraphics[width=\linewidth]{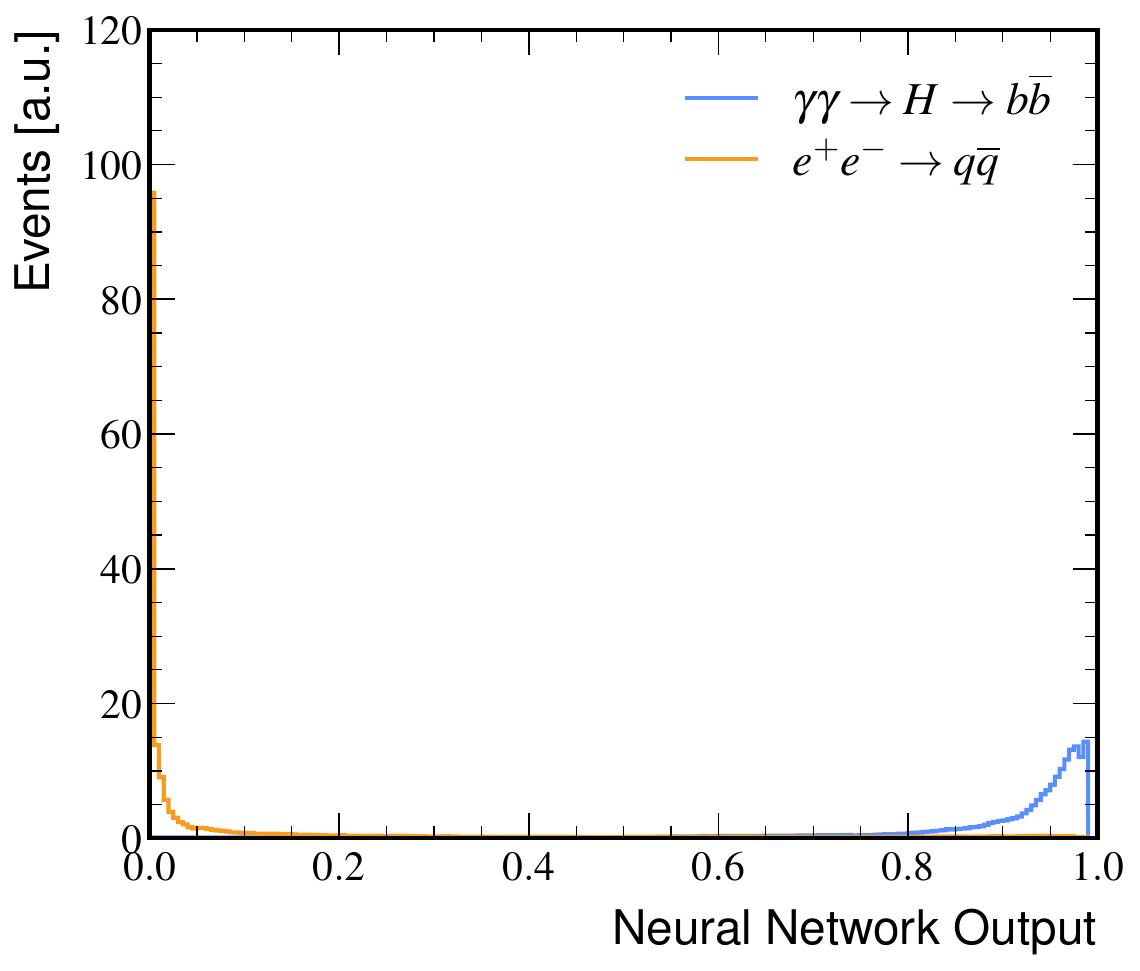}
    \end{minipage}
    \caption{Output distributions for the set transformer networks discriminating between the $H\to b\overline{b}$ signal and the $\gamma\gamma \to q\overline{q}$ (left) and $e^+e^-\to q\overline{q}$ (right) backgrounds. The distributions are normalized such that the area under the curve is unity.}
    \label{fig:NN_outs}
\end{figure}

The machine learning algorithm's output on the independent test set, i.e.~data that is completely unseen during model training and validation, is used for all subsequent significance calculations, ensuring unbiased performance estimates. For each physics channel analyzed (with the exception of $H \to \mu^+ \mu^-$, $H \to \gamma\gamma$, and $H \to Z\gamma(\to \ell\ell/\nu \nu)$) for which we use BDTs, we train an ensemble of set transformer classifiers for signal-background discrimination. Specifically, for each background process, a separate set transformer is trained to discriminate between that particular background and the signal, resulting in one binary classifier per background. The outputs of these independent classifiers define a multi-dimensional feature space where each coordinate corresponds to the discriminant score of one signal-vs-background classifier. Fig.~\ref{fig:NN_outs} displays neural network output distributions for the signal and two dominant backgrounds. The output ranges from 0 to 1, representing the likelihood of an event being either signal-like (output near 1) or background-like (output near 0).

A genetic algorithm is subsequently applied to this multi-dimensional space to determine optimal selection thresholds that maximize the signal significance. The genetic algorithm operates on threshold vectors $\mathbf{\theta} = [\theta_1, \theta_2, \ldots, \theta_K]$, where $K$ is the number of background processes and each component represents the required minimum output score for the corresponding classifier. An event is classified as signal if and only if it passes simultaneous threshold requirements on all classifiers: i.e.~the neural network output from each signal-vs-background discriminant exceeds its corresponding threshold. Events failing to meet any threshold are classified as background. The genetic algorithm's fitness function is the signal significance metric $ S / \sqrt{S + B}$, where $S$ is the weighted signal count and $B$ is the total weighted background count surviving the selection thresholds. Event weights are derived from expected 10-year scaled yields normalized by the event sample size, properly accounting for different cross-sections and branching ratios across processes.

The genetic algorithm employs a population of 120 individuals, each encoding a complete threshold vector. Tournament selection with tournament size 5 is used to identify parents. Offspring are generated through blend crossover with blend factor $\alpha = 0.5$, with probability 0.7, and Gaussian mutation with standard deviation 0.3, with probability 0.3 per individual and 0.3 per coordinate. The algorithm runs for 80 generations, maintaining a Hall of Fame to track the best threshold configuration encountered. Upon convergence, the optimal threshold vector is applied to all datasets to determine final signal and background event survival rates. This approach naturally accommodates the large differences in cross-section between the backgrounds and the multi-dimensional nature of the selection problem and automatically discovers threshold combinations that may not be obvious through manual analysis.

Training and evaluation of all models were conducted in a high-performance computing environment utilizing NVIDIA A100 GPUs. The canonical \textsc{PyTorch} deep learning library was employed for implementing, configuring, training, and evaluating the set transformer classifiers while the genetic algorithm was implemented using the Distributed Evolutionary Algorithms in Python (DEAP) framework. \textsc{PyTorch} and DEAP are well-established in their reputation for flexibility, ease-of-use, and computational efficiency, this motivating their use in this study.

\section{Results}\label{sec:res}

\begin{table}[t]
    \centering
    \small
    \setlength{\tabcolsep}{6pt}
    \renewcommand{\arraystretch}{1.1}
    \begin{tabular}{l l l l l l}
        \hline
        \textbf{Channel} & $\boldsymbol{S}$ & $\boldsymbol{B}$ & $\boldsymbol{S/B}$ & $\boldsymbol{S/\sqrt{S+B}}$ & $\boldsymbol{\delta(\sigma\times \mathrm{Br})}$ \\
        \hline
        $H\to b\bar b$                 & 342,144 & 34,292 & 9.98  & 557.7 & 0.18 \\
        $H\to c\bar c$                 & 16,642  & 34,437 & 0.483 & 73.6  & 1.4 \\
        $H\to gg$                      & 53,283  & 85,359 & 0.624 & 143.1 & 0.70 \\
        $H\to s\overline{s}$                      & 164     & 1,543  & 0.106 & 3.97  & 26 \\
        $H\to WW^*\to qqqq$            & 33,094  & 32,566 & 1.02  & 129.2 & 0.77 \\
        $H\to ZZ^*\to qqqq$            & 3,719   & 34,419 & 0.108 & 19.0  & 5.3 \\
        $H\to \mu^+\mu^-$              & 98      & 17     & 5.76  & 9.14  & 11 \\
        $H\to \tau^+\tau^-$            & 27,088  & 10,023 & 2.70  & 140.6 & 0.71 \\
        $H\to WW^*\to \ell\nu qq$      & 15,420  & 5,565  & 2.77  & 106.4 & 0.94 \\
        $H\to \gamma\gamma$ (with DRO)   & 2,177   & 201    & 10.8  & 44.6  & 2.2 \\
        $H\to \gamma\gamma$ (no DRO) & 1,781   & 299    & 5.96  & 39.1  & 2.6 \\
        $H\to \gamma Z(\to \ell\ell)$  & 69      & 103    & 0.67  & 5.26  & 19 \\
        $H\to \gamma Z(\to qq)$        & 212     & 228    & 0.93  & 10.1  & 9.9 \\
        $H\to \gamma Z(\to \nu\nu)$    & 116     & 32     & 3.62  & 9.54  & 10.5 \\
        $H\to WW^*\to \ell\ell\nu\nu$  & 9,157   & 32,084 & 0.285 & 45.1  & 2.2 \\
        $H\to ZZ^*\to \ell\ell\nu\nu$  & 73      & 43     & 1.70  & 6.78  & 15 \\
        $H\to ZZ^*\to qq\nu\nu$        & 1,718   & 111    & 15.5  & 40.2  & 2.5 \\
        $H\to ZZ^*\to qq\ell\ell$      & 793     & 1,216  & 0.652 & 17.7  & 5.7 \\
        $H\to ZZ^*\to 4\ell$           & 125     & 118    & 1.06  & 8.02  & 12 \\
        \hline
    \end{tabular}
    \caption{Summary of post-selection signal ($S$) and background ($B$) yields for all Higgs decay channels in the 10-year XCC scenario, together with $S/B$, the signal significance $S/\sqrt{S+B}$, and the projected statistical precision on $\sigma(\gamma\gamma\to H)\times\mathrm{Br}$ quoted in $\%$ per channel.}
    \label{tab:xcc_results_summary}
\end{table}

Table~\ref{tab:xcc_results_summary} summarizes the final post-selection yields and projected sensitivities for all Higgs decay channels considered in this study. For each channel we report the expected number of selected signal events ($S$) and the corresponding total selected background yield ($B$) in the 10-year XCC scenario. We also quote the resulting purity, $S/B$, signal signfificance $S/\sqrt{S+B}$, and $\sigma(\gamma\gamma\to H)\times \mathrm{Br}(H\to X)$, denoted $\delta(\sigma\times \mathrm{Br})$, as obtained from the corresponding channel-specific selection. The dominant channel, $H \to b\bar{b}$, achieves a post-selection purity of $S/B \approx 10$ and a signal significance of $S/\sqrt{S+B} \approx 558$, corresponding to a projected statistical precision of $\delta(\sigma \times \mathrm{Br}) \approx 0.18\%$. The $H \to \tau^+\tau^-$ and $H \to gg$ channels follow closely, with precisions of $0.71\%$ and $0.70\%$, respectively. Notably, the $H \to c\bar{c}$ channel reaches a precision of $1.4\%$, a result that is unprecedented for any $\gamma\gamma$ collider concept. Most remarkably, the $H \to ss$ channel, with a branching fraction of $\mathcal{O}(10^{-4})$, reaches a significance of $\sim 4\sigma$ and a precision of $26\%$, representing, to our knowledge, the first demonstration that the Higgs-to-strange coupling can be probed at any collider---a result that enabled by the charge-suppressed $\gamma\gamma \to s\bar{s}$ background at a $\gamma\gamma$ collider.

The $H \to \gamma\gamma$ channel is particularly noteworthy at a $\gamma\gamma$ collider. Without a Dual-Readout (DRO) calorimeter, the analysis achieves $\delta(\sigma \times \mathrm{Br}) \approx 2.2\%$ with $S/B \approx 10.8$. With DRO calorimetry, which sharpens the photon energy resolution and thus the diphoton mass peak, the signal purity changes to $S/B \approx 6.0$ while the significance shifts to $S/\sqrt{S+B} \approx 39$, yielding $\delta(\sigma \times \mathrm{Br}) \approx 2.6\%$. Although the DRO configuration retains fewer signal events due to the tighter energy window enabled by the improved resolution, both scenarios demonstrate the unique advantage of resonant $s$-channel Higgs production at a $\gamma\gamma$ collider over $e^+e^-$ Higgs factories for this channel.

\begin{table}[h]
\centering
    \begin{tabular}{ l l l}
\hline
 \textbf{Coupling} & $\boldsymbol{\mathrm{HL{-}LHC}\ @\ 14\ \mathrm{TeV}}$ & $\boldsymbol{\mathrm{XCC\ 0.5\ ab}^{-1}@\ 125\ \mathrm{GeV}}$ \\ \hline
 $HZZ$ & 1.6 & 0.68 \\
   $HWW $    & 1.6 & 0.79 \\
   $Hbb $ & 3.6    & 0.71 \\
   $H\tau\tau $  &  1.9  & 0.78 \\
   $Hgg $   & 2.4  & 0.67 \\
   $Hcc $  &   -   & 1.0 \\
   $H\gamma\gamma $  &  1.8  & 0.095 \\
   $H\gamma Z $  & 6.8    & 3.1 \\
   $H\mu\mu $   & 3.0   & 2.8 \\ 
   $Hss $  &   -   & 13.0 \\
 \hline
\end{tabular}
\caption{Projected uncertainties in the Higgs boson couplings, in percent, for ATLAS and CMS combined at HL-LHC\cite{ATLAS:2025eii} and the XCC 
using the $\kappa$-framework  \cite{LHCHiggsCrossSectionWorkingGroup:2012nn} with the Higgs branching fraction to BSM final states fixed to zero.  HL-LHC results are included in the XCC kappa fit.The assumed luminosity for ATLAS and CMS is 3~ab$^{-1}$ each. The XCC luminosity corresponds to the integrated $\gamma\gamma$ luminosity defined in Section~\ref{sec:XCC}. \label{tab:couplingILC}}
\end{table}

\begin{table}[h]
\centering
    \begin{tabular}{ l l l}
\hline
 \textbf{Coupling} & $\boldsymbol{\mathrm{LCF\ 3\ ab}^{-1}@\ 250\ \mathrm{GeV}}$ & $\boldsymbol{+\ \mathrm{XCC\ 0.5\ ab}^{-1}@\ 125\ \mathrm{GeV}}$ \\ \hline
 $HZZ$ & 0.34 & 0.34 \\
   $HWW $    & 0.34 & 0.34 \\
   $Hbb $ & 0.72    & 0.41 \\
   $H\tau\tau $  &  0.83  & 0.50 \\
   $Hgg $   & 1.31  & 0.53 \\
   $Hcc $  &  1.45  & 0.74 \\
   $H\gamma\gamma $  &  1.02  & 0.22 \\
   $H\gamma Z $  & 7.51    & 3.1 \\
   $H\mu\mu $   & 3.87   & 3.1 \\ 
   $Hss $  &   -   & 13.0 \\
 \hline
   $\Gamma_\mathrm{tot} $    & 1.39 &  1.02 \\
   $\Gamma_\mathrm{inv} $ (95\% CL)    & 0.36 & 0.36 \\
   $\Gamma_\mathrm{unclassified} $ (95\% CL)    & 1.53 & 1.48 \\   
 \hline
\end{tabular}
\caption{Projected uncertainties in the Higgs boson couplings, in percent, for LCF at 250 GeV from a leading order SMEFT fit~\cite{Barklow:2017suo} without and with the XCC $\sigma\times BR$ measurements included. HL-LHC results are included in the SMEFT fit.  The XCC luminosity corresponds to the integrated $\gamma\gamma$ luminosity defined in Section~\ref{sec:XCC}. \label{tab:couplingLCF}}
\end{table}

The $H \to \mu^+\mu^-$ channel achieves a significance of $S/\sqrt{S+B} \approx 9.1$ and a statistical precision of $11\%$, limited primarily by the small branching fraction ($\sim 2.2 \times 10^{-4}$) but benefiting from the exceptionally clean final state and narrow mass resolution. The three $H \to \gamma Z$ channels, reconstructed via the leptonic, hadronic, and invisible $Z$ decay modes, achieve precisions ranging from $\sim 10\%$ to $\sim 19\%$, consistent with the relatively small branching fraction of $H \to \gamma Z$ ($\sim 0.15\%$).

We translate the $\sigma \times \mathrm{Br}$ measurements into projected constraints on the Higgs boson couplings using the $\kappa$-framework~\cite{LHCHiggsCrossSectionWorkingGroup:2012nn} and leading-order SMEFT fits~\cite{Barklow:2017suo}. Table~\ref{tab:couplingILC} presents projected uncertainties on the Higgs boson couplings from a $\kappa$-framework fit combining the expected HL-LHC results from ATLAS and CMS (with $3~\mathrm{ab}^{-1}$ each at $\sqrt{s} = 14$~TeV)~\cite{LHCHiggsCrossSectionWorkingGroup:2012nn} with the XCC $\sigma \times \mathrm{Br}$ measurements at $\mathcal{L}_{\rm Compton} = 0.5~\mathrm{ab}^{-1}$. The fit assumes the Higgs branching fraction to BSM final states is fixed to zero. The XCC delivers transformative improvements over the HL-LHC alone across all couplings. Most notably, the $H\gamma\gamma$ coupling precision improves from $1.8\%$ at the HL-LHC to $0.095\%$ with the XCC, an improvement by a factor of $\sim 19$, owing to the resonant $s$-channel production that directly probes the $H\gamma\gamma$ vertex. The $Hgg$ coupling improves from $2.4\%$ to $0.67\%$, the $Hbb$ coupling from $3.6\%$ to $0.71\%$, and the $HWW$ and $HZZ$ couplings from $1.6\%$ to $0.79\%$ and $0.68\%$, respectively. The $H\tau\tau$ coupling reaches $0.78\%$, improving upon the HL-LHC projection of $1.9\%$ by more than a factor of two. The $H\gamma Z$ coupling, projected at $6.8\%$ by the HL-LHC, reaches $3.1\%$ at the XCC. The $H\mu\mu$ coupling improves from $3.0\%$ to $2.8\%$, a more modest gain reflecting the statistics-limited nature of this channel for XCC versus the higher-statistics measurement at HL-LHC.  Crucially, the XCC provides first access to the $Hcc$ and $Hss$ couplings at $1.0\%$ and $13\%$, respectively.

\begin{table}[h]
\centering
    \begin{tabular}{ l l l}
\hline
 \textbf{Coupling} & $\boldsymbol{\mathrm{FCCee}\ @\ 240\ \&\ 365\ \mathrm{GeV}}$ & $\boldsymbol{+\ \mathrm{XCC}\ 0.5\ \mathrm{ab}}^{-1}@\ 125\ \mathrm{GeV}$ \\ \hline
 $HZZ$ & 0.19 & 0.16 \\
   $HWW $    & 0.19 & 0.15 \\
   $Hbb $ & 0.32    & 0.23 \\
   $H\tau\tau $  &  0.42  & 0.31 \\
   $Hgg $   & 0.42  & 0.32 \\
   $Hcc $  &  0.79  & 0.55 \\
   $H\gamma\gamma $  &  1.0  & 0.088 \\
   $H\gamma Z $  & 4.1    & 2.5 \\
   $H\mu\mu $   & 3.9   & 3.2 \\ 
   $Hss $  &  58.0   & 13.0 \\
   $\Gamma_\mathrm{tot} $    & 0.50 &  0.38 \\  
 \hline
\end{tabular}
\caption{Projected uncertainties in the Higgs boson couplings, in percent, for FCC-ee at 240 \& 365 GeV from a leading order SMEFT fit~\cite{Barklow:2017suo} without and with XCC $\sigma\times BR$ measurements included. It is assumed in the SMEFT ftts that there are no invisible or exotic or Higgs decays. Expected HL-LHC $\sigma\times \mathrm{BR}$ results \cite{deBlas:2022ofj} are included in the SMEFT fits.
The assumed luminosity for FCCee is 10.8~ab$^{-1}$ at 240~GeV and 3.12~ab$^{-1}$ at 365~GeV with expected FCCee $\sigma\times \mathrm{BR}$ precision from~\cite{Selvaggi:2025kmd}. The FCCee Higgs couplings obtained with the SMEFT fit agree well with the couplings quoted in the European Strategy Briefing Book \cite{deBlas:2025gyz} and the 2025 FCC Feasibility Study Report \cite{FCC:2025lpp}.   The XCC luminosity corresponds to the integrated $\gamma\gamma$ luminosity defined in Section~\ref{sec:XCC}. \label{tab:couplingfcc}}
\end{table}

Table~\ref{tab:couplingLCF} compares the projected Higgs coupling uncertainties from a leading-order SMEFT fit~\cite{Barklow:2017suo} for a 250~GeV linear collider facility (LCF) operating at $3~\mathrm{ab}^{-1}$, with and without the addition of XCC $\sigma \times \mathrm{Br}$ measurements. HL-LHC results are included in both fits. The LCF alone, through the recoil-mass measurement in $e^+e^- \to ZH$, provides model-independent access to the $HZZ$ coupling at $0.34\%$ and a total width determination at $1.39\%$. The XCC complements the LCF in several important ways. The $Hbb$ coupling improves from $0.72\%$ to $0.41\%$, reflecting the much larger $H \to b\bar{b}$ event yield at the XCC due to the absence of the associated $Z$ boson. Similarly, $H\tau\tau$ improves from $0.83\%$ to $0.50\%$, $Hgg$ from $1.31\%$ to $0.53\%$, and $Hcc$ from $1.45\%$ to $0.74\%$. The $H\gamma\gamma$ coupling sees a dramatic improvement from $1.02\%$ to $0.22\%$, again owing to the direct sensitivity of the $\gamma\gamma \to H$ production cross section to this vertex. The total width determination improves from $1.39\%$ to $1.02\%$. The $HZZ$ and $HWW$ couplings remain at $0.34\%$, as these are primarily constrained by the LCF recoil measurement. The XCC additionally provides unique access to $Hss$ at $13\%$, a measurement that is traditionally considered challenging at the LCF. These results demonstrate that the XCC and a 250~GeV $e^+e^-$ collider are highly complementary, with the combination yielding a Higgs coupling program that substantially exceeds the reach of either machine alone.

Table~\ref{tab:couplingfcc} presents the analogous comparison for FCC-ee operating at $\sqrt{s} = 240$ and $365$~GeV, with integrated luminosities of $10.8~\mathrm{ab}^{-1}$ and $3.12~\mathrm{ab}^{-1}$, respectively, using expected $\sigma \times \mathrm{Br}$ precisions from Ref.~\cite{Selvaggi:2025kmd}. The SMEFT fit includes HL-LHC inputs and assumes no invisible or exotic Higgs decays. The FCC-ee coupling projections obtained here agree well with those quoted in the European Strategy Briefing Book~\cite{deBlas:2022ofj} and the 2025 FCC Feasibility Study Report~\cite{FCC:2025lpp}.
 
Even against the formidable baseline of FCC-ee, the XCC provides significant improvements. The $Hbb$ coupling improves from $0.32\%$ to $0.23\%$, $H\tau\tau$ from $0.42\%$ to $0.31\%$, $Hgg$ from $0.42\%$ to $0.32\%$, and $Hcc$ from $0.79\%$ to $0.55\%$. The most striking gain is again in $H\gamma\gamma$, which improves from $1.0\%$ to $0.088\%$, a factor of $\sim 11$ improvement, making the XCC the most powerful probe of the $H\gamma\gamma$ coupling among all proposed facilities. The $H\gamma Z$ coupling improves from $4.1\%$ to $2.5\%$, and the total width from $0.50\%$ to $0.38\%$, representing a $\sim 24\%$ relative improvement. The $HZZ$ and $HWW$ couplings see modest improvements from $0.19\%$ to $0.16\%$ and $0.15\%$, respectively, as the FCC-ee recoil measurement is already highly constraining. The $Hss$ coupling, projected at $58\%$ by FCC-ee alone, dramatically improves to $13\%$ with the XCC, highlighting the unique sensitivity of a $\gamma\gamma$ collider to this elusive coupling through the charge-suppressed background mechanism. These results establish that the XCC provides a powerful and complementary addition to even the most ambitious circular $e^+e^-$ collider programs.

\section{Discussion and Conclusion}\label{sec:conc}

We have presented a comprehensive analysis of single Higgs boson production in $\gamma\gamma$ collisions at $\sqrt{s} = 125$~GeV at the XFEL Compton $\gamma\gamma$ collider, targeting all major decay modes, including the nearly traditionally inaccessible $H \to s\bar{s}$ channel. This constitutes, to our knowledge, the most complete single-Higgs analysis for any proposed $\gamma\gamma$ collider concept and the first to employ transformer-based deep learning for event classification at such a facility.

The XCC's near-monochromatic luminosity spectrum, achieved through Compton back-scattering at very large $x$ values ($x \geq 1000$), concentrates the collision energy near the Higgs resonance mass, yielding approximately 1.1 million $\gamma\gamma \to H$ events in a 10-year run. The resulting signal topology with energetic, central Higgs decay products unaccompanied by an associated vector boson is considerably cleaner than Higgsstrahlung. A key result of this work is the demonstration that $\gamma\gamma$ collider backgrounds, historically cited as a limitation of photon colliders, can be effectively mitigated through a combination of this near-monochromatic luminosity spectrum and the set transformer architecture with a genetic algorithm optimizer for cut optimization.

Our parallel analysis at an optical $\gamma\gamma$ collider, conducted with an identical simulation and analysis chain, shown in Appendix \ref{app:OCCANA}, directly quantifies the XCC improvement. For $H \to b\bar{b}$, the XCC achieves $\delta(\sigma \times \mathrm{Br}) \approx 0.18\%$ compared to $\approx 0.4\%$ at the OCC, nearly a factor of two improvement driven by the $\sim 5\times$ larger Higgs yield. For $H \to c\bar{c}$ the improvement is a factor of $\sim 3.5$ ($1.4\%$ vs.\ $5\%$), and for $H \to gg$ a factor of $\sim 2$ ($0.70\%$ vs.\ $1.5\%$). These comparisons confirm that the XCC paradigm fundamentally transforms the physics case for $\gamma\gamma$ Higgs factories relative to earlier optical collider concepts.

Our $e^+e^- \to ZH(\to b\bar{b})$ analysis at $\sqrt{s} = 250$~GeV (Appendix~\ref{app:eeZH}), performed with the same simulation and ML workflow, yields $\delta[\sigma \cdot \mathrm{Br}(H \to b\bar{b})] \approx 0.84\%$, consistent with ILD full-simulation studies. The XCC achieves $0.18\%$ for the same decay mode, which is a factor of $\sim 4.7$ improvement arising from the larger signal yield, the absence of an associated $Z$ boson, and the clean dijet topology at $\sqrt{s} = m_H$.

The coupling fits in Tables~\ref{tab:couplingILC}--\ref{tab:couplingfcc} demonstrate that the XCC provides leading sensitivity in several key couplings. The $H\gamma\gamma$ coupling reaches $0.095\%$ ($0.088\%$) in combination with the HL-LHC (FCC-ee) is an order of magnitude better than any other proposed facility and sensitive to BSM physics at scales well above 10~TeV through virtual corrections to the $H\gamma\gamma$ loop. The fermionic couplings $Hbb$, $H\tau\tau$, $Hcc$, and $Hgg$ are all substantially improved by adding XCC data to either a linear collider or FCC-ee. The $Hss$ coupling, measured at $13\%$, provides the best direct probe of the Higgs-to-strange Yukawa coupling of any collider.

While this study represents a substantial advancement for future $\gamma\gamma$ Higgs factories, it employs fast detector simulation via \textsc{Delphes}, which cannot fully capture the effects of enhanced incoherent $e^+e^-$ pair production at the XCC. As such, full \textsc{Geant4}-based simulation is required to validate detector occupancies and flavor-tagging performance. 

A full detector simulation incorporating realistic beam-induced backgrounds is essential to validate these projections and determine achievable flavor-tagging performance, as is a detailed study of systematic and theory uncertainties.
The exploitation of beam polarization observables for direct measurement of Higgs CP properties in the $H\gamma\gamma$ vertex, and the inclusion of differential $\sigma\times BR$ measurements within an EFT framework, represent unique physics opportunities that merit dedicated study. The forward acceptance cut ($|\cos\theta| < 0.95$) is also conservative; dedicated algorithms, planned as future work, are expected to extend reconstruction to $|\cos\theta| < 0.98$.

In summary, this study demonstrates that the XCC concept, combining XFEL technology with modern machine learning techniques, offers a Higgs physics program of extraordinary precision that substantially improves upon and complements proposed $e^+e^-$ Higgs factories. The unique access to the $H\gamma\gamma$ vertex and $Hss$ coupling establishes the XCC as a compelling component of the future collider landscape.

\acknowledgments

The authors express their gratitude to Gudrid Moortgat-Pick, Marten Berger, Caterina Vernieri, Dong Su, and Michael Peskin for insightful discussions and useful feedback on the manuscript. Additional thanks are extended to Dimitris Ntounis for providing the \textsc{Delphes} SiD configurations used in earlier iterations of this study. This work used the resources of the SLAC Shared Science Data Facility (S3DF) at SLAC National Accelerator Laboratory. SLAC is operated by Stanford University for the U.S. Department of Energy's Office of Science. This work is sponsored by the U.S. Department of Energy, Office of Science under Contract No. DE-AC02-76SF00515.


\appendix
\section{Detector Dependence of Heavy-Flavor Tagging at the XCC}
\label{app:XCCFTAG}
While the large $x$ values of the XCC give it the characteristic nearly-monochromatic luminosity distribution that substantially improves the physics case over OCCs, it also results in considerably increased beam-induced background. In particular incoherent $e^+e^-$ pair production (IPP) from Bethe–Heitler $\gamma\gamma^* \to e^+e^-$, Breit–Wheeler $\gamma \gamma \to e^+e^-$, and Landau–Lifshitz $\gamma^*\gamma^* \to e^+e^-$ processes constitute a significant background. While full detector simulation studies are required to accurately determine the occupancy induced by this background and thus the compatibility of existing SiD and IDEA detector designs, the worst-case end result will be to push the beam pipe and innermost vertex detector to larger radii to maintain acceptable levels of occupancy as determined by studies at the ILC. Since the innermost vertex detector is predominantly responsible for vertexing resolution, which is a crucial specification for heavy-flavor tagging, one expects the performance of flavor tagging to degrade as a result.

Accurate flavor-tagging of jets originating from the hadronic decays of the Higgs boson is central to our study, we therefore study the extent of this degradation using the latest developments in \textsc{Delphes}. In particular, we leverage the \textsc{DetectorGeometry} and \textsc{TrackCovariance} modules in \textsc{Delphes}, similar to Sec.~\ref{sec:sampandsims}. For this study, we use the \textsc{ParticleNet} algorithm and evaluate the flavor tagging performance for the detector SiD concept, as a function of the radial location of innermost vertex detector, which we refer to as simply the ``inner radius'' for brevity henceforth. 

Samples are simulated considering the $\gamma\gamma \to H \to j\overline{j}$ processes, where $j = b, c, g, q$, with $q=\{u,d,s\}$ considering a center-of-mass-energy of $\sqrt{s}=125$ GeV using the same workflow described in Sec~\ref{sec:sampandsims}. For each jet flavor, we generate samples corresponding to inner radii of $r=14\,\mathrm{mm}, 15\,\mathrm{mm}, 16\,\mathrm{mm}, 17\,\mathrm{mm}$, and 18\,mm. The beam pipe location is also varied accordingly, maintaining a 1~mm clearance between the beam pipe and the innermost vertex layer.

\begin{figure}
    \centering
    \begin{minipage}{0.495\textwidth}
        \includegraphics[width=\linewidth]{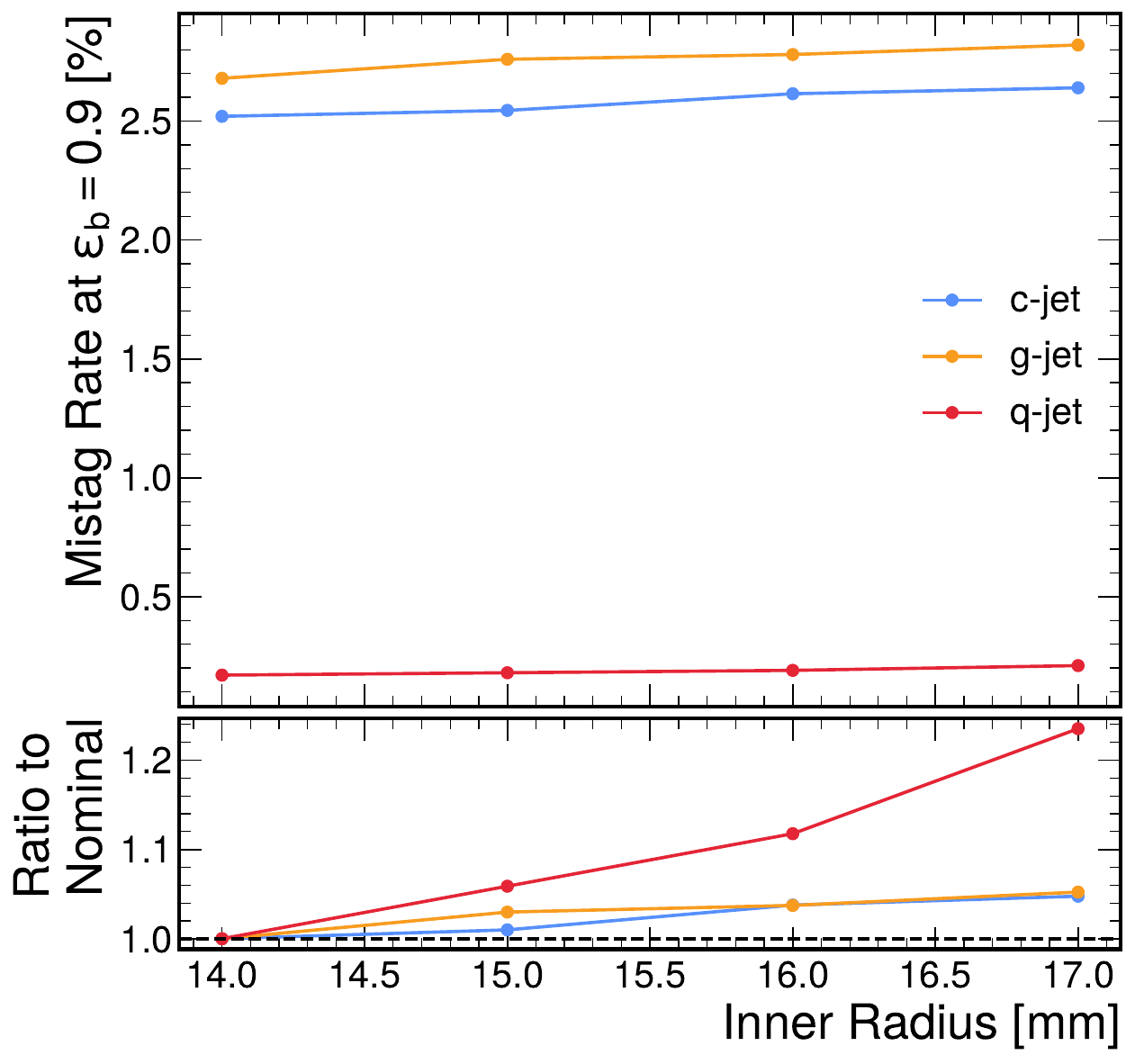}
    \end{minipage}
    \begin{minipage}{0.495\textwidth}
        \includegraphics[width=\linewidth]{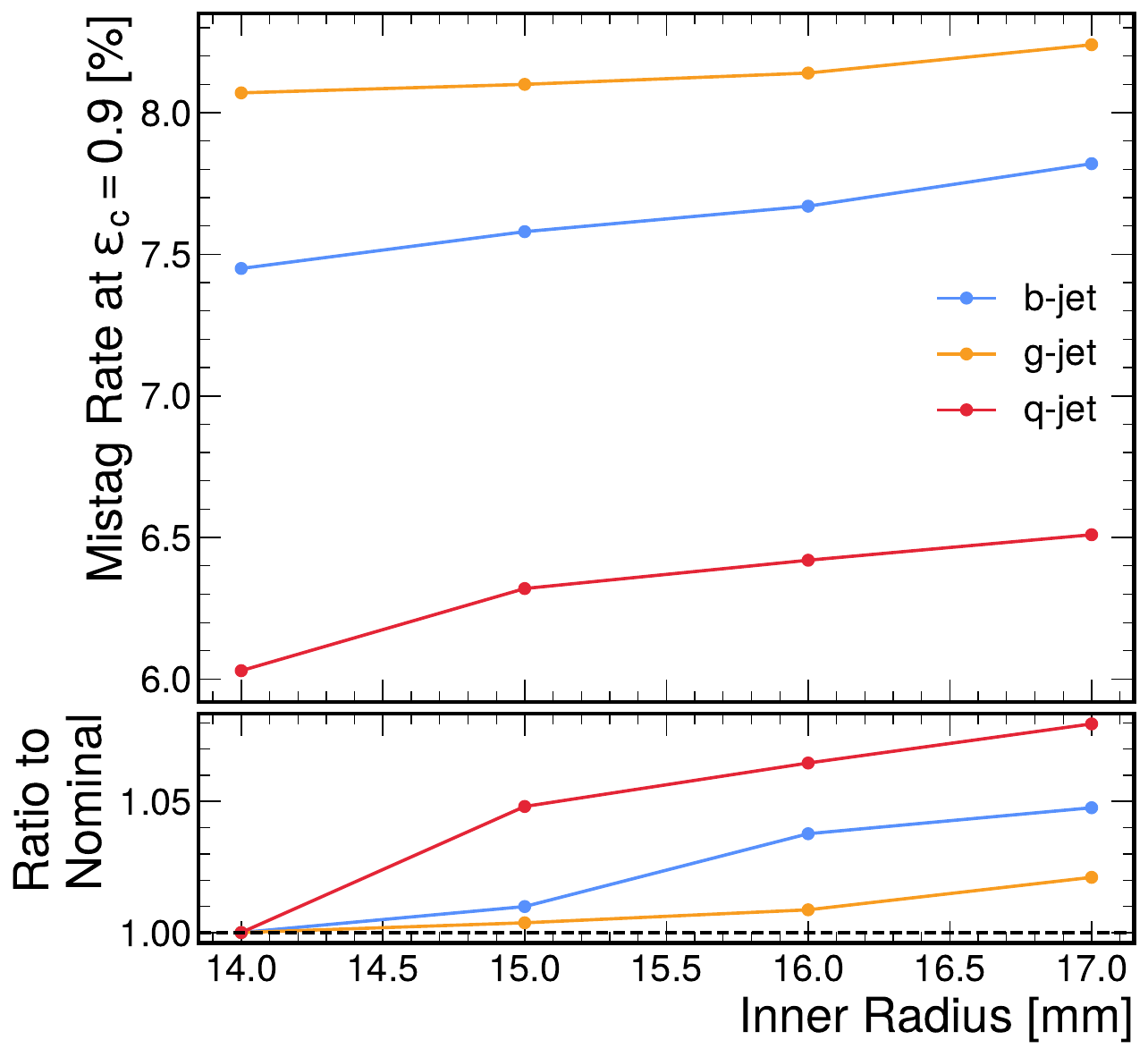}
    \end{minipage}
    \caption{Flavor tagging performance of the ParticleNet algorithm on the SiD detector concepts. The mistag rates for $b$ $(c)$ (blue), $g$ (orange), and $q$ (red) jets are shown on at a fixed $b$ $(c)$-jet efficiency of $90\%$ on the left (right) panels.}
    \label{fig:bc_tag_XCC}
\end{figure}

Fig.~\ref{fig:bc_tag_XCC} presents the flavor tagging performance of the \textsc{ParticleNet} algorithm on the SiD detector concept as a function of the inner radius. The left panel shows the mistag rates for $c$-jets (blue), $g$-jets (orange), and light quark $q$-jets (red) at a fixed $b$-jet tagging efficiency of $\varepsilon_b = 90\%$, while the right panel shows the corresponding mistag rates for $b$-jets, $g$-jets, and $q$-jets at a fixed $c$-jet tagging efficiency of $\varepsilon_c = 90\%$. Below each panel, the ratio to the nominal inner radius ($r = 14$~mm) is displayed.
 
For $b$-tagging at $\varepsilon_b = 90\%$ (left panel), the $c$-jet mistag rate increases from approximately $2.5\%$ at $r = 14$~mm to approximately $2.7\%$ at $r = 17$~mm, representing roughly $8\%$ degradation. The $g$-jet mistag rate rises from ${\sim}2.6\%$ to ${\sim}2.8\%$ over the same range, while the $q$-jet mistag rate increases from $\sim 0.15\%$ to ${\sim}0.21\%$. These increases, while measurable, remain modest in absolute terms. The ratio plot shows that the degradation is approximately $20$--$25\%$ at $r = 17$~mm relative to the nominal configuration for all three mistag categories. Most importantly, even at $r = 17$~mm, the inner radius adopted for the nominal XCC detector configuration in this study, the mistag rates remain substantially below those achieved at the LHC, thus validating the conservative nature of the flavor-tagging assumptions employed throughout our analysis.
 
For $c$-tagging at $\varepsilon_c = 90\%$ (right panel), the overall mistag rates are naturally higher than for $b$-tagging, reflecting the inherently more challenging nature of charm jet identification. The $b$-jet mistag rate is relatively stable at approximately $7.5$--$8.0\%$ across the full range of inner radii, showing less than $5\%$ degradation from $r = 14$~mm to $r = 17$~mm. The $g$-jet and $q$-jet mistag rates at the $c$-tagging working point show a mild increase of $\sim 3$--$5\%$ over the range studied.

These results have two important implications. First, they validate the feasibility of the XCC detector concept with a slightly enlarged inner radius ($r = 17$~mm) relative to the nominal SiD design ($r = 14$~mm), demonstrating that the flavor-tagging performance degradation is manageable and well within the margin accommodated by our conservative working point assumptions. Second, they indicate that the actual performance of the \textsc{ParticleNet} algorithm at $r = 17$~mm is substantially better than the ``Medium'' and ``Loose'' working points adopted in this study, which were intentionally degraded to approximate worst-case LHC-like mistag rates. Specifically, at the nominal $r = 17$~mm configuration and $\varepsilon_b = 85\%$, the observed mistag rates are $< 1\%$ for $c$-jets and $< 0.5\%$ for light quark jets, compared to the $5\%$ and $1\%$ mistag rates assumed in our analysis. This conservative margin ensures that our projections are robust against potential further degradation from IPP-induced occupancy effects that can only be fully quantified with a complete \textsc{Geant4}-based detector simulation, which is the subject of ongoing work.

\section{\texorpdfstring{Analysis of Hadronic \boldmath $\gamma\gamma \to H $ with an Optical $\gamma\gamma$ Collider}{}}
\label{app:OCCANA}
To fully demonstrate the improvements afforded by XCC over existing optical $\gamma\gamma$ collider concepts of yesteryear, we present analyses targeting all major hadronic decay modes of the Higgs boson. While there exist a myriad of $\gamma\gamma \to H \to b\overline{b}$ analyses, none, to our knowledge, use machine learning to the same sophistication as we have in our XCC study. Therefore, to provide a fair comparison, we use the same analysis workflow for an optical $\gamma\gamma$ collider.

As before, we simulate beam–beam and non-linear QED effects for $\gamma \gamma$, $e^\pm \gamma$, and $e^\pm e^\mp$ using \textsc{Cain}. For optical wavelengths, we use $x=4.82$, $2P_c\lambda_e = -0.9 $, the photon waist radius $a_\gamma = 8.6\, \mu\mathrm{m}$, and the longitudinal electron (photon) r.m.s.~$\sigma_{ez} =20\, \mu\mathrm{m} $ $(\sigma_{\gamma z}=  300\, \mu\mathrm{m})$. The rest of the simulation chain is identical to the XCC, with the exception that instrumentation is not limited to $\abs{\cos \theta} < 0.95$ due to the absence of a large Compton background. Tracking and calorimetry are instead restricted to the nominal XCC values of $\abs{\cos \theta} < 0.984$ for tracking and $\abs{\cos \theta} < 0.995$ for calorimetry. For backgrounds, we consider the $\gamma\gamma \to qq\overline{qq}$, $\gamma\gamma \to q\overline{q}$, $\gamma e \to q \overline{q} e$, $\gamma e \to q\overline{q} \nu $, and $e^+e^-\to q\overline{q}$ processes.

\subsection{\texorpdfstring{\boldmath $H\to b\overline{b}$}{}}

For the $H\to b\overline{b}$ channel, we require two $b$-tagged jets using the same working point for the \textsc{ParticleNet} algorithm as the XCC. In addition, we veto events with an isolated lepton. Further, we require modest missing energy with $E_T^\mathrm{miss} < 18$ GeV. Finally, we require the $b$-dijet system to be consistent with Higgs decays, so we enforce cuts on Durham distance $\sqrt{d_{12}}> 75$ GeV and dijet invariant mass $m(jj) > 100$. Events passing these preselections are passed to the set transformer-genetic algorithm combination as described in Sec.~\ref{sec:MLAnalysis}.

For the optical $\gamma\gamma \to H \to b\overline{b}$ analysis, the pre-selection retains $S = 132{,}935$ of the initial $231{,}200$ signal events, while the continuum backgrounds are reduced from a total of $B = 60{,}833{,}246$ events before preselection to $B = 1{,}108{,}449$ after pre-selection. Imposing the GA-optimized cuts on the set-transformer outputs further tightens the selection, keeping $S = 87{,}307$ signal events (about two thirds of the preselected sample) while suppressing the backgrounds by more than an order of magnitude to a total of $B = 31{,}699$ events. The final selection thus achieves $S/B \simeq 2.8$ and a signal significance of $S/\sqrt{S+B} \simeq 250$, corresponding to a projected relative statistical precision on $\sigma(\gamma\gamma \to H)\cdot \mathrm{Br}(H\to b\overline{b})$ of $\approx 0.4\%$ in the 10-year optical photon–collider scenario.

\begin{table}
\centering
\begin{tabular}{l l l l}
\hline
\textbf{Process} & \textbf{Before Preselection} & \textbf{After Preselection} & \textbf{After GA} \\ \hline
$\gamma\gamma \to H \to b\overline{b} $ & 231{,}200 & 132{,}935 &87,307 \\ \hline
$\gamma\gamma \to q\overline{q}$                    & 15{,}439{,}063 & 30{,}023  & 13,946 \\
$\gamma e \to q\overline{q}e$                     & 17{,}692{,}550 & 289{,}857 & 9,688 \\
$\gamma e \to q\overline{q}\nu$                   & 19{,}896{,}867 & 107{,}857 & 3,672 \\
$e^+e^- \to q\overline{q}$                          & 7{,}804{,}766  & 680{,}712 & 4,393 \\ \hline
\end{tabular}
\caption{Number of events before and after the optical $\gamma\gamma \to H\to b\overline{b}$  pre-selection filters for the signal and dominant backgrounds, and after the genetic algorithm cuts.  \label{tab:optical_H_bb_selection}}
\end{table}

\subsection{\texorpdfstring{\boldmath $H\to c\overline{c}$}{}}
The pre-selections for the $H\to c\overline{c}$ remain largely unchaged from the XCC, with two $c$-tagged jets following the same ``Loose'' working point of the \textsc{ParticleNet} algorithm, modest missing energy with $E_T^\mathrm{miss} < 18$ GeV and a lepton veto. Selected events are vetted with a Higgs mass resonance cut requiring $m(jj) > 100$ GeV and a Durham distance $\sqrt{d_{12}}<75$ GeV. As shown in Table \ref{tab:optical_H_cc_selection}, the preselection filters retain $S=8{,}255$ signal events (about $69\%$ of the initial signal) while reducing the background to $B=1{,}918{,}943$. Applying the GA-optimized cuts on the set transformer outputs, keeps $S=3{,}797$ signal events (roughly $46\%$ of the preselected signal and $32\%$ of the original) and reducing the background to $B=36{,}360$ (a factor $\sim 53$ reduction relative to the preselected sample). The final selection thus achieves $S/B\simeq 0.10$ and a significance $S/\sqrt{S+B}\simeq 19$, corresponding to a statistical precision on $\sigma(\gamma\gamma\to H)\times\mathrm{Br}(H\to c\overline{c})$ of $\delta(\sigma\times\mathrm{Br})\approx 1/19\simeq 5\%$ in the 10-year running scenario. 

\begin{table}
\centering
\begin{tabular}{l r r r}
\hline
\textbf{Process} & \textbf{Before Preselection} & \textbf{After Preselection} & \textbf{After GA} \\ \hline
$\gamma\gamma \to H \to c\overline{c}$  & 12{,}000       & 8{,}255   & 3,797 \\ \hline
$\gamma\gamma \to qq\overline{qq}$                  & 771{,}803      & 10{,}848  & 312 \\
$\gamma\gamma \to q\overline{q}$       & 15{,}439{,}063 & 631{,}957 & 28,159 \\
$\gamma e \to q\overline{q}e$          & 17{,}692{,}550 & 428{,}814 & 3,632 \\
$\gamma e \to q\overline{q}\nu$        & 19{,}896{,}867 & 163{,}766 & 1,611 \\
$e^+e^- \to q\overline{q}$             & 7{,}804{,}766   & 683{,}558 & 2,646 \\ \hline
\end{tabular}
\caption{Number of events before and after the optical $\gamma\gamma \to H\to c\overline{c}$ pre-selection filters for the signal and dominant backgrounds, and after the genetic algorithm cuts. \label{tab:optical_H_cc_selection}}
\end{table}

\subsection{\texorpdfstring{\boldmath $H\to gg$}{}}
For the $H\to gg$ channel, we target final states with two jets, moderate missing transverse energy with $E_T^\mathrm{miss} < 18$ GeV, and an isolated lepton veto. Heavy-flavor contamination is removed with a veto on $(b/c)$-jets. To suppress backgrounds with light quark jets, we require $N_\mathrm{track}> 25$. Finally, we impose a dijet invariant mass cut of $m(jj)>100$ and Durham distance $\sqrt{d_{12}}>75$ GeV. 

For the optical $\gamma\gamma \to H\to gg$ selection, applying the pre-selection, the signal is reduced to $S = 23{,}214$ events while the background is suppressed to $B = 2.78\times 10^6$, improving the purity to $S/B \simeq 8.4\times 10^{-3}$ and the significance to $S/\sqrt{S+B} \simeq 13.9$. The final genetic–algorithm cuts on the set–transformer outputs retain $S = 16{,}049$ signal events against $B = 40{,}948$ background events, yielding a much cleaner sample with $S/B \simeq 0.39$ and a signal significance of $S/\sqrt{S+B} \simeq 67$, which corresponds to a projected statistical precision on $\sigma(\gamma\gamma\to H)\times\mathrm{Br}(H\to gg)$ of about $1.5\%$.

\begin{table}
\centering
\begin{tabular}{l r r r}
\hline
\textbf{Process} & \textbf{Before Preselection} & \textbf{After Preselection} & \textbf{After GA} \\ \hline
$\gamma\gamma \to H \to gg$                               & 34{,}400       & 23{,}214       & 16,049 \\ \hline
$\gamma\gamma \to qq\overline{qq}$                                     & 771{,}803      & 80{,}569       & 2,067 \\
$\gamma\gamma \to q\overline{q}$                          & 15{,}439{,}063 & 332{,}609      & 14,556\\
$\gamma e \to q\overline{q}e$                             & 17{,}692{,}550 & 1{,}084{,}359  & 1,054 \\
$\gamma e \to q\overline{q}\nu$                           & 19{,}896{,}867 & 597{,}701      & 5,110 \\
$e^+e^- \to q\overline{q}$                                & 7{,}804{,}766  & 680{,}712      & 18,161 \\ \hline
\end{tabular}
\caption{Number of events before and after the optical $\gamma\gamma \to H\to gg$ pre-selection filters for the signal and dominant backgrounds, and after the genetic algorithm (GA) cuts. \label{tab:optical_H_gg_selection}}
\end{table}

\section{\texorpdfstring{Analysis of  \boldmath $e^+e^- \to ZH(\to b\overline{b})$ with a Linear $e^+e^-$ Collider}{}}
\label{app:eeZH}
Existing Higgs analyses for ILD and other $e^+e^-$ linear collider concepts almost exclusively rely on full detector simulation rather than a parameterized fast simulation employed in our study. Frequently, the aforementioned full simulation studies include a more thorough consideration of physics effects such as beam-induced backgrounds and detector response than what is possible with fast simulation. Thus, in an effort to provide a more direct comparison, we report an $e^+e^-\to ZH(\to b \overline{b})$ analysis performed using a near-identical setup to our XCC analyses to better isolate genuine physics differences between an $e^+e^-$ linear collider and the XCC. We consider all possible decays of the $Z$ boson and compare our results to full simulation ILD studies for consistency and to our XCC numbers for performance comparisons.

As before, signal and background samples are produced using \textsc{Whizard} for all matrix-element and parton-level event generation. We use the official ILD Set-A set of $e^+e^-$ beam spectra for $\sqrt{s}=250$ GeV operation, generated using \textsc{GuineaPig}, which is converted to the \textsc{Circe2} format and subsequently interfaced with \textsc{Whizard} for convolution. Parton-level events are then passed to \textsc{Pythia6} for parton-showering and hadronization. The production cross-sections for $e^+e^- \to Z (\to q\overline{q})H (\to b \overline{b})$, $e^+e^- \to Z (\to \ell\overline{\ell})H (\to b \overline{b})$, and $e^+e^- \to Z (\to \nu \overline{\nu})H (\to b \overline{b})$ are 154.7 fb, 29.4 fb, and 58.1 fb respectively. The dominant backgrounds for an $e^+e^- \to ZH$ analysis the are the $e^+e^-\to ZZ$ and $WW$ processes. Therefore, we consider backgrounds with final states ending in $\nu \overline{\nu} q\overline{q}$, $\nu\overline{ \ell} q\overline{q}$, $\ell \overline{\ell} q\overline{q}$, $\nu \overline{\nu} \ell \overline{\ell} $, $q q \overline{qq}$, and $\ell \ell \overline{\ell\ell}$. In addition we also consider the $e^+e^-\to q\overline{q}$ process.

Fast detector simulation is finally performed with \textsc{Delphes} using the SiD detector geometry and parameters, identical to the XCC. Unlike the XCC and similar to the OCC, instrumentation is not limited to $\abs{\cos \theta} < 0.95$ due to the absence of a large Compton background. Tracking is instead restricted to $\abs{\cos \theta} < 0.984$ and calorimetry to $\abs{\cos \theta} < 0.995$. Pileup from additional $\gamma\gamma \to \mathrm{hadrons}$ interactions due to Bremsstrahlung and Beamstrahlung photons is negligible for $\sqrt{s}=250$ GeV operation, particularly for the $H\to b\overline{b}$ channel, and is not considered for this study. Finally, jets are clustered using the $e^+e^-$ inclusive anti-$k_\mathrm{T}$ algorithm using $R=1.5$ and $E>8$ GeV with \textsc{FastJet}. 

\subsection{\texorpdfstring{$\boldsymbol{e^+e^- \to Z (\to q\overline{q})H (\to b \overline{b})}$}{}}
We enact preselection criteria similar to the OCC and XCC analyses. Jets are clustered with the Durham algorithm requiring $n_\mathrm{jets}=4$, of which we require at least 2 to be $b$-tagged. We veto events with isolated, well-identified leptons with $E>10$ GeV to remove leptonic and semi-leptonic backgrounds. We require modest $E_T^\mathrm{miss} < 15$ GeV to remove backgrounds with prompt neutrinos. The Higgs dijet candidate is formed by considering the $b$-jet pair with the highest dijet invariant mass, which is required to be consistent with Higgs decays i.e.~$m(bb) > 95$ GeV. Events passing the preselection criteria are passed to the machine learning and genetic algorithm workflow for further signal-background discrimination and signal significance computation as detailed in Sec.~\ref{sec:MLAnalysis}. The signal and dominant background counts before preselection, after preselection, and after the genetic algorithm cuts are listed in Table \ref{tab:ee_zh_qq_selection}. After all cuts, the signal yield is $S = 14{,}168$ events, while the total background is $B = 16{,}093$ events. This corresponds to a signal–to–background ratio of $S/B \simeq 0.88$ and an expected significance of $S/\sqrt{S+B} \simeq 81$. This leads to a statistical precision $\delta(\sigma\times\mathrm{Br}) \simeq 1.2\%$ in a 10-year scenario.

\begin{table}
\centering
\begin{tabular}{ l l l l}
\hline
\textbf{Process} & \textbf{Before Preselection} & \textbf{After Preselection} & \textbf{After GA} \\ \hline
$e^+e^- \to ZH \to q\overline{q}\,b\overline b$                            & 38,675    & 16,976 & 14,168 \\ \hline
$e^+e^- \to \ell \overline{\ell}\,q\overline{q}$                & 452,973   & 690 & 15 \\
$e^+e^- \to q q \overline{q q}$                                 & 401,049   & 19,006 & 12,055 \\
$e^+e^- \to q\overline{q}$                                      & 1,962,744 & 32,244 &  4,023\\ \hline
\end{tabular}
\caption{Number of events before and after the $e^+e^- \to Z (\to q\overline{q})H (\to b \overline{b})$ pre-selection filters and the genetic algorithm cuts on the set transformer output distributions for the signal and dominant backgrounds. \label{tab:ee_zh_qq_selection}}
\end{table}

\subsection{\texorpdfstring{$\boldsymbol{e^+e^- \to Z (\to \nu\overline{\nu})H (\to b \overline{b})}$}{}}

\begin{table}
\centering
\begin{tabular}{ l l l l }
\hline
\textbf{Process} & \textbf{Before Preselection} & \textbf{After Preselection} & \textbf{After GA} \\ \hline
$e^+e^- \to ZH \to \nu\bar\nu\,b\bar b$                         & 14{,}530      & 6{,}801 & 6,693 \\ \hline
$e^+e^- \to \nu \overline{\nu}\,q\overline{q}$                  & 43{,}211      & 970 & 222 \\
$e^+e^- \to \ell \overline{\ell}\,q\overline{q}$                & 452{,}973     & 103 & 10 \\
$e^+e^- \to q q \overline{q q}$                                 & 401{,}049     & 403 & 3 \\
$e^+e^- \to q\overline{q}$                                      & 1{,}962{,}744 & 26{,}203 & 112 \\ \hline
\end{tabular}
\caption{Number of events before and after the $e^+e^- \to Z (\to \nu\overline{\nu})H (\to b \overline{b})$ pre-selection filters and the genetic algorithm cuts on the set transformer output distributions for the signal and dominant backgrounds. \label{tab:ee_zh_nunu_selection}}
\end{table}

For the invisible decays of the $Z$ boson in the $e^+e^- \to Z (\to \nu\overline{\nu})H (\to b \overline{b})$ channel, we target a large missing energy with a $b$-dijet topology. Jets are clustered with the Durham algorithm, with $n_\mathrm{jets}=2$, as such we require events to have exactly two $b$-tagged jets with a mass $m(bb)>95$ GeV to be consistent with Higgs decays. Further, we require large missing energy $E_\mathrm{T}^\mathrm{miss}>25$ GeV, characterized by the prompt neutrinos from the $Z\to \nu \overline{\nu}$ decay. Moreover, we veto events with isolated leptons. Table \ref{tab:ee_zh_nunu_selection} summarizes the final event counts before preselection, after preselection, and after the genetic algorithm cuts.

Starting from $14{,}530$ signal events before preselection in Table~\ref{tab:ee_zh_nunu_selection}, the final GA selection retains $S = 6{,}693$ signal events, while the total surviving background from $e^+e^- \to \nu\bar\nu,q\bar q$, $e^+e^- \to \ell\bar\ell,q\bar q$, $e^+e^- \to qq\bar q\bar q$, and $e^+e^- \to q\bar q$ is reduced to $B = 347$ events. This corresponds to a signal-to-background ratio of $S/B \simeq 19$ and a signal significance of $S/\sqrt{S+B} \simeq 79$. In turn, the relative statistical precision on the production cross section times branching ratio in this channel is projected to be $\delta(\sigma\times\mathrm{Br}) \approx 1.3\%$ over the 10-year $e^+e^-$ running scenario.

\subsection{\texorpdfstring{$\boldsymbol{e^+e^- \to Z (\to \ell\overline{\ell})H (\to b \overline{b})}$}{}}

\begin{table}
\centering
\begin{tabular}{ l l l l }
\hline
\textbf{Process} & \textbf{Before Preselection} & \textbf{After Preselection} & \textbf{After GA} \\ \hline
$e^+e^- \to ZH \to \ell \overline{\ell}\,b\overline b$                  & 7{,}348       & 3{,}440 & 2,805 \\ \hline
$e^+e^- \to \ell \overline{\ell}\,q\overline{q}$                & 452{,}973     & 5{,}620  & 1,015 \\
$e^+e^- \to q  \overline{q}$                                 & 401{,}049     & 145 & 84 \\ \hline
\end{tabular}
\caption{Number of events before and after the $e^+e^- \to Z (\to \ell\overline{\ell})H (\to b \overline{b})$ pre-selection filters and the genetic algorithm cuts on the set transformer output distributions for the signal and dominant backgrounds. \label{tab:ee_zh_ll_selection}}
\label{tab:ee_zh_ll_selection}
\end{table}

For the leptonic decays of the $Z$ boson in the $e^+e^- \to Z (\to \ell\overline{\ell})H (\to b \overline{b})$ channel, we target a di-lepton with a $b$-dijet topology. As before, we require events to have exactly two $b$-tagged Durham jets with a mass $m(bb)>95$ GeV. But now, we require modest missing energy $E_\mathrm{T}^\mathrm{miss}<15$ GeV. Further, we require two isolated leptons or two hadronically decaying taus $\tau_h$. Table \ref{tab:ee_zh_nunu_selection} summarizes the final event counts before preselection, after preselection, and after the genetic algorithm cuts.  

The preselection already provides a relatively clean sample, reducing the signal from $7{,}348$ to $3{,}440$ events while suppressing the inclusive $e^+e^- \to \ell\overline{\ell},q\overline{q}$ and $e^+e^- \to q\overline{q}$ backgrounds to $5{,}620$ and $145$ events, respectively. The subsequent GA-optimized cuts on the set transformer outputs further refine this sample, yielding a final signal yield of $S = 3{,}805$ events and a total post-selection background of $B = 1{,}099$ events, dominated by residual $e^+e^- \to \ell\overline{\ell},q\overline{q}$ with a small contribution from mis-tagged $e^+e^- \to q\overline{q}$. This corresponds to a signal-to-background ratio of $S/B \simeq 3.5$ and a signal significance of $S/\sqrt{S+B} \simeq 54.3$, which in turn implies a projected statistical precision on the $ZH$ production rate in this channel of $\delta\left(\sigma \times\mathrm{Br}\right) \approx 1.8\%$.

Combining all channels yields a signal significance $S/\sqrt{S+B}\approx 120$, yielding an error $\delta \left(\sigma (e^+e^-\to ZH)\cdot \mathrm{Br}(H\to b\overline{b})\right)\approx 0.84\%$. Similar to the OCC, our results are consistent with existing ILD analyses which quote a precision of $ \left(\sigma (e^+e^-\to ZH)\cdot \mathrm{Br}(H\to b\overline{b})\right) \sim 0.9{-}1.1\%$, reflecting a reasonable improvement, which is to be expected since we employ sophisticated machine learning in addition to a fast detector simulation; thus, validating our methodology.


\bibliographystyle{JHEP}
\bibliography{biblio.bib}






\end{document}